\documentclass[11pt]{article}
\pdfoutput=1

\usepackage[usenames,dvipsnames,table]{xcolor}
\usepackage[utf8]{inputenc}

\usepackage{jheppub} 

\usepackage{tikz-cd} 
\usepackage{circuitikz}
\usepgfmodule{shapes}
\usetikzlibrary{decorations.pathmorphing}
\usetikzlibrary{decorations.pathreplacing,decorations.markings,snakes}
\usetikzlibrary{backgrounds, arrows,calc,shapes,decorations.pathreplacing, automata,positioning}
\usetikzlibrary {arrows.meta}
\usepackage{cancel} 
\usepackage{xcolor}

\usetikzlibrary{shapes.misc}

\tikzset{cross/.style={cross out, draw=black, minimum size=5*(#1-\pgflinewidth), inner sep=0pt, outer sep=0pt},
cross/.default={2pt}}

\tikzset{snake it/.style={decorate, decoration=snake}}

\tikzset{mid arrow/.style={postaction={decorate,decoration={
        markings,
        mark = at position .55 with {\arrow[#1]{Straight Barb[width=5pt]}}
      }}}}
      
\tikzset{mid arrowsm/.style={postaction={decorate,decoration={
        markings,
        mark = at position .55 with {\arrow[#1]{Straight Barb[width=3pt]}}
      }}}}
      
\tikzset{middx arrowsm/.style={postaction={decorate,decoration={
        markings,
        mark = at position .7 with {\arrow[#1]{Straight Barb[width=3pt]}}
      }}}}
      
\tikzset{midsx arrowsm/.style={postaction={decorate,decoration={
        markings,
        mark = at position .4 with {\arrow[#1]{Straight Barb[width=3pt]}}
      }}}}

\newcommand{\wigE}{\draw[snake=zigzag]}

\newcommand{\wigM}{\draw[snake=coil,,segment aspect=0,segment length=6pt, double]}
\newcommand{\wigMc}{\draw[->,snake=coil,segment aspect=0,segment length=6pt]}
\newcommand{\wigMB}{\draw[->,snake=zigzag,segment aspect=0,segment length=6pt]}
\newcommand{\wigT}{\draw[dashed]}

\newcommand{\chir}{\draw[-,mid arrow]}

\usepackage{graphicx}
\usepackage{amsmath, amssymb}
\usepackage{youngtab}
\usepackage{changepage}

\newcommand{\be}{\begin{eqnarray}}
\newcommand{\ee}{\end{eqnarray}}
\newcommand{\ba}{\begin{array}}
\newcommand{\ea}{\end{array}}
\newcommand{\bea}{\begin{eqnarray}}
\newcommand{\eea}{\end{eqnarray}}
\newcommand{\bpic}{\begin{tikzpicture}}
\newcommand{\epic}{\end{tikzpicture}}

\newcommand{\nn}{\nonumber}
\newcommand{\bn}{\begin{enumerate}}
\newcommand{\en}{\end{enumerate}}

\def\Ge{\Gamma_e}

\def\CF{{\cal F}}

\def\CI{{\cal I}}

\def\cN{{\cal N}}
\def\CN{{\cal N}}
\def\CO{{\cal O}}

\def\CW{{\cal W}}
\def\cW{{\cal W}}

\def\Tr{\mathop{\text{Tr}}\nolimits}

\def\b{\beta}

\def\d{\delta}

\def\r{\rho}

\def\s{\sigma}

\def\D{\Delta}
\def\M{\mathfrak{M}}

\title{Star-triangle dualities and supersymmetric improved bifundamentals}

\author[a]{Sergio Benvenuti,}
\author[b,c]{Riccardo Comi,}
\author[b,c]{Sara Pasquetti}

\affiliation[a]{INFN, Sezione di Trieste, SISSA, via Bonomea 265, 34136 Trieste, Italy}
\affiliation[b]{Dipartimento di Fisica, Università di Milano-Bicocca, Piazza della Scienza 3, I-20126 Milano, Italy}
\affiliation[c]{INFN, Sezione di Milano-Bicocca, Piazza della Scienza 3, I-20126 Milano, Italy}

\emailAdd{benve79@gmail.com, r.comi2@campus.unimib.it, sara.pasquetti@gmail.com}

\abstract{
Recently it was shown that  mirror duals of $3d$ and $4d$  theories with four supercharges
can be described by \emph{generalized} quiver theories,  constructed using strongly coupled SCFTs as  elementary building blocks that replace and \emph{improve}  standard bifundamentals.
In this work we study and extend the family of  such \emph{improved bifundamentals}
and discuss the network of \emph{star-triangle} dualities they satisfy.
We  provide a field theoretic proof of the star-triangle dualities, which only assumes the basic Seiberg dualities, using the sequential deconfinement technique. 

}

\begin{document}

\maketitle

\flushbottom

\section{Introduction and summary}

An effective strategy to  construct SCFTs  starts from defining a set  elementary building blocks that can be then glued together via the identification and gauging of global symmetries. 
These elementary building blocks can be taken to be generic strongly coupled SCFTs, so the resulting theory may be a non-Lagrangian theory described by a \emph{generalized} quiver theory such as in  the class S construction of $4d$ SCFTs \cite{Gaiotto:2009we}.
In class S, Riemann surfaces over which the $6d$ theories are compactified, can be decomposed into simpler ones, such as pairs of pants, which are associated to usually strongly coupled non-Lagrangian SCFTs that are glued together via the gauging of global symmetries reflecting the way the surfaces are glued together.
Generalized quivers were also used to construct  $6d$ SCFTs \cite{DelZotto:2014hpa} with elementary building blocks given by $(G,G)$ bifundamental matter blocks, strongly coupled non-Lagrangian SCFTs associated to an M5-brane probing an ADE-type singularity.
This approach has been extended also to $5d$ SCFTs in \cite{DeMarco:2023irn}.

Recently in \cite{BCP2} it was shown that  mirror  duals of $4d$ and $3d$  theories with four supercharges  are captured by generalized quivers involving  \emph{improved bifundamentals}, strongly coupled SCFTs that contain a bifundamental  in the operator spectrum. 
Contrary to the class S trinions  and the conformal matter blocks, the improved bifundamentals do   admit a  Lagrangian  UV completion but at the moment  do  not always admit a stringy realization.
The generalized quivers obtained by gluing together these improved bifundamental  are {\it quasi}-Lagrangian, since the gluing is performed by gauging global symmetries which possibly emerge only in the IR.

In this paper we study and extend  the family of such improved bifundamentals.
The progenitor of this family is the $4d$  $\cN=1$ SCFT $FE_N$, introduced in \cite{Pasquetti:2019hxf}\footnote{In other works, as for example \cite{Pasquetti:2019hxf,Hwang:2020wpd,Bottini:2021vms,Comi:2022aqo}, the $FE_N$ theory is called $FE[USp(2N)]$.}
which we represent in compact form as two square flavor nodes connected by a zig-zag line:
\be\label{FE_symbol}
\begin{tikzpicture}[thick,node distance=3cm,gauge/.style={circle,draw,minimum size=5mm},flavor/.style={rectangle,draw,minimum size=5mm}]
\path (-1.5,0) node[flavor,blue](y1) {$\!2N\!$} -- (1.5,0) node[flavor,red](y3) {$\!2N\!$};    \wigE (y1) -- (y3); 
\end{tikzpicture}\ee
The compact form help us visualizing the fact that 
the global symmetry contains an  ${\color{blue} USp(2N)} \times  {\color{red} USp(2N)}$ non abelian component in addition to a $U(1)^2$ symmetry.
This theory generalizes and improves a standard Wess-Zumino model of a $USp(2N)^2$ bifundamental chiral multiplet in the sense that the spectrum contains a bifundamental chiral operator and the theory supports an extra $U(1)$ global symmetry. In addition, the $FE_N$ theory admits a superpotential deformation which triggers a Renormalization Group (RG) flow leading to the standard bifundamental chiral multiplet.
The $FE_N$ SCFT has been identified as the fundamental building block of the E-string tube compactification \cite{Pasquetti:2019hxf} and, more recently, it has been shown to appear in the $4d$ $\mathcal{N}=1$ mirror-like dual of the $USp(2N)$ SQCD \cite{BCP2}. 
Interestingly, the $FE_N$ SCFTs also appears in disguise in a completely different context: The superconformal index of the $FE_N$ theory coincides with the integral definition of the \emph{interpolation kernel} introduced and discussed in \cite{Rains_2018}. 

It was observed in \cite{Pasquetti:2019hxf} that the $FE_N$ theory satisfies a star-triangle relation:\footnote{On the l.h.s.~there is a superpotential coupling the antisymmetric chiral, represented by the arc, and operators in the spectrum of the two $FE_N$ theories. On the r.h.s.~we instead have a cubic coupling between the two chirals and the bifundamental operator in the spectrum of the $FE_N$.}
\be\label{braid0} 
    \includegraphics[]{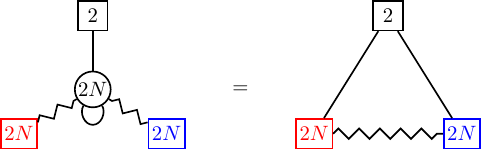}
\ee
 
This duality plays a role in consistency checks for the compactification of the E-string and is used in the construction of $4d$ $\mathcal{N}=1$ mirror-like dualities. The integral identity corresponding
to the equality of the superconformal indices of the two theories appearing in the star-triangle duality 
was proven in  \cite{Rains_2018}.
One of the main results presented in this paper is a complete field theoretical proof of this duality which works by induction in the integer parameter $N$. This proof assumes only the $N=1$ case, which is the confining Seiberg duality for $SU(2)$ \cite{Seiberg:1994pq}, and uses Intriligator-Pouliot duality \cite{Intriligator:1995ne} to prove the inductive step. Let us stress that this proof is structurally different from the argument given in \cite{Rains_2018} and it is similar in spirit to the sequential confinement and deconfinement technique of \cite{Benvenuti:2017kud, Giacomelli:2017vgk, Giacomelli:2019blm, Pasquetti:2019uop,Pasquetti:2019tix, Sacchi:2020pet,Benvenuti:2020gvy, Benvenuti:2021nwt, Bottini:2022vpy,Bajeot:2022kwt, Amariti:2022wae, Bajeot:2022lah, Amariti:2022tbd, Bajeot:2022wmu, Bajeot:2023gyl, Amariti:2023wts, Amariti:2024sde, Amariti:2024gco,Amariti:2024usp,Benvenuti:2024glr}.

Using the star-triangle duality in \eqref{braid0}, we show that it is possible to generate star-star dualities that generalize the web of self-dualities for the antisymmetric $USp(2N)$ SQCD with eight fundamentals \cite{Csaki:1997cu,Spiridonov_2010}.

\vspace{0.2cm}

\begin{table}[b!]
    \centering
    \begin{tabular}{|c|c|c|c|}
    \hline
    Name & Symbol & Global symmetry & Operators \\ 
    \hline
    $FE^{3d}_N$ 
    & \resizebox{0.11\hsize}{!}{\begin{tikzpicture}[thick,node distance=2cm,gauge/.style={circle,draw,minimum size=3mm}, flavor/.style={rectangle,draw,minimum size=3mm}] \path (0,0) node[flavor,blue] (x) {$\!C_N\!$}  -- (1.5,0) node[flavor,red] (y) {$\!C_N\!$};\wigE (x) -- (y);\draw (0.7,0.3) node {$\,$};\end{tikzpicture}}
    & ${\color{blue} USp(2N)}\! \times \! {\color{red} USp(2N)} \! \times \! U(1)^2$  
    & {\color{blue} Asym}, {\color{red} Asym}, ({\color{blue} $\bf 2N$}, {\color{red} $\bf 2N$}), Sing. \\ 
    \hline
    $FM_N$ 
    & \resizebox{0.11\hsize}{!}{\begin{tikzpicture}[thick,node distance=2cm,gauge/.style={circle,draw,minimum size=3mm}, flavor/.style={rectangle,draw,minimum size=3mm}] \path (0,0) node[flavor,blue] (x) {$\!U_N\!$}  -- (1.5,0) node[flavor,red] (y) {$\!U_N\!$};\wigM (x) -- (y); \draw (0.7,0.35) node {$\,$}; \end{tikzpicture}}
    & $S [{\color{blue} U(N)} \times {\color{red} U(N)} ] \times U(1)^2 $  
    & {\color{blue} Adj}, {\color{red} Adj}, ({\color{blue} $\bf N$}, {\color{red} $\bf \bar{N}$}), ({\color{blue} $\bf \bar{N}$}, {\color{red} $\bf N$}), Sing. \\ 
    \hline
    $FH_N$ 
    & \resizebox{0.11\hsize}{!}{\begin{tikzpicture}[thick,node distance=2cm,gauge/.style={circle,draw,minimum size=3mm}, flavor/.style={rectangle,draw,minimum size=3mm}] \path (0,0) node[flavor,blue] (x) {$\!C_N\!$}  -- (1.5,0) node[flavor,red] (y) {$\!U_N\!$};\wigMB (x) -- (y); \draw (0.7,0.35) node {$\,$}; \end{tikzpicture}}
    & $ {\color{blue} USp(2N)} \times {\color{red} U(N)} \times U(1) $ 
    & {\color{blue} Asym}, {\color{red} Adj}, ({\color{blue} $\bf 2N$}, {\color{red} $\bf \bar{N}$}) \\ 
    \hline
    $FC^{\pm}_N$ 
    & \resizebox{0.11\hsize}{!}{\begin{tikzpicture}[thick,node distance=2cm,gauge/.style={circle,draw,minimum size=3mm}, flavor/.style={rectangle,draw,minimum size=3mm}] \path (0,0) node[flavor,blue] (x) {$\!U_N\!$}  -- (1.5,0) node[flavor,red] (y) {$\!U_N\!$};\wigMc (x) -- (y); \draw (0.7,0.22) node {$\pm$}; \end{tikzpicture}}
    & $S [{\color{blue} U(N)} \times {\color{red} U(N)} ] \times U(1) $ 
    & {\color{blue} Adj}, {\color{red} Adj}, ({\color{blue} $\bf N$}, {\color{red} $\bf \bar{N}$}) \\ 
    \hline
    $FT_N$ 
    & \resizebox{0.11\hsize}{!}{\begin{tikzpicture}[thick,node distance=2cm,gauge/.style={circle,draw,minimum size=3mm}, flavor/.style={rectangle,draw,minimum size=3mm}] \path (0,0) node[flavor,blue] (x) {$\!N\!$}  -- (1.5,0) node[flavor,red] (y) {$\!N\!$};\wigT (x) -- (y); \draw (0.7,0.3) node {$\,$}; \end{tikzpicture}}
    & ${\color{blue} SU(N)} \times {\color{red} SU(N)} \times U(1) $ 
    & {\color{blue} Adj}, {\color{red} Adj} \\ 
    \hline
    \end{tabular}
    \caption{Summary of the $3d$ improved bifundamentals. In each line we give the name and the short symbol representing the theory, the global symmetry group and a schematic list of the chiral operators in the spectrum.}
    \label{tab:impbif0}
\end{table}
We then perform a $3d$ circle reduction of the $FE_N$ theory followed by suitable real mass deformations, along the lines of \cite{Aharony:2013dha,Benini:2017dud}. With this strategy we introduce a family of $3d$ $\mathcal{N}=2$ \emph{improved bifundamentals} that are summarized in Table \ref{tab:impbif0}.
The SCFT in the first line is simply the circle reduction of the $4d$ $FE_N$ and has the same spectrum and properties of its $4d$ ancestor.

In the second line of Table \ref{tab:impbif0} there is the $FM_N$ SCFT, which was first introduced in \cite{Pasquetti:2019tix}. The $FM_N$ theory has $S[{\color{blue} U(N)} \times {\color{red} U(N)}] \times U(1)^2$ global symmetry and it contains in the spectrum:
two {\it moment map} operators in the adjoint representation of each $U(N)$ global symmetry, a pair of ${\color{blue}U(N)} \times {\color{red}U(N)}$ bifundamentals and some singlets of the non-abelian global symmetries. The $FM_N$ theory improves a standard ${\color{blue}U(N)} \times {\color{red} U(N)}$ bifundamental hypermultiplet. 
In \cite{BCP2}, it was shown that the $FM_N$ theory can be used to construct linear improved quivers effectively associated to brane setups with four supercharges consisting of $N$ $D3$ branes stretching along a sequence of $NS$ and $D5'$ branes. Such brane setups have chiral symmetry, that is a sequence of $F$ $D5'$ provide $U(F)^2$ instead of $U(F)$ global symmetry, as in the eight supercharges case. Accordingly, a sequence of $K$ $NS$ branes provides $U(K)^2$ emergent symmetry which is obtained from the topological symmetries and the extra $U(1)$'s carried by the $FM_N$ theories.

The theory in the third line of Table \ref{tab:impbif0} is the $FH_N$ theory, where $"H"$ stands for \emph{hybrid}, having a 
${\color{blue} USp(2N)} \times  {\color{red} U(N)} \times U(1)$ global symmetry. Its BPS spectrum contains a bifundamental operator, an $USp(2N)$ antisymmetric and a $U(N)$ adjoint. We expect this theory to play an important role in mirror dualities for symplectic $\mathcal{N}=2$ theories.

In the second to last line of Table \ref{tab:impbif0} we have the $FC^\pm_N$ SCFTs, that are two new theories
with $S[{\color{blue} U(N)} \times  {\color{red} U(N)} ] \times U(1)$ global symmetry, which spectrum contains a single bifundamental operator and two {\it moment maps} in the adjoint reresentation.
These two theories improve the bifundamental chiral.\footnote{The $FC^+_N$ and $FC^-_N$ theories are distinguished only by their behavior under gauging of the global symmetries. In some sense, these two theories act as a bifundamental chiral with different choices of background BF couplings.  We comment more in this difference in Subsection \ref{sec:FC}.} 
This new theories are necessary to construct the mirror dual of chiral $\mathcal{N}=2$ theories that might also have non-zero Chern-Simons level \cite{BCP3}.

The last item in Table \ref{tab:impbif0} is the $FT_N$ theory, a flipped version of the $3d$ $\cN=4$ $T[SU(N)]$ \cite{Gaiotto:2008ak}. There are no bifundamental operators in the spectrum in this case, nevertheless  we include it in this list since also the $FT_N$ SCFT can be regarded as a building block to construct generalized quiver theories as in the case of the $3d$ S-folds constructions \cite{Assel:2018vtq,Garozzo:2018kra,Garozzo:2019hbf,Garozzo:2019ejm}, in the context of the 3d-3d correspondence \cite{Terashima_2011,Gang_2016,Gang_2018,Gang2_2018}
and in the mirror dual of  theories with non-zero Chern-Simons level \cite{BCP3}. 

\begin{figure}[t!]
    \centering
    \includegraphics[width=\linewidth]{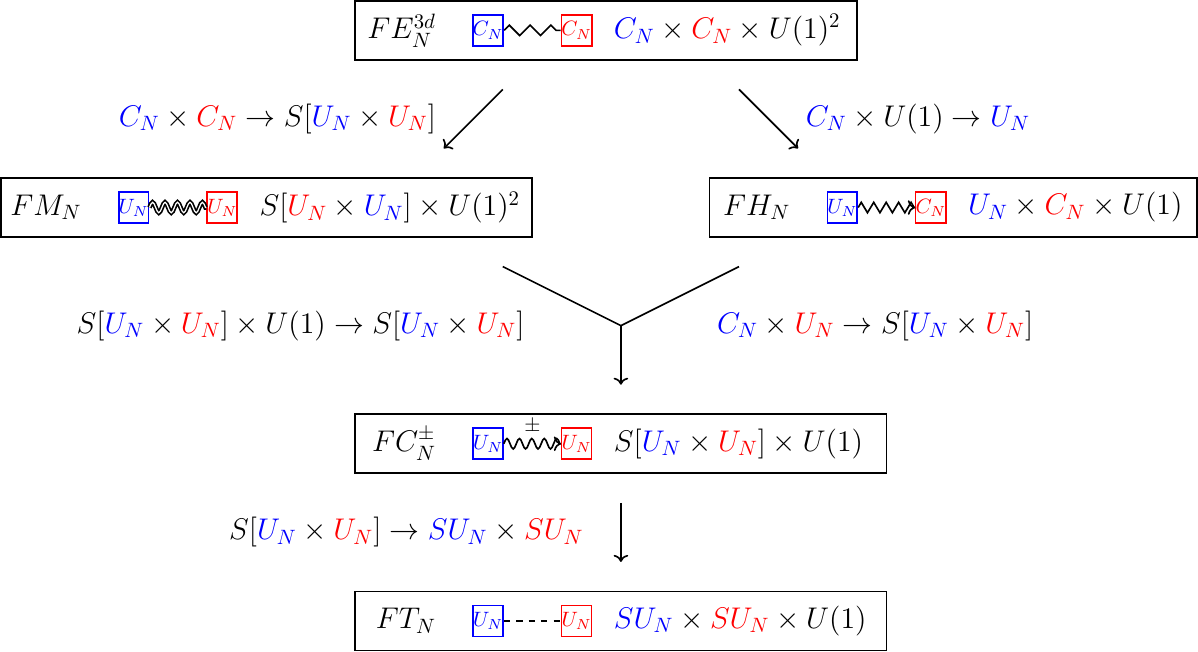}
    \caption{Summary of the relation between $3d$ improved bifundamentals via real mass deformations. Each arrow represents a possible deformation and it is labelled by the global symmetry breaking pattern.}
    \label{fig:3dimpbif_realmass}
\end{figure}

The improved bifundamentals are related to each other via real mass deformations that have the effect of breaking part of the global symmetry, as summarized in the diagram in Figure \ref{fig:3dimpbif_realmass}. 

Starting from the $4d$ star-triangle relation involving the $FE_N$ theory, it is possible to perform a $3d$ circle reduction, followed by real mass deformations, and generate a class of $3d$ star-triangle relations that involve $3d$ improved bifundamentals. These star-triangle dualities include and generalize all the basic moves appearing in the dualization algorithm approach to $\mathcal{N}=4$ and $\mathcal{N}=2$ mirror dualities \cite{Hwang:2021ulb, Comi:2022aqo, BCP2, BCP3} and play an important role to prove consistencies in these dualities.

\vspace{0.5cm}

The paper is organized as follows.

In Section \ref{sec:FE} we give the definition and discuss the most important properties of the $4d$ $FE_N$ theory. Extending the results of \cite{Pasquetti:2019hxf}, in section \ref{sec:FE} we discuss the chiral ring of the $FE_N$ theories. In Section \ref{sec:braid} we discuss the star-triangle duality, with a particular focus on the mapping of holomorphic operators. In Section \ref{sec:SD} we show how to obtain star-star dualities by composing star-triangle dualities. In Section \ref{sec:proof} we provide a purely field theoretical proof of the star-triangle duality. 

In section \ref{3dGB} we discuss the $3d$ reduction of the $FE_N$ theory and introduce the $3d$ improved bifundamentals\footnote{See \cite{BenvenutiPedde:2024} for a computation of the Hilbert Series of these five classes of improved bifundamentals.}. With the same strategy, in Section \ref{3dST} we derive $3d$ star-triangle dualities from the reduction of the $4d$ one.

\section{The \boldmath{$FE_N$} SCFT: an improved bifundamental}\label{sec:FE}
In this section we review the definition 
and some of the interesting properties of the $FE_N$ theory.
We mostly collected results from \cite{Pasquetti:2019hxf} 
and
\cite{Hwang:2020wpd,Comi:2022aqo},
but we add more details, including a discussion on the chiral ring generators of the theory and some of their quantum relations.

The $FE_N$ theory is a $4d$ $\cN=1$ SCFT denoted by the following symbol:
\be\label{fig:FE_symbol}
    \includegraphics[]{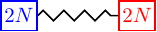}
\ee
which makes explicit the non-abelian part of its global symmetry group which is
\begin{align}
    {\color{red}USp(2N)} \times {\color{blue}USp(2N)} \times U(1)_\tau \times U(1)_\pi \,,
\end{align}
in addition to the $\CN=1$ $U(1)$ R-symmetry,
and also that the spectrum contains a bifundamental of ${\color{red}USp(2N)} \times {\color{blue}USp(2N)}$, as we will discuss in more detail later on. 
The $FE_N$ theory was introduced in \cite{Pasquetti:2019hxf} (and  further studied in \cite{Hwang:2020wpd,Bottini:2021vms}) and shown to be the building block for the construction of 
rank-Q E-string compactifications  on $2d$ Riemann surfaces \cite{Pasquetti:2019hxf,Hwang:2021xyw}.

\subsection{UV completions, symmetries and operators}
\begin{figure}
\centering
    \includegraphics[]{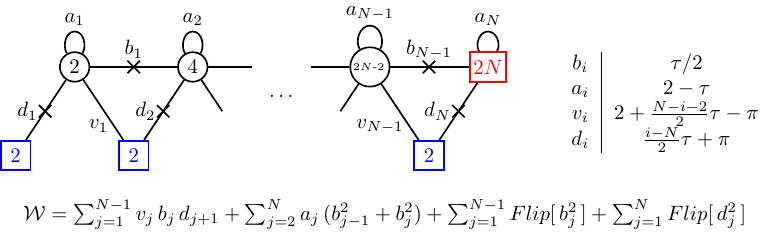}
\caption{Quiver representation of the UV completion of the $FE_N$ SCFT. Each node, square or round, labeled with a number $2n$, represents respectively a gauge or flavor $USp(2n)$ group. Each line is a field in the fundamental representation of the nodes to which is attached, except for arch lines that are fields in the traceless antisymmetric representation. Crosses denotes the presence of flipping fields for which we will adopt the following notation: Given an operator which is a singlet under all the gauge and non-abelian flavor symmetries $X$ and the singlet $\CO_X$, we will denote the superpotential term $\CW = \CO_X X$ as $Flip[X]$ and the singlet $\CO_X$ as $\CF[X]$. Lastly, all the superpotential terms are written in short by omitting the traces which also includes an appropriate antisymmetric tensor to obtain $USp$ invariants. Also, on the right, we give the table with the R-charge of all the fields in the theory.}
\label{fig:FE_quiver}
\end{figure}
The $FE_N$ theory is the SCFT sitting at the infrared fixed point of a linear quiver theory with $N-1$ symplectic gauge nodes of increasing ranks given in Figure \ref{fig:FE_quiver}\footnote{Other possible UV completions are discussed later in \eqref{fig:FE_part_open} and \eqref{fig:FEmir}.}. For $N=1$ it is simply a chiral in the  ${\color{red}USp(2)} \times {\color{blue}USp(2)}$ bifundamental, with a singlet flipping its square.
The theory in Figure \ref{fig:FE_quiver} has the UV global symmetry group
\begin{align}
	{\color{red}USp(2N)} \times {\color{blue}USp(2)}^N \times U(1)_\tau \times U(1)_\pi \,,
\end{align}
however at the IR fixed point the theory is characterized by the enhanced global symmetry of the $FE_N$ theory:
\begin{align}
	{\color{red}USp(2N)} \times {\color{blue}USp(2N)} \times U(1)_\tau \times U(1)_\pi \,.
\end{align}
The gauge invariant operators indeed reorganize into representations of the IR symmetry group. The list of the chiral ring generators of the $FE_N$ SCFT, along with their charges and representations, is given in Table \ref{tab:FE_operators}.
\begin{table}[]
\renewcommand{\arraystretch}{1.2}
\centering
\begin{tabular}{|c|cc|c|}\hline
{} & ${\color{red}USp(2N)}$ & ${\color{blue}USp(2N)}$ & R-charge\\ \hline
${\color{red}\mathsf{A}}$ & ${\bf N(2N-1)-1}$ & $\bf 1$  & $2-\tau$ \\
${\color{blue}\mathsf{A}}$ & $\bf1$ & ${\bf N(2N-1)-1}$  & $2-\tau$ \\
$\Pi$ & $\bf N$ & $\bf N$  & $\pi$ \\
$\mathsf{B}_{n, m}$ & $\bf1$ & $\bf1$  & $2n-2\pi+(m-n)\tau$ \\
 \hline
\end{tabular}
\caption{List of all the gauge invariant operators that compose the spectrum of the $FE_N$ SCFT. The R-charge is given as a trial value mixed with the other two abelian symmetries of the theory, $U(1)_\tau$ and $U(1)_\pi$, whose mixing values are given by the two real variables $\tau$ and $\pi$.}
\label{tab:FE_operators}
\end{table}
From the UV perspective in Figure \ref{fig:FE_quiver} these operators are constructed in the following way\footnote{Often, we will refer to the operators ${\color{blue}\mathsf{A}}$ and ${\color{red}\mathsf{A}}$ as moment maps.}:
\begin{itemize}
	\item ${\color{red}\mathsf{A}}$ is simply the gauge singlet field $a_N$, in the traceless antisymmetric representation of $\color{red}USp(2N)$.

	\item ${\color{blue}\mathsf{A}}$ is in the traceless antisymmetric representation of ${\color{blue}USp(2N)}$ and is built collecting all the mesonic operators in the bifundamental of any pair of ${\color{blue}USp(2)}$ groups:  $d_i b_i b_{i+1} \ldots b_{j-1} v_j$, for $i,j=1,\ldots,N-1$ and $i < j$, together with the $N-1$ flippers $\CF[b_i^2]$.
	
	\item $\Pi$ is a ${\color{red}USp(2N)} \times {\color{blue}USp(2N)}$ bifundamental built by collecting the mesons in the bifundamental of ${\color{red}USp(2N)} \times {\color{blue}USp(2)}$: $d_i b_i \ldots b_{N-1}$, for $i=1,\ldots,N$.
	
	\item $\mathsf{B}_{n, m}$ ($n=1,...,N$ and $m=1,...,N+1-n$) is a collection of operators charged only under $U(1)_\pi$ and $U(1)_\tau$: the singlets $\CF[d_{N+1-m}^2] = \mathsf{B}_{1, m}$ and the dressed mesons  $\Tr(a_{N-m}^{n-2} v_{N-m}^2) = \mathsf{B}_{n, m}$ for $n \geq 2$. For example, for $N=4$, the singlet matrix $\mathsf{B}_{n,m}$ is given as:
\be
\{ \mathsf{B}_{n, m} \} = \left(
\begin{tabular}{cccc}
$\CF[d_4^2]$ & $\CF[d_3^2]$ & $\CF[d_2^2]$ & $\CF[d_1^2]$ \\
$v_3^2$ & $v_2^2$ & $v_1^2$ & 0 \\
$v_3^2a_3 $ & $v_2^2a_2 $ & 0 & 0 \\
$v_3^2a_3^2 $ & 0 & 0 & 0
\end{tabular}\right)
\ee

\end{itemize}
A few comments regarding the chiral operators are in order. 

If we perform a-maximization we see that in the isolated $FE_N$ theory, some operators in Table \ref{tab:FE_operators} fall below the unitarity bound (for instance many singlets $\mathsf{B}_{n,m}$)\footnote{This fact has implications for the conformal manifold. The isolated $FE_N$ SCFT has $(N(2N+1))^2 + (N(2N-1)-1)^2$ marginal directions given by the operators $\Pi^2 \mathsf{B}_{1,1}$. However, this conformal manifold is \emph{virtual}. If we perform a-maximization we see that in the isolated $FE_N$ theory, the singlet $\mathsf{B}_{1,1}$ falls below the unitarity bound for all $N$. Therefore, $\mathsf{B}_{1,1}$ decouples in the IR and the aforementioned operators $\Pi^2 \mathsf{B}_{1,1}$ cannot be constructed. This is the reason we say that such marginal directions are only virtual.}. So at the quantum level Table \ref{tab:FE_operators} contains operators which are not in the chiral ring. 
In general throughout this paper we do not concern ourselves with such issues, we keep the operators falling below the unitarity bound, some composite operators and some operators which might be set to zero by e.o.m. in our discussions. The reason is that in this way we can discuss a mapping of the chiral ring generators which contains more information, and can be useful when the $4d$ $\cN=1$ QFT's we study here as isolated are coupled to other QFT's, and some of the operators that in the isolated theory are only \emph{virtual} chiral ring operators become \emph{real} chiral ring operators, and knowing their mapping is useful. We will come back to this point later in some examples.

Indeed it is  convenient to introduce two \emph{virtual} moment map  operators  ${\color{red}\mathsf{V}}$  and ${\color{blue}\mathsf{V}}$ transforming respectively in the asymmetric of   ${\color{red}USp(2N)}$ and ${\color{blue}USp(2N)}$.
${\color{red}\mathsf{V}}$ coincides with the  meson $b^2_{N-1}$ 
set to zero  by the F-term of ${\color{red}\mathsf{A}}$, which we recall is identified with the gauge singlet $a_N$ in the UV completion in Figure \ref{fig:FE_quiver}.
If the $FE_N$ theory is coupled to other matter through some interaction involving the operator $\color{red}\mathsf{A}$, the F-terms of  ${\color{red}\mathsf{A}}=a_N$ are modified and ${\color{red}\mathsf{V}}$, rather than set to zero, can be identified with some operator and become a real chiral ring operator.
The other virtual moment map ${\color{blue}\mathsf{V}}$ can be similarly defined considering the second (mirror dual) UV completion of the $FE_N$ theory discussed later below \eqref{fig:FEmir}.

While we do not attempt to analyze in complete detail all the relations obeyed by the chiral ring generators listed in Table \ref{tab:FE_operators}, let us comment on some of them. We give our analysis, performed mainly using the superconformal index, in Appendix \ref{app:FE_relations}.
\begin{itemize}
    \item The operator obtained by squaring $\Pi$ generically decomposes into representations of ${\color{red}USp(2N)} \times {\color{blue}USp(2N)}$ as:
    \begin{align}
        (\text{symm},\text{symm}) \oplus (\text{asymm},\text{asymm}) \oplus (\text{asymm},1) \oplus (1,\text{asymm}) \oplus (1,1) \,.
    \end{align}
    The only representations that are not set to zero in the chiral ring are $(\text{symm},\text{symm})$ and $(\text{asymm},\text{asymm})$. The singlet obtained as $\Tr(\Pi^2)$ is set to zero in the chiral ring and can be restored by flipping the operator $B_{1,1}$.
    
	\item For the operator $\Pi \,\mathsf{A}^j$, where $\mathsf{A}$ can be any of the two antisymmetric operators and $j=1,\ldots,N-1$ we claim that for any $j$  the contribution of the (Fund,Fund) representation  is  set to zero and the only non-zero irreducible representation is (Symm,Fund), when considering $\color{red}\mathsf{A}$ or (Fund,Symm) for $\color{blue}\mathsf{A}$.
 
	\item The traces of the two antisymmetric operators are identified as: $\Tr ({\color{red}\mathsf{A}}^j) = \Tr ({\color{blue}\mathsf{A}}^j)$, for any $j=2,\ldots,N$, recalling that classically $\Tr ({\color{red}\mathsf{A}}) = \Tr ({\color{blue}\mathsf{A}}) = 0$ since the two antisymmetrics are traceless.
\end{itemize}

Let us also comment on why we consider the $FE_N$ theory the improved version of a standard $USp(2N)^2$ bifundamental chiral multiplet. First, we observe that the global symmetry of the standard bifundamental chiral is $USp(2N)^2 \times U(1)$, where the abelian factor is an axial symmetry. The $FE_N$ theory, on the other hand, has a $USp(2N)^2 \times U(1)^2$ global symmetry, meaning that it carries an extra $U(1)$. 
The spectrum of the theories is also very similar. A  standard bifundamental chiral $b$ has, obviously, $4N^2$ chiral ring generators that are the components of $b$, but we can also construct composite operators. In particular we can obtain an operator in the antisymmetric representation of one $USp(2N)$ by taking $b^2$ and tracing over one of the two $USp(2N)$ global symmetries with the insertion of an antisymmetric tensor. 
The $FE_N$ theory, similarly, has a spectrum consisting of a bifundamental chiral operator $\Pi$ that we can square to obtain composite operators. However, notice from the previous discussion, that we can not obtain antisymmetric operators by squaring $\Pi$ since these representations are set to zero in the chiral ring. Instead, the antisymmetric operators are independently given by $\color{red}\mathsf{A}$ and $\color{blue}\mathsf{A}$.

\paragraph{Recursive definition}
It is  convenient to introduce a UV completion of the $FE_N$ SCFT in terms of the $FE_{N-1}$ SCFT, as follows:
\be\label{fig:FE_part_open}
    \includegraphics[]{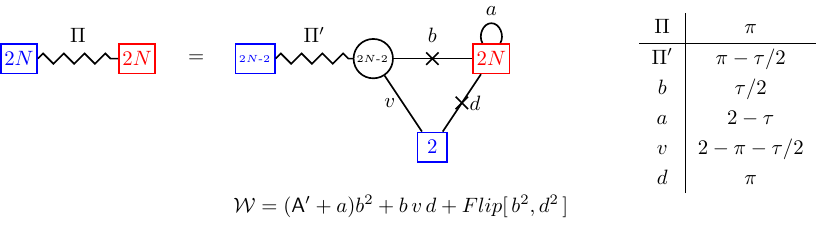}
\ee
Where $\mathsf{A}'$ is the moment map of the $FE_{N-1}$ theory which is charged under the gauge symmetry. In the pictures we label the $FE_N$ theories by only specifying the name and R-charge of the bifundamental operator. However, in order to completely specify the parametrization of a $FE_N$ theory is necessary to fix also the R-charge of the two antisymmetric operators. Unless noted otherwise, the R-charge of the two antisymmetric operators is canonically assumed to be $2-\tau$\footnote{We choose to not specify the operators $\color{red}\mathsf{A}$, $\color{blue}\mathsf{A}$ and the rest of the $FE_N$ operators in the pictures to avoid cluttering. Their presence is always implied and we label them using sub/superscripts when necessary. For example, in figure \eqref{fig:FE_part_open}, all the operators of the $FE_{N-1}$ are primed: ${\color{blue}\mathsf{A}}'$, $\mathsf{A}'$, $\mathsf{B}'_{n,m}$.}.

The spectrum of gauge invariant operators given in Table \ref{tab:FE_operators} can be built from the UV completion given in figure \eqref{fig:FE_part_open} as:
\begin{itemize}
	\item ${\color{red}\mathsf{A}}$ is simply the gauge singlet field $a$.
	
	\item ${\color{blue}\mathsf{A}}$ decomposes into $\{{\color{blue}\mathsf{A}}', \Pi' \, v, \CF[b^2]\}$.
	
	\item $\Pi$ decomposes into $\{\Pi' \, b, d\}$.
	
	\item $\mathsf{B}_{n,m}$ splits as: 
	\begin{align}
	\begin{cases}
		\CF[d^2] \quad & \text{for} \quad n=m=1 \nn \\
		\Tr(v^2 (\mathsf{A}')^{n-2}) \quad & \text{for} \quad n > 1 \quad \text{and} \quad m=1 \nn \\
		\mathsf{B}'_{n, m-1} \quad & \text{for} \quad n > 1 \quad \text{and} \quad m > 1
	\end{cases}
	\end{align}
	
\end{itemize}

\paragraph{Second (mirror dual) UV completion}

The $FE_N$ SCFT admits a second self-dual Lagrangian UV completion in which the manifest and emergent $USp(2N)$ global symmetries are swapped, as it is schematically summarized below.
\begin{align}\label{fig:FEmir}
    \includegraphics[scale=0.8]{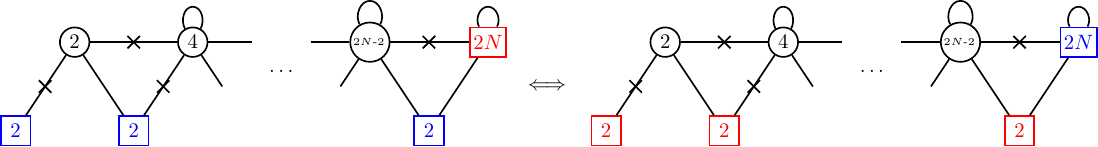}
\end{align}
Notice that in the second UV completion, on the r.h.s.~of the above figure, the R-charges and the charge under all the abelain global symmetries of all the fields follow the same rules presented in Figure \ref{fig:FE_quiver}. 

We can view the two possible UV completions as a choice of which $USp(2N)$ global symmetry is realized manifestly in the UV, without changing the Lagrangian description:
\begin{align}
    & \text{UV completion 1:} \qquad {\color{red}USp(2N)}\times {\color{blue}USp(2)}^N\times U(1)_\tau \times U(1)_\pi \nn \\
    & \text{UV completion 2:} \qquad {\color{red}USp(2)}^N\times {\color{blue}USp(2N)} \times U(1)_\tau \times U(1)_\pi \,.
\end{align}
As far as we know, there is no UV completion of the $FE_N$ theory in which both the $USp(2N)$ global symmetries are manifest at the same time.

The two UV completions flow to the same IR SCFT
and are related by a \emph{mirror}-like duality. Under this duality all the operators are mapped to themselves, except for the two antisymmetric operators ${\color{red}\mathsf{A}}$ and ${\color{blue}\mathsf{A}}$ that are exchanged. Moreover, in the mirror UV completion it is natural to introduce the virtual moment map
${\color{blue}\mathsf{V}}$ identified  with the  meson $b^2_{N-1}$ 
set to zero by the F-term of ${\color{blue}\mathsf{A}}=a_N$. Therefore, since the properties of the IR theory should not depend on the choice of the UV completion we claim that the $FE_N$ SCFT possesses both the $\color{blue}\mathsf{V}$ and $\color{red}\mathsf{V}$ virtual moment maps, as already stated at the beginning of the section.

As discussed in \cite{Hwang:2021ulb} the self-mirror property can be demonstrated inductively using the mirror dualization algorithm and the manifest self-mirror property of the $FE_1$ theory.

\paragraph{Superconformal Index of the theory}
Inspired by the recursive definition for the $FE_N$ theory given in \eqref{fig:FE_part_open}, we define the Superconformal Index of the theory starting the index of the $FE_{N-1}$\footnote{Notice that with respect with the notation for the $FE_N$ theory used in other works, as for example in \cite{Pasquetti:2019hxf,Hwang:2020wpd,Bottini:2021vms,Comi:2022aqo}, here we are considering all the antisymmetric chirals to be traceless.
Notice also that  \eqref{eq:FE_SCI} coincides with the
interpolation kernel appearing in  \cite{Rains_2018}.}.
From the UV completion in figure \eqref{fig:FE_part_open} we write the following index:
\begin{align}\label{eq:FE_SCI}
	\CI_{FE,1}^{(N)} (\vec{x},\vec{y},t,c) = & \Ge \big( pqt^{-1} \big)^N \prod_{j<k}^N \Ge \big( pq t^{-1} y_j^\pm y_k^\pm \big) \Ge \big( pq c^{-2} \big) \prod_{a=1}^{N} \Ge \big( c y_N^\pm x_a^\pm \big) \times \nn \\
	& \times \oint d\vec{z}_{N-1} \D_{N-1}(\vec{Z})
	\prod_{j=1}^{N-1} \left[ \, \prod_{a=1}^{N} \Ge \big( t^{\frac{1}{2}}z_j^\pm x_a^\pm \big) \Ge \big( pq t^{-\frac{1}{2}}c^{-1} y_N^\pm z_j^\pm \big) \right] \times \nn \\
	& \times \CI_{FE}^{(N-1)} \big( \vec{z},\{ y_1,\ldots,y_{N-1}\},t,t^{-\frac{1}{2}}c \big) \,,
\end{align}
with the base of the recursion:
\begin{align}
	\CI_{FE,1}^{(1)} (x,y,t,c) = \Ge(pq c^{-2}) \Ge(c x^\pm y^\pm)  \,.
\end{align}
We name this the first UV completion with respect to the second UV completion, which is the self-mirror dual theory described in \eqref{fig:FEmir}.
The fugacities for the $U(1)$ symmetries are related to the R-charge mixing as:
\begin{align}
	c = (pq)^{\pi/2} \quad, \quad t = (pq)^{\tau/2} \,.
\end{align}
Also, the vectors $\vec{x}$ and $\vec{y}$ are the fugacities for the manifest and emergent $USp(2N)$ symmetries respectively, and $\vec{z}$ is the fugacity for the gauge group $USp(2N-2)$. The convention used to write $4d$ superconformal indexes can be found in Appendix \ref{app:conventions}.
Notice that in the definition in eq.~\eqref{eq:FE_SCI} the first entry is associated to the $USp(2N)$ global symmetry which is manifest in the UV completion in Figure \ref{fig:FE_quiver}, while the second to the emergent $USp(2N)$ symmetry.

As already mentioned, the second UV completion is given by the mirror-dual theory described in \eqref{fig:FEmir} in which the emergent and manifest $USp(2N)$ global symmetries are swapped. The partition function of the second UV description can be thus obtained from \eqref{eq:FE_SCI} by swapping the two sets $\vec{x} \leftrightarrow \vec{y}$:
\begin{align}
  \CI_{FE,2}^{(N)} (\vec{x}, \vec{y}, t, c ) = \CI_{FE,1}^{(N)} (\vec{y}, \vec{x}, t, c) \,,
\end{align}
where now the first entry of the partition function is associated to the emergent $USp(2N)$ symmetry while  the second to the manifest $USp(2N)$ symmetry.

Since the two UV completions are IR mirror  dual, i.e.~they flow to the same IR SCFT, the two apparently different superconformal indexes match. We can therefore define the partition function of the $FE_N$ SCFT simply as:
\begin{align}\label{}
\CI_{FE}^{(N)}(\vec{x}, \vec{y}, \tau, \pi) \overset{\text{def}}{=} \CI_{FE,1}^{(N)} (\vec{x}, \vec{y}, \tau, \pi ) = \CI_{FE,2}^{(N)} (\vec{x},\vec{y}, \tau, \pi) 
    \,.
\end{align}

\paragraph{Flip-flip self-duality}
This duality maps the $FE_N$ theory to a copy of itself where the R-charges are redefined shifting the $U(1)_\tau$ symmetry as: $\tau \to 2 - \tau$. It also adds two antisymmetric fields. Graphically it is depicted as:
\be\label{fig:FE_flipflip}
    \includegraphics[]{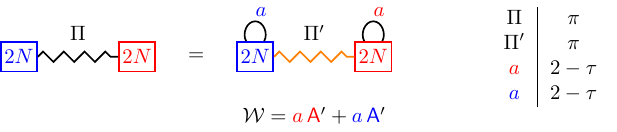}
\ee
Where ${\color{blue}\mathsf{A}}'$ and ${\color{red}\mathsf{A}}'$ are the two moment maps of the flipped $FE_N$ theory that are charged under the two $USp(2N)$ flavor symmetries. In this picture we have introduced a new notation: In a generic $FE_N$ theory depicted with a black line it is implied that the two antisymmetric operators have R-charge $2-\tau$, instead in a $FE_N$ denoted with an orange line, the two antisymmetric operators have R-charge $\tau$. We can think of an orange $FE_N$ theory as obtained from a black one with the redefinition: $\tau \to 2-\tau$, and vice-versa\footnote{In this paper we never have to deal with situations where in the same theory there are $FE_N$ theories with antisymmetrics with R-charges different from $\tau$ or $2-\tau$.}.

As an identity between SCI, this duality can be written as follows:
\begin{align}\label{eq:FE_flipflip}
	\CI_{FE}^{(N)}(\vec{x},\vec{y},t,c) = \Ge(pqt^{-1})^{2N-2}\prod_{j<k}^{N} \Ge(pq t^{-1} x_j^\pm x_k^\pm) \Ge(pq t^{-1} y_j^\pm y_k^\pm) \CI_{FE}^{(N)}(\vec{x},\vec{y},pqt^{-1},c) \,.
\end{align}
The flip-flip duality can be proven by iterative applications of the Intriligator-Pouliot (IP) duality \cite{Intriligator:1995ne} as shown in \cite{Hwang:2020wpd}. 

The map between the gauge invariant operators is:
\begin{align}\label{eq:flipflip_map}
\renewcommand{\arraystretch}{1.1}
\begin{tabular}{c|c|c}
	l.h.s.~of \eqref{fig:FE_flipflip} & r.h.s.~of \eqref{fig:FE_flipflip} & R-charge \\
	\hline
    $\Pi$ & $\Pi'$ & $\pi$ \\
    ${\color{blue}\mathsf{A}}$ & ${\color{blue}a}$ & $2-\tau$ \\
    ${\color{red}\mathsf{A}}$ & ${\color{red}a}$ & $2-\tau$ \\
    $\mathsf{B}_{n, m}$ & $\mathsf{B}'_{m, n}$ & $2n-2\pi+(m-n)\tau$
\end{tabular}
\end{align}
Notice that in the Flip-Flip frame the virtual moment maps become real chiral ring generators. This is due to the superpotential terms involving the moment maps on the r.h.s.~of figure \eqref{fig:FE_flipflip}. Following the same reasoning presented at the beginning of the section, shortly after the discussion of the chiral operators below Table \ref{tab:FE_operators}, we have ${\color{blue}\mathsf{V}}'={\color{blue}a}$ and ${\color{red}\mathsf{V}}'={\color{red}a}$. Moreover considering the $FE_N$ relation
$\Tr({\color{blue}\mathsf{A}}^j) \sim \Tr({\color{red}\mathsf{A}}^j)$
and the operator map we also argue that:
\begin{align}
	\Tr({\color{blue}a}^j) \sim \Tr({\color{red}a}^j) \,.
\end{align}

\subsection{Fusion to identity}
An interesting property of the $FE_N$ theory is that the operation of  gluing together two of them triggers an RG flow that leads to a singular Identity-wall theory with a quantum deformed moduli space:
\be\label{fig:FEdelta}
    \includegraphics[]{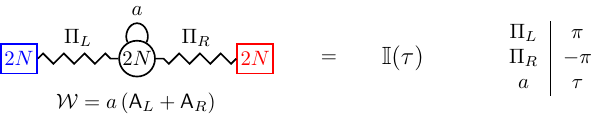}
\ee
Notice that by fixing the R-charge of $a$ to be $\tau$, the R-charge of $\mathsf{A}_L$ and $\mathsf{A}_R$ are forced to be $2-\tau$, as it is in the canonical $FE_N$ theory. Moreover, the vanishing of the NSVZ $\b$-function imposes that the R-charges of the $\Pi_L$ and $\Pi_R$ operator are opposite. Therefore the operator built as $\Pi_L \Pi_R$ has exactly zero R-charge, thus it  acquires a VEV which causes the spontaneous breaking of
the global $USp(2N)^2$  to its diagonal subgroup.

 At the level of the SCI this property becomes:
\begin{align}\label{eq:FEdelta}
	\oint d \vec{z}_N \Delta_N(\vec{z},t) \mathcal{I}_{FE}^{(N)} (\vec{x}, \vec{z}, t,c) \mathcal{I}_{FE}^{(N)} (\vec{z}, \vec{y}, t,c^{-1}) =
  {}_{\vec{x}} \mathbb{I}_{\vec{y}} (t)
 \,,
\end{align}
where the identity operator is defined as:
\begin{align}\label{eq:4d_identity}
{}_{\vec{x}} \mathbb{I}_{\vec{y}} (t) = \frac{\prod_{j=1}^{N} 2\pi i x_j}{\Delta_N(\vec{x},t)} \prod_{\sigma \in S_N} \prod_{j=1}^{N} \delta \big( x_j - y_{\sigma(j)}^\pm \big) \,.
\end{align}
As shown in \cite{Bottini:2021vms} this property can be demonstrated by iterative applications of the IP duality. 

\subsection{Interesting deformations}\label{FEdeform}
We can organize the superpotential deformations of the $FE_N$ theory in two categories: those that are ${\color{red}USp(2N)}\times {\color{blue}USp(2N)}$ preserving and those that are not. 
We report below some interesting deformation discussed in 
in \cite{Hwang:2020wpd} and  \cite{Comi:2022aqo}.

\paragraph{\boldmath{$USp(2N)^2$} preserving deformations} An interesting set of deformations preserving the ${\color{red}USp(2N)} \times {\color{blue}USp(2N)}$ global symmetry consist in turning on linear superpotential terms for the singlets $\mathsf{B}_{1, m}$, that, having R-charge $2-2\pi+(m-1)\tau$, have the effect of breaking only the $U(1)_{\pi} \times U(1)_{\tau}$ symmetry down to the subgroup given by the constraint $\pi = \frac{m-1}{2}\tau$ or, in terms of the fugacities $c = (pq)^{\pi/2}$ and $t = (pq)^{\tau /2}$, it consist in the specialization $c = t^{\frac{m-1}{2}}$. The singlet $\mathsf{B}_{1,m}$ is the flipping field for the meson squared built from the field $d_{N+1-m}$, thus a linear superpotential term for $\mathsf{B}_{1,m}$ has the effect of giving a VEV to $d_{N+1-m}^2$. This VEV triggers an RG flows that leads to a resulting theory which depends on the choice of $m$. Throughout this paper we will be interested only in the cases $m=1,2$. 

Let us focus on the case $m=1$ first. The linear superpotential term for $\mathsf{B}_{1,1}$ has the effect of breaking $U(1)_\pi$ while it preserves $U(1)_\tau$. It consists in the specialization $\pi = 0$, or analogously $c=1$. 
Under this deformation the improved bifundamental behaves as a singular delta-theory. The corresponding SCI identity is:
\begin{align}\label{eq:FE_c=1}
	\mathcal{I}_{FE}^{(N)} (\vec{x}, \vec{y}, t, c=1) = {}_{\vec{x}} \mathbb{I}_{\vec{y}} (t)\,,
\end{align}
where the identity operator is defined as in eq. \eqref{eq:4d_identity}. Indeed, as we break the $U(1)_{\pi}$ symmetry the R-charge of the $\Pi$ operator is fixed to be zero. Therefore it can acquire a VEV with the effect of spontaneously breaking the $USp(2N)^2$ global symmetry down to $USp(2N)$, as encoded by the identity operator on the r.h.s.~of eq.~\eqref{eq:FE_c=1}.

The next case is $m=2$, this deformation has the effect of breaking $U(1)_\pi$ and $U(1)_\tau$ to their combination expressed by the constraint: $\pi = \tau/2$, or analogously $c=t^{1/2}$. Due to this deformation the improved bifundamental flows to a standard ${\color{red}USp(2N)}\times {\color{blue}USp(2N)}$ bifundamental with R-charge $\tau/2$, coupled via a cubic superpotential to two antisymmetric fields and a flipping singlet:
\be\label{fig:FE_c=t/2}
    \includegraphics[]{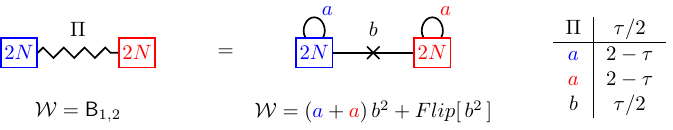}
\ee
As a superconformal index identity this limit can be written as:
\begin{align}
	\CI_{FE}^{(N)} \big( \vec{x},\vec{y},t,c = t^{1/2} \big) = & \Ge(pqt^{-1})^{2N-1} \prod_{j<k}^N \left[ \Ge( pqt^{-1} x_j^\pm x_k^\pm) \Ge( pqt^{-1} y_j^\pm y_k^\pm) \right] \times \nn \\
    & \times \prod_{j,k=1}^N \Ge( t^{1/2} x_j^\pm y_k^\pm ) \,.
\end{align}

Since flip-flip duality maps $\mathsf{B}_{n,m}$ into $\mathsf{B}_{m,n}$, we then know for free the effect of deformations given by linear $\mathsf{B}_{n,1}$ terms. For example, turning on $\d \CW=\mathsf{B}_{2,1}$ has the effect of making the $FE_N$ theory into a flipped ${\color{red}USp(2n)}\times {\color{blue}USp(2N)}$ bifundamental, analogously to the identity in \eqref{fig:FE_c=t/2}, but without the extra antisymmetric fields as:
\be\label{fig:FE_c=1-t/2}
    \includegraphics[]{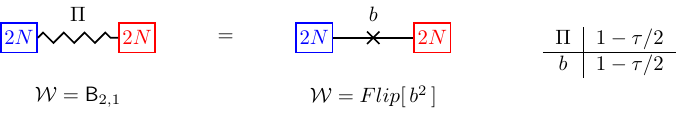}
\ee
As a superconformal index identity this limit can be written as:
\begin{align}
	\CI_{FE}^{(N)} \big( \vec{x},\vec{y},t,c = (pq/t)^{1/2} \big) = & \Ge(t) \prod_{j,k=1}^N \Ge( (pq/t)^{1/2} x_j^\pm y_k^\pm ) \,.
\end{align}

\paragraph{\boldmath{$USp(2N)^2$} breaking deformations} A class of deformations breaking the ${\color{red}USp(2N)}\times {\color{blue}USp(2N)}$ global symmetry consists in giving masses to the two antisymmetric operators in the form of Jordan matrices\footnote{Notice that a mass for the antisymmetric operator for the manifest $USp(2N)$ global symmetry induces a VEV for the mesonic operator $b_{N-1}^2$ which triggers a sequential Higgsing along the quiver tail.}.
The two deformations are specified uniquely by a pair of partitions $(\rho,\sigma)$ of $N$.
Throughout this paper we will be only interested in the case of a maximal deformations for one of the two antisymmetric fields which has the effect of breaking the relative $USp(2N)$ symmetry down to $SU(2)$. In practice, those kind of deformations can be implemented by specializing the fugacities of the $USp(2N)$ symmetry, let it be a $N$ dimensional vector $\vec{x}$, in terms of $\tau$, associated to the $U(1)_\tau$ symmetry, and a new fugacity $v$ which parameterize the remaining $SU(2)$ symmetry. The specialization is as follows:
\begin{align}
	x_i = \frac{N+1-2j}{2}\tau + v \,.
\end{align}
In order to obtain a theory with a well defined superpotential, let us add an antisymmetric field for the unbroken $USp(2N)$ symmetry, the result of the deformation is then given by:
\be\label{fig:FEtoFlav}
    \includegraphics[]{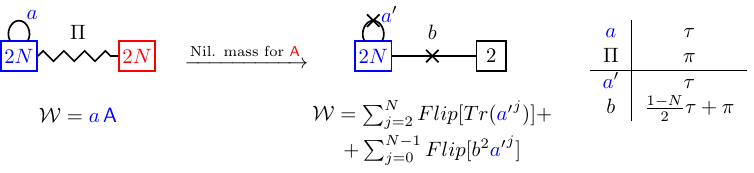}
\ee

As a superconformal index identity we write:
\begin{align}
	& \CI_{FE}^{(N)} ( \vec{x}, \{ t^{\frac{N-1}{2}} y, \ldots, t^{\frac{1-N}{2}} y \}, t, c ) =
    \prod_{j=1}^N \Ge( t^{\frac{1-N}{2}} c x_j^\pm v^\pm ) \prod_{j=2}^N \Ge(pq t^{-j} ) \prod_{j=0}^{N-1} \Ge( pq t^{N-1-j} c^{-2} ) \,.
\end{align}

\section{The braid or star-triangle duality}\label{sec:braid}

In this section we discuss the star-triangle or braid duality relating the star theory on the l.h.s. in  figure \eqref{fig:braid},
to the triangle theory on the r.h.s. in  figure \eqref{fig:braid}, an \emph{improved} Wess-Zumino model
 composed by an  improved bifundamental coupled to two singlets.
\be\label{fig:braid}
\resizebox{0.95\hsize}{!}{
    \includegraphics[]{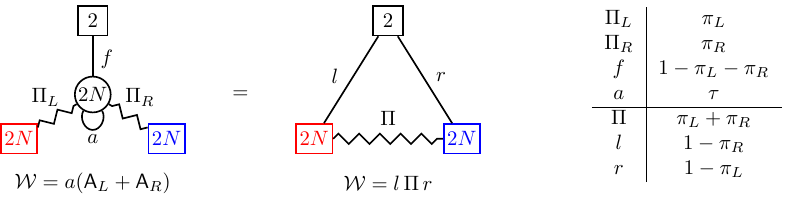}
}
\ee
The two antisymmetric fields $\mathsf{A}_L$ and $\mathsf{A}_R$ appearing in the superpotential are the moment maps charged under the gauge symmetry of the left and right $FE_N$ theories on the star side. 

The global symmetry of the two dual theories is given by:
\begin{align}
    {\color{red}USp(2N)} \times {\color{blue}USp(2N)} \times USp(2) \times U(1)_{\pi_L} \times U(1)_{\pi_R} \times U(1)_\tau \,,
\end{align}
where the number of abelian global symmetries can be computed by recalling that each $FE_N$ theory provides two $U(1)$'s. On the l.h.s.~we then have six $U(1)$'s, where two are provided by $f$ and $A$, and there is an NSVZ constraint plus two superpotential terms, leaving only three abelian global symmetries on top of the $\mathcal{N}=1$ R-symmetry. An analogous computation for the r.h.s.~provides the same answer.

For $N=1$ the braid duality in figure \eqref{fig:braid} reduces simply to the S-confinement of the $SU(2)$ SQCD with 6 fundamentals \cite{Seiberg:1994pq}. Regarding the $FE_N$ theory as an improved bifundamental, it is natural to think of the braid duality as an \emph{improved} S-confinement for $N>1$.

\paragraph{Operator map}
It is important to know how chiral BPS operators map across the duality.
The operators transforming in non trivial representations of the ${\color{red}USp(2N)} \times {\color{blue}USp(2N)} \times USp(2)$ global symmetry and some singlets are known to map as in the first $6$ lines of \eqref{braidmap} \cite{Pasquetti:2019hxf}. In this paper we propose the mapping of the remaining $USp(2N)^2 \times USp(2)$ singlets, as follows:
\begin{align}\label{braidmap}
\begin{tabular}{c|c|c}
	Star & Triangle & R-charge \\
	\hline
    $\Pi_L \Pi_R$ & $\Pi$ & $\pi_L +\pi_R$ \\
    $\Pi_L f$ & $r$ & $1-\pi_R$ \\
    $\Pi_R f$ & $l$ & $1-\pi_L$ \\
    ${\color{red}\mathsf{A}}_L$ & ${\color{red}\mathsf{A}}$ & $2-\tau$ \\
    ${\color{blue}\mathsf{A}}_R$ & ${\color{blue}\mathsf{A}}$ & $2-\tau$ \\
    $\Tr(f^2 a^m \mathsf{A}^{n})$ & $\mathsf{B}_{n+1, m+1}$ & $2(n+1)-2(\pi_L+\pi_R)+(m-n)\tau$ \\
    $(\mathsf{B}_{n+1, m+1})_{L}$ & $\Tr( r^2 {\color{blue}\mathsf{A}}^n {\color{blue}\mathsf{V}}^m )$ & $2(n+1)-2\pi_L+(m-n)\tau$ \\
    $(\mathsf{B}_{n+1, m+1})_{R}$ & $\Tr( l^2 {\color{red}\mathsf{A}}^n {\color{red}\mathsf{V}}^m )$ & $2(n+1)-2\pi_R+(m-n)\tau$ \\
    $\Tr(a^j)$ & $\Tr {\color{blue}\mathsf{V}}^j \sim \Tr {\color{red}\mathsf{V}}^j$ & $j \tau$ 
\end{tabular}
\end{align}
The F-terms for the adjoint field $a$, on the l.h.s.~of the braid duality \eqref{fig:braid}, imply a relation between the two moment maps: $\mathsf{A}_L \sim \mathsf{A}_R$. We therefore define as $\mathsf{A}$ the combination $(\mathsf{A}_L - \mathsf{A}_R)/2$. We also recall that $\color{red}\mathsf{V}$ and $\color{blue}\mathsf{V}$ are the virtual chiral operators defined in Section \ref{sec:FE} after the construction of the chiral ring generators below Table \ref{tab:FE_operators}. Comments regarding the operator map are in order.

We see that $(\mathsf{B}_{n+1, m+1})_{L}$ and $(\mathsf{B}_{n+1, m+1})_{R}$ for $m>0$ are trivial in the chiral ring  since they are mapped to operators built with the virtual moment maps. As we will see later when theories are embedded into a larger theory, the relation setting these operators to zero might change and identify these \emph{virtual} operators to other \emph{physical} operators. For this reason it is useful to study a larger operator map which might include operators that are trivial in the chiral ring. For example, the full mapping will be useful in Section \ref{sec:SD} when we construct star-star dualities.

Similarly, in the last line of the map in \eqref{braidmap} we observe that the traces of the adjoint chiral $a$ on the l.h.s.~are mapped to traces built from the virtual moment map of the $FE_N$ theory on the r.h.s. The relation that identifies $\Tr {\color{blue}\mathsf{V}}^j$ and $\Tr {\color{red}\mathsf{V}}^j$ was observed already in the flip-flip duality for the $FE_N$ theory \eqref{fig:FE_flipflip}, we now conjecture that this relation is true also in the theory on the r.h.s.~of the braid duality \eqref{fig:braid}. 

Notable absent from the list in \eqref{braidmap} are the operators  $\Tr \Pi_L^2, \Tr \Pi_R^2$ and also dressed ``mesons" involving the improved $FE_N$ bifundamental: $\Tr(\Pi_{L} a^m \mathsf{A}^n \Pi_{R})$, $\Tr(\Pi_{L/R} a^m \mathsf{A}^n f)$. These operators cannot be mapped to the dual theory because in the triangle theory there are no operators in the same representation of the global symmetries and with the same R-charge. In fact, these operators are set to zero by relations of the $FE_N$ theory, as discussed in Section \ref{sec:FE} (see also Appendix \ref{app:FE_relations}). Analogously, similar operators on the triangle side given by $\Pi$ and $l$ or $r$ dressed with powers of the antisymmetric operators 
${\color{red}\mathsf{A}}$, ${\color{blue}\mathsf{A}}$
 are absent from the list. 

\paragraph{Superconformal index identity} 
As an identity between $4d$ $\mathcal{N}=1$ superconformal indexes, the braid duality \eqref{fig:braid} can be written as:
\begin{align}\label{eq:braid_SCI}
	& \oint d\vec{z}_N \Delta_N(\vec{z},t) \mathcal{I}^{(N)}_{FE}(\vec{x},\vec{z},t,c_L) \mathcal{I}^{(N)}_{FE}(\vec{z},\vec{y},t,c_R) \prod_{j=1}^N \Gamma_e ( \sqrt{pq}(c_L c_R)^{-1} v^\pm z_j^\pm ) = \nn \\
	& = \mathcal{I}^{(N,N)}_{FE}(\vec{x},\vec{y},t,c_L c_R) \prod_{j=1}^N \left[ \Gamma_e ( \sqrt{pq}c_R^{-1}v^\pm x_j^\pm ) \Gamma_e ( \sqrt{pq}c_L^{-1}v^\pm y_j^\pm ) \right] \,.
\end{align}
The dictionary between the superconformal index fugacities $c_L,c_R,t$ and the R-charges appearing in \eqref{fig:braid} is:
\be
c_L = (pq)^{\pi_L/2}\,, \qquad c_R = (pq)^{\pi_R/2}\,, \qquad t = (pq)^{\tau/2} \,.
\ee
The identity \eqref{eq:braid_SCI} has been proven in the math literature in \cite{Rains_2018}. Such proof is structurally different from the proof we present in Section \ref{sec:proof}, where we provide an inductive proof of identity \eqref{eq:braid_SCI} which only relies on Intriligator-Pouliot duality \cite{Intriligator:1995ne} and therefore is a genuine field-theoretical proof.


\subsection{Interesting deformations}

The braid duality can be seen as the ancestor of many dualities that can be recovered by turning on various relevant deformations. The results presented in this section have been already discussed in \cite{Comi:2022aqo}.

The first class of deformation that we present is obtained turning on a superpotential term in the star side of the braid \eqref{fig:braid} of the type: 
\begin{align}\label{eq:BraidWf2Ak}
	\cW = f^2 \mathsf{A}^k \qquad \text{for} \,\,k=0, \cdots,N-1 \,.
\end{align}
Let us start from the case $k=0$, this corresponds to giving a mass to the flavor leaving two improved bifundamentals glued without the addition of matter.  This deformations leads to the fusion to identity property in \eqref{fig:FEdelta}.
Indeed on the l.h.s.~the superpotential is a mass term for the flavor $f$, which can be then integrated out. 
On the r.h.s.~this deformation is mapped into the term $\cW = \mathsf{B}_{1,1}$ which has the effect of transforming the improved bifundamental theory into a singular delta theory as \eqref{eq:FE_c=1}. As already pointed out, we can see the $\mathsf{B}_{1,1}$ deformation as a VEV for the bifundamental operator $\Pi$, therefore the superpotential $\cW = l \Pi r$ becomes a mass term for both $l$ and $r$.
The resulting duality is then:
\be
    \includegraphics[]{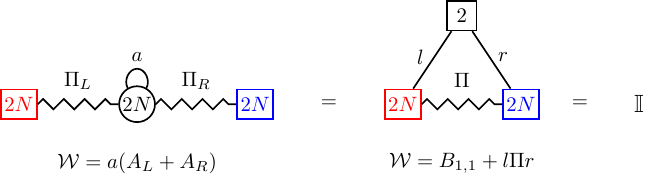}
\ee

The next case is given by superpotential \eqref{eq:BraidWf2Ak} for $k=1$. On the triangle side of the duality this deformation is mapped into: $\cW = \mathsf{B}_{2,1}$, which has the effect of ironing the improved bifundamental to a $USp(2N)^2$ standard one (see \eqref{fig:FE_c=t/2}). The resulting duality is given by:
\be\label{fig:Basic_S-move}
    \includegraphics[]{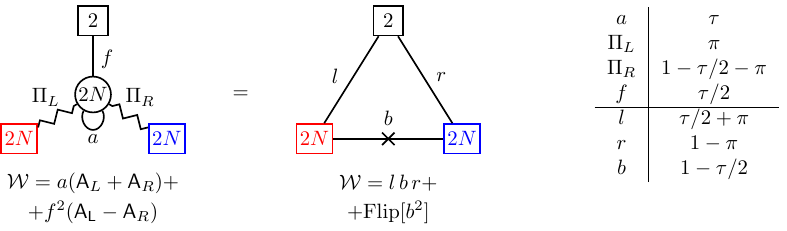}
\ee
This property was first derived in \cite{Bottini:2021vms} by iterating IP dualities and then it was interpreted as the basic 
non-abelian mirror duality move in \cite{Hwang:2021ulb}.

In principle, one could also study cases with higher $k$, since we will not be interested to this type of deformations we leave this analysis to future work. 

Another class of deformations consists in nilpotent masses for the antisymmetric operators, similar to those described in section \ref{FEdeform}. The maximal nilpotent mass breaks one of the $USp(2N)$ symmetries down to $USp(2)$, doing so in the braid duality \eqref{fig:braid}, let's say for the antisymmetric field charged under ${\color{blue}USp(2N)}$, we obtain the following duality:
\be\label{fig:Braidnilvev}
\resizebox{.95\hsize}{!}{ 
    \includegraphics[]{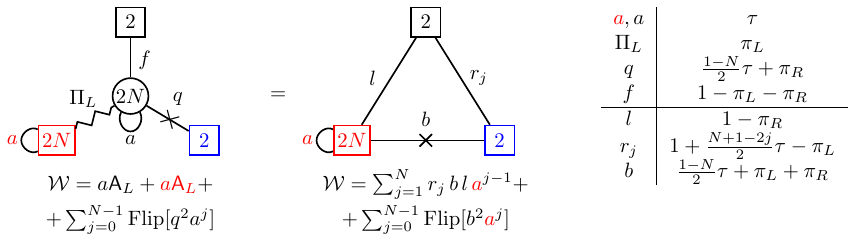}
}
\ee
Notice that we added an extra antisymmetric field with respect to the original braid duality defined in \eqref{fig:braid}, this is needed in order to write a consistent superpotential on the r.h.s.~of the duality \eqref{fig:Braidnilvev}. 

We can also consider the case where we turn on a nilpotent mass for both antisymmetric operators.
This deformation leads to the duality in \eqref{fig:USp(2N)w6}
\be
\resizebox{.95\hsize}{!}{
    \includegraphics[]{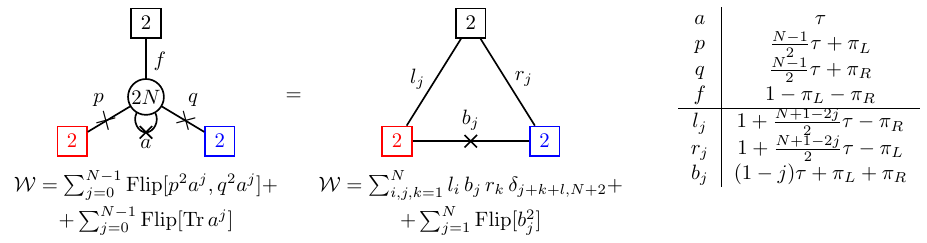}
}
\ee
which is a known s-confining duality for the $USp(2N)$ SQCD with 6 chirals and an antisymmetric field \eqref{fig:USp(2N)w6} \cite{Cs_ki_1997}. The full superpotential in the Wess-Zumino side contains $\sim N^2$ terms, it was proposed in \cite{Benvenuti:2018bav} and derived via sequential deconfinement in \cite{Bottini:2022vpy,Bajeot:2022kwt}.

\section{From the star-triangle duality to the star-star triality}\label{sec:SD}

In this section we derive and study the following star-star triality:
\be \label{fig:4d_starstar}
\resizebox{.95\hsize}{!}{
    \includegraphics[]{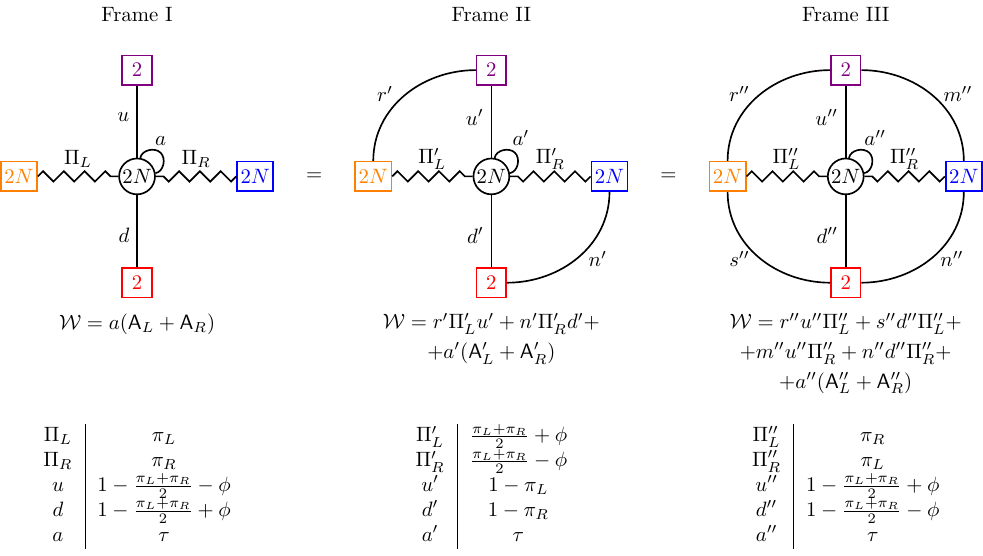}
}\ee
The IR global symmetry of the three theories figure \eqref{fig:4d_starstar} is
\begin{align}
    {\color{orange}USp(2N)} \times {\color{blue}USp(2N)} \times 
    SU(4) \times U(1)_{\pi_L} \times U(1)_{\pi_R} \times U(1)_\tau \,.
\end{align}
In frames I and frame III this symmetry is manifest in the UV since the two flavors and the flippers are on the same footing and therefore there we can recollect the global symmetries as:
\begin{align}
    {\color{violet}USp(2)} \times {\color{red}USp(2)} \times U(1)_{\phi} \to SU(4) \,,
\end{align}
in frame II this global symmetry is not manifest in the UV since it is manifestly broken by the flippers. Indeed in frame II the global symmetry enhances in the IR to match the manifest global symmetry of the two other frames. 

The duality web in figure \eqref{fig:4d_starstar} generalizes, and can be deformed to, the self-duality web of the $USp(2N)$  SQCD with an antisymmetric field and $8$ fundamental chirals. By turning on nilpotent VEVs for both the antisymmetrics charged under the orange and blue $USp(2N)$ global symmetries in each frame, we land on the triality:
\be\label{fig:4d_starstar_limit}
\resizebox{.95\hsize}{!}{
    \includegraphics[]{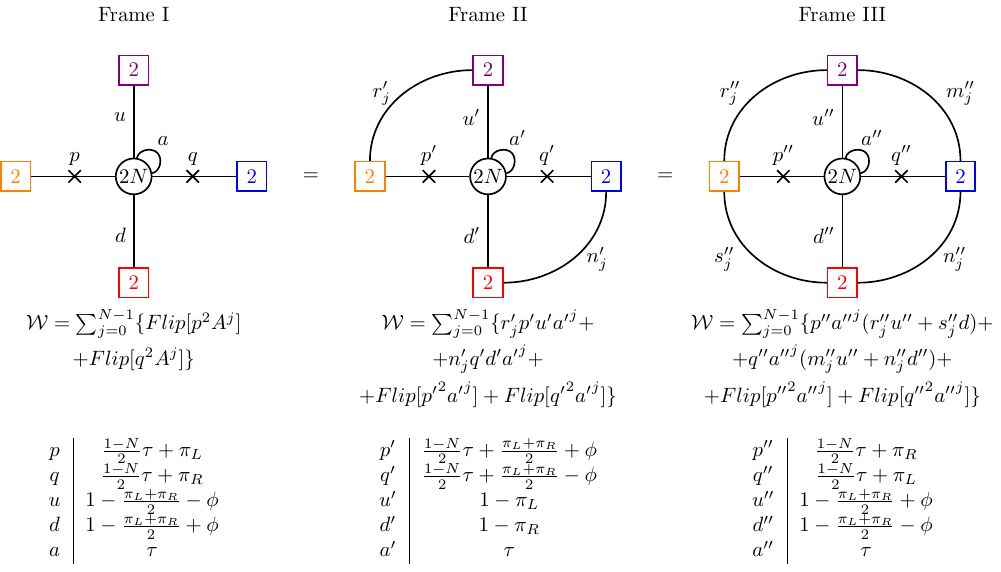}}
\ee
This triality can be used to generate the whole set of 72 duality frames for the $USp(2N)$ SQCS with 8 chirals and an antisymmetric field (see Appendix \ref{app:USp2Nw/8}). \\
The dualities between frame I and II and that between frame I and III in figure \eqref{fig:4d_starstar_limit} were both presented in \cite{Spiridonov_2010} (modulo flips) and we refer to them respectively as  CSST-like and Seiberg-like dualitiy. Analogously, we name the dualities between frame I and II and that between frame I and III in figure \eqref{fig:4d_starstar} as generalized-CSST and generalized-Seiberg dualities.

\section*{Superconformal index identity}
The superconformal indices associated to each frame of the improved star-star triality in \eqref{fig:4d_starstar} are given by:
\begin{align}\label{eq:starstar_ind_frame1}
	\CI_{\text{Frame I}} = & \oint d\vec{z}_N \D_N(\vec{z},t) \CI_{FE}^{(N)}(\vec{x},\vec{z},t,c_L) \CI_{FE}^{(N)}(\vec{y},\vec{z},t,c_R) \prod_{j=1}^N \prod_{a=1}^2 
    \Ge \big( (\tfrac{pq}{c_L c_R})^{1/2} f^{-1} z_j^\pm v_a^\pm \big)  \,,
\end{align}

\begin{align}\label{eq:starstar_ind_frame2}
	\CI_{\text{Frame II}} = & \oint d\vec{z}_N \D_N(\vec{z},t) \CI_{FE}^{(N)}(\vec{x},\vec{z},t,(c_Lc_R)^{1/2}f) \CI_{FE}^{(N)}(\vec{y},\vec{z},t,(c_Lc_R)^{1/2}f^{-1}) \times  \nn \\
	& \times \prod_{j=1}^N \big[ \Ge \big( (pq)^{1/2} c_L^{-1} z_j^\pm v_1^\pm \big) \Ge \big( (pq)^{1/2} c_R^{-1} z_j^\pm v_2^\pm \big) \times \nn \\
	& \times  \Ge \big( (\tfrac{pq \, c_R}{c_L})^{1/2}  f^{-1} x_j^\pm v_1^\pm \big) \Ge \big( (\tfrac{pq \, c_L}{c_R})^{1/2} f y_j^\pm v_2^\pm \big) \big] \,,
\end{align}

\begin{align}\label{eq:starstar_ind_frame3}
	\CI_{\text{Frame II}} = & \oint d\vec{z}_N \D_N(\vec{z},t) \CI_{FE}^{(N)}(\vec{x},\vec{z},t,c_R) \CI_{FE}^{(N)}(\vec{y},\vec{z},t,c_L) \prod_{j=1}^N \prod_{a=1}^2 \big[ 
    \Ge \big( (\tfrac{pq}{c_L c_R})^{1/2} f z_j^\pm v_a^\pm \big) \times \nn \\
	& \times \Ge \big( (\tfrac{pq \, c_L}{c_R})^{1/2} f^{-1} x_j^\pm v_a^\pm \big) \Ge \big( (\tfrac{pq \, c_R}{c_L})^{1/2} f^{-1} y_j^\pm v_a^\pm \big) \big] \,.
\end{align}
The triality consist in the following chain of identities:
\begin{align}
	\CI_{\text{Frame I}} = \CI_{\text{Frame II}} = \CI_{\text{Frame III}} \,,
\end{align} 
which can be proven as we explain in the next subsections.
To write the superconformal indexes we defined $\vec{x},\vec{y}$ to be the fugacities of the orange and blue $USp(2N)$ flavor groups; $v_1,v_2$ are the fugacities for the violet and red  $USp(2)$ flavors and $\vec{z}$ is the set of gauge fugacities. The map between $U(1)$ fugacities and R-charge mixings is:
\begin{align}
	t = (pq)^{\tau/2} \qquad , \qquad c_L = (pq)^{\pi_L/2} \qquad , \qquad c_R = (pq)^{\pi_R/2} \qquad , \qquad f = (pq)^{\phi/2} \,.
\end{align}
As already explained below figure \eqref{fig:4d_starstar}, in frame I and III there is a bigger global symmetry. The fugacities for the $SU(4)$ global symmetry can be defined from $v_1,v_2$ and $\phi$ as:
\begin{align}
    \vec{w} = \{ \phi v_1, \phi v_1^{-1}, \phi^{-1} v_2, \phi^{-1} v_2^{-1} \} \quad : \quad \prod_{j=1}^4 w_j = 1 \,.
\end{align}

\subsection{Frame I = Frame II: the generalized CSST-like duality}
Let's focus on the duality relating frames I and II:
\be\label{fig:wigCSST}
    \includegraphics[]{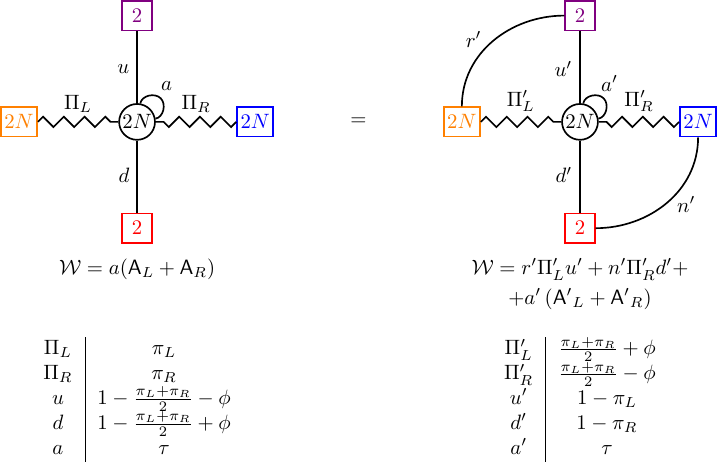}
\ee
The operator map for this duality reads:
\begin{align}\label{mapCSST}
\begin{tabular}{c|c|c}
	Frame I & Frame II & R-charge \\
	\hline
    $\Pi_L \Pi_R$ & $\Pi'_L \Pi'_R$ & $\pi_L +\pi_R$ \\
    $\Pi_L u$ & $r'$ & $1+\pi_L/2 -\pi_R/2-\phi$ \\
    $\Pi_L d$ & $\Pi'_L d'$ & $1+\pi_L/2 -\pi_R/2+\phi$ \\
    $\Pi_R u$ & $\Pi'_R u'$ & $1-\pi_L/2 +\pi_R/2-\phi$ \\
    $\Pi_R d$ & $n'$ & $1-\pi_L/2 +\pi_R/2+\phi$ \\
    ${\color{orange}\mathsf{A}}_L$ & ${\color{orange}\mathsf{A}}'_L$ & $2-\tau$ \\
    ${\color{blue}\mathsf{A}}_R$ & ${\color{blue}\mathsf{A}}'_R$ & $2-\tau$ \\
    $u d a^m \mathsf{A}^{n}$ & $u'\, d'\, {a'}^{m}\, \mathsf{A'}^{n}$ & $2(n+1)-\pi_L-\pi_R+(m-n)\tau$ \\
    $u^2 a^m \mathsf{A}^{n}$ & $(\mathsf{B}_{m,n})'_L$ & $2(n+1)-\pi_L-\pi_R+(m-n)\tau-2\phi$ \\
    $d^2 a^m \mathsf{A}^{n}$ & $(\mathsf{B}_{m,n})'_R$ & $2(n+1)-\pi_L-\pi_R+(m-n)\tau+2\phi$ \\
    $(\mathsf{B}_{n,m})_L$ & ${u'}^{2} {a'}^{m}{u'}^{2} {a'}^{m} \mathsf{A'}^{n}$ & $2(n+1)-2\pi_L+(m-n)\tau$ \\
    $(\mathsf{B}_{n,m})_R$ & ${d'}^2 {a'}^m \mathsf{A'}^{n}$ & $2(n+1)-2\pi_R+(m-n)\tau$ \\
    $\Tr(a^j)$ & $\Tr({a'}^j)$ & $j \, \tau$ \\
    $\Tr(\mathsf{A}^j)$ & $\Tr(\mathsf{A'}^j)$ & $j \, (2-\tau)$
\end{tabular}
\end{align}
Again the F-terms for the adjoint field $a$, $a'$ imply a relation between the two moment maps: $\mathsf{A}_L \sim - \mathsf{A}_R$ and  $\mathsf{A'}_L \sim - \mathsf{A'}_R$ which we identify respectively with 
 $\mathsf{A}$ and  $\mathsf{A'}$.

Notice that the operators listed in the map \eqref{mapCSST} recollect into representations of $SU(4)$. In particular for Frame I:
\begin{align}\label{eq:SU(4)repmap}
    \{ \Pi_L u , \Pi_L d \} &\quad \to \quad {\bf \bar{4}} \qquad 1 + \pi_L/2 - \pi_R/2 \nn \\
    \{ \Pi_R u , \Pi_R d \} &\quad \to \quad {\bf \bar{4}} \qquad 1 - \pi_L/2 + \pi_R/2 \nn \\
    \{ uda^m\mathsf{A}^n , u^2a^m\mathsf{A}^n , d^2a^m\mathsf{A}^n \} &\quad \to \quad {\bf \bar{6}} \qquad 2(n+1) - \pi_L - \pi_R + (m-n)\tau \nn \\
\end{align}
while the rest are singlets for $SU(4)$. In Frame II the operators recollect accordingly to the map \eqref{mapCSST}.

We can prove the duality \eqref{fig:wigCSST} in a couple of simple steps. Let us start from the following auxiliary quiver: 
\be\label{wigCSSTproof}
    \includegraphics[]{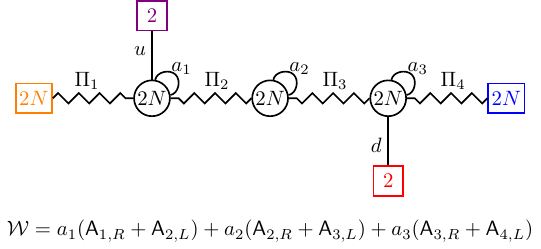}
\ee

If we use the fusion to identity property
for the two central improved bifundamentals with charges $\Pi_2,\Pi_3$ we obtain the theory on the l.h.s.~of \eqref{fig:wigCSST}. If we instead we apply twice the star-triangle duality \ref{fig:braid} to the 
two star-shaped parts of the quiver \eqref{wigCSSTproof}
involving  respectively the improved bifundamentals and flavor with charges $\Pi_1,\Pi_2,u$ and $\Pi_3,\Pi_4,d$, 
we land on the theory on the r.h.s.~of \eqref{fig:wigCSST}.

It is immediate to retrace these steps at the level of the superconformal index to prove the triality identities.
The index associated to quiver \eqref{wigCSSTproof} theory is given by:
\begin{align}\label{eq:starstar_ind_start}
	\oint & \prod_{a=1}^3 \big[ d\vec{z}^{\,(a)}_N \D_N(\vec{z}^{\,(a)},t) \big] \CI_{FE}^{(N)}(\vec{x},\vec{z}^{(1)},t,c_L) \prod_{j=1}^N \Ge \big( (\tfrac{pq}{c_L c_R})^{1/2} f^{-1} v_1^\pm z_j^{(1)\pm} \big) \times \nn \\
	& \times \CI_{FE}^{(N)}(\vec{z}^{\,(1)},\vec{z}^{\,(2)},t,(c_R/c_L)^{1/2}f) \CI_{FE}^{(N)}(\vec{z}^{\,(2)},\vec{z}^{\,(3)},t,(c_R/c_L)^{-1/2}f^{-1}) \times \nn \\
	& \times \prod_{j=1}^N \Ge \big( (\tfrac{pq}{c_L c_R})^{1/2} f v_2^\pm z_j^{(3)\pm} \big) \CI_{FE}^{(N)}(\vec{z}^{\,(3)},\vec{y},t,c_R ) \,,
\end{align}
where for the three $USp(2N)$ gauge groups we turned on fugacities by $\vec{z}^{\,(1)},\vec{z}^{\,(2)},\vec{z}^{\,(3)}$ from left to right. We can now recognize that in the second line of the superconformal index \eqref{eq:starstar_ind_start} once we integrate over 
$\vec{z}^{\,(2)}$ we produce an  identity wall, implementing it has the effect of identifying the first and third gauge node. If we baptize the set of fugacities for the remaining gauge node as $\vec{z}$ we land precisely on the superconformal index of the frame I in \eqref{eq:starstar_ind_frame1}. 

If we instead use twice the star-trinagle dualilty \eqref{eq:braid_SCI} after relabelling $\vec{z}^{(2)}$ as $\vec{z}$ we obtain precisely the superconformal index of frame II \eqref{eq:starstar_ind_frame2}. All in all, we establish the duality between frame one and two as:
\begin{align}
	\CI_{\text{Frame I}} = \CI_{\text{Frame II}} \,.
\end{align}

The operator map in \eqref{mapCSST} can be  worked out starting from
the auxiliary quiver  and composing  the map for the braid duality \eqref{braidmap}.
The first $6$ lines in \eqref{mapCSST} are simply the shortest path between the two flavor symmetry nodes connected by the meson. For example $\Pi_L d$ is mapped to $\Pi_1 \Pi_2 \Pi_3 d$ in the auxiliary quiver \eqref{wigCSSTproof}. Using the braid map \eqref{braidmap} we see  that $\Pi_1 \Pi_2\to \Pi'_L$ while $\Pi_3 d \to d'$  so we conclude that $\Pi_L d \to \Pi'_L  d'$.
The mapping of the singlets is trickier but still can be inferred from the map of the braid \eqref{braidmap}. For example, $(\mathsf{B}_{n,m})_L$ is mapped to $u^{'2} \mathsf{A}^n \mathsf{V'_L}^m\sim u^{'2} \mathsf{A}^n a^{'m} $ in Frame II  since  the Virtual moment maps are now real and identified as $\mathsf{V'_L}\sim \mathsf{V'_R}\sim a'$.

\subsection{Frame I = Frame III: the generalized Seiberg-like duality}\label{sec:genSeib}
Let us focus now on the duality relating frame I and III:
\be\label{wigSeiberg}
    \includegraphics[]{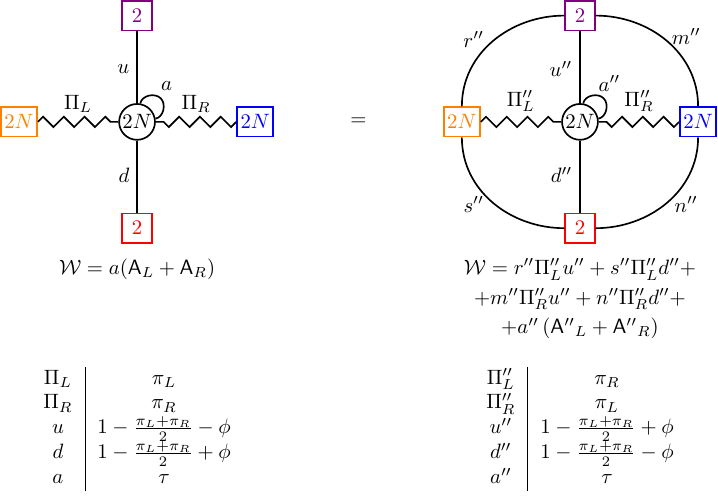}
\ee
The operator map for this duality reads:
\begin{align}\label{mapSeiberg}
\begin{tabular}{c|c|c}
	Frame I & Frame III & R-charge \\
	\hline
    $\Pi_L \Pi_R$ & $\Pi''_L \Pi''_R$ & $\pi_L +\pi_R$ \\
    $\Pi_L u$ & $r''$ & $1+\pi_L/2 -\pi_R/2-\phi$ \\
    $\Pi_L d$ & $s''$ & $1+\pi_L/2 -\pi_R/2+\phi$ \\
    $\Pi_R u$ & $m''$ & $1-\pi_L/2 +\pi_R/2-\phi$ \\
    $\Pi_R d$ & $n''$ & $1-\pi_L/2 +\pi_R/2+\phi$ \\
    ${\color{orange}\mathsf{A}}_L$ & ${\color{orange}\mathsf{A}}''_L$ & $2-\tau$ \\
    ${\color{blue}\mathsf{A}}_R$ & ${\color{blue}\mathsf{A}}''_R$ & $2-\tau$ \\
    $Tr(u d a^m (\mathsf{A})^{n})$ & $Tr(u'' {d''} {a''}^m (\mathsf{A''})^{n})$ & $2(n+1)-\pi_L-\pi_R+(m-n)\tau$ \\
    $Tr(u^2 a^m (\mathsf{A})^{n})$ & $Tr({d''}^2 {a''}^m (\mathsf{A''})^{n})$ & $2(n+1)-\pi_L-\pi_R+(m-n)\tau-2\phi$ \\
    $Tr(d^2 a^m (\mathsf{A})^{n})$ & $Tr({u''}^2 {a''}^m (\mathsf{A''})^{n})$ & $2(n+1)-\pi_L-\pi_R+(m-n)\tau+2\phi$ \\
    $(\mathsf{B}_{n,m})_L$ & $(\mathsf{B}''_{n,m})_R$ & $2(n+1)-2\pi_L+(m-n)\tau$ \\
    $(\mathsf{B}_{n,m})_R$ & $(\mathsf{B}''_{n,m})_L$ & $2(n+1)-2\pi_R+(m-n)\tau$
\end{tabular}
\end{align}
As already noticed, both frame I and frame III exibit a manifest $SU(4)$ global symmetry. The duality \eqref{wigSeiberg} can be rewritten in an $SU(4)$-invariant notation as follows
\be\label{wigSeibergsu4}
    \includegraphics[]{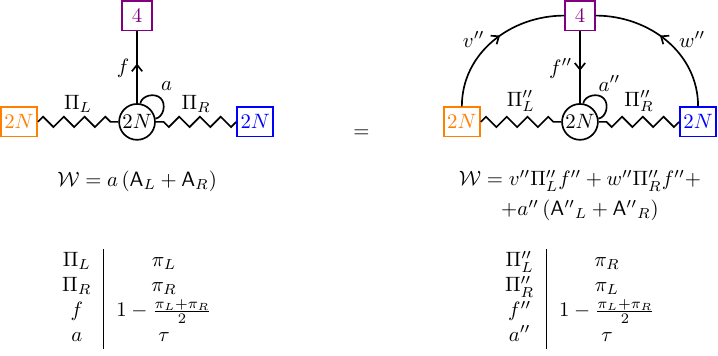}
\ee
Also the operator map \eqref{mapSeiberg} can be rewritten recollecting operators into representation of $SU(4)$ as in \eqref{eq:SU(4)repmap}.

We can prove the duality following the strategy of \cite{Benvenuti:2018bav} where it was shown how to derive the Seiberg-like self-duality  
relating frames I and III in \eqref{fig:4d_starstar_limit}, by iterating the  CSST-like duality relating frames I and II in \eqref{fig:4d_starstar_limit}.
Consider the generalized CCST-like duality in  \eqref{fig:wigCSST}, clearly there is a similar duality where the two $USp(2N)$ flavor symmetries are exchanged, that is on the r.h.s.~the flippers $s'$ and $m'$ flip the mesons $\Pi'_L d', \Pi'_R u'$ instead of $\Pi'_L u', \Pi'_R d'$:
\be\label{wigCSSTp} 
    \includegraphics[]{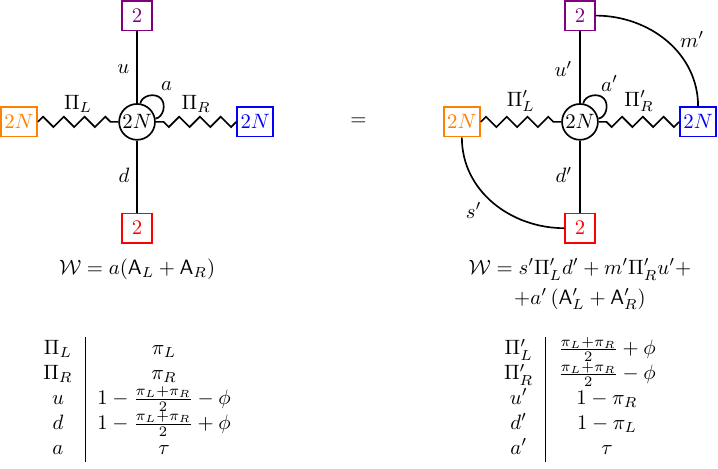}
\ee
Now we can use sequentially the two dualities \eqref{fig:wigCSST} and \eqref{wigCSSTp} starting from frame I. The net effect of this two dualities is to generate the full set of flippers present in frame III, thus leading to the generalized Seiberg-like duality in \eqref{wigSeiberg}.

\section{An inductive field theory proof of the Braid duality}\label{sec:proof}
We now present a QFT proof of the Braid or star-triangle duality.

\begin{figure}[h!]
\centering
\resizebox{.75\hsize}{!}{
    \includegraphics[]{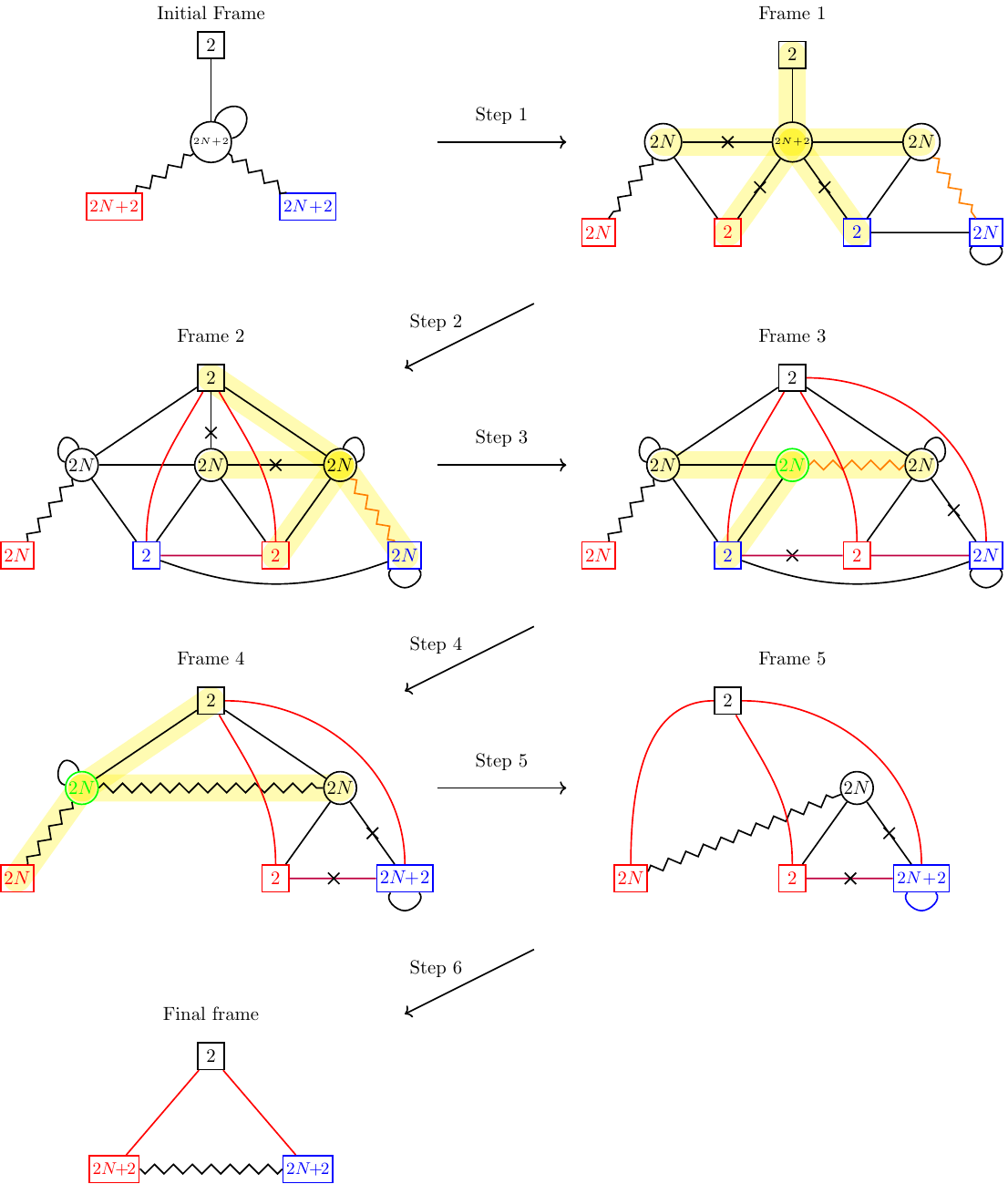}
}
\caption{Schematic summary of the inductive proof of the braid duality. In each frame the fields on which we act with a duality are highlighted in yellow.}
\label{fig:braid_proof_summary}
\end{figure}

The strategy is to prove inductively that the braid duality for rank $N+1$ is a consequence of the the braid duality for rank $N$ and of the IP duality, as summarized in Figure \ref{fig:braid_proof_summary}. Since the braid duality for rank $1$ is just the s-confining duality for  $USp(2)$ with 6 chirals, we can conclude that braid duality is a consequence of basic Seiberg-like dualities. \\
During the proof we will employ deformed versions of both the braid duality and the generalized-Seiberg duality, the specific deformed dualities are reported for reference in Appendix \ref{app:aux_dualities}.  Since these dualities  at rank $N$ are consequences of the star-triangle duality at rank $N$ we are allowed to use them in the inductive step.
	
\paragraph{Step 1}
In the initial frame we have the theory:
\be\label{frame0}
    \includegraphics[]{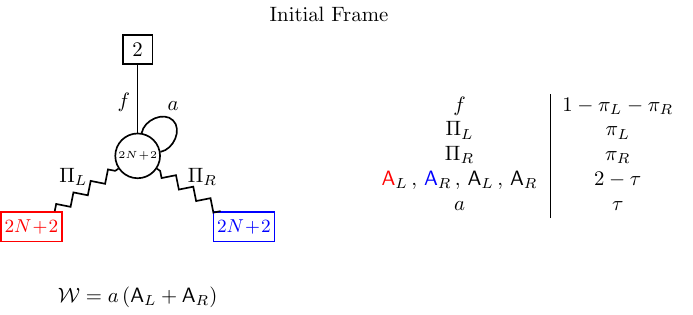}
\ee
In the first step we perform a flip-flip duality on the right $FE_N$. The antisymmetric of the $USp(2N+2)$ gauge node becomes massive and we each the theory:
\be\label{framep1} 
    \includegraphics[]{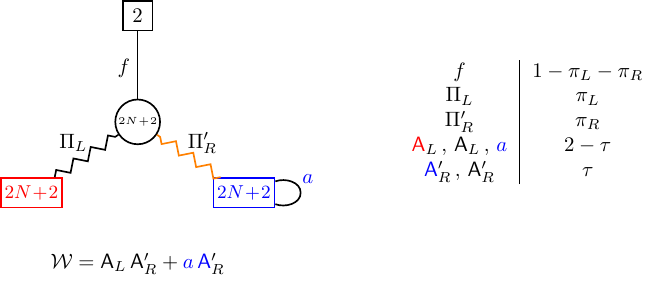}
\ee
Notice that by integrating out the antisymmetric field $a$ in the presence of the superpotential $a(\mathsf{A}_L + \mathsf{A}_R)$ \eqref{frame0} generates the new term $\mathsf{A}_L \mathsf{A}'_R$ in the figure above.
We then use the recursive defintion of the  $FE_{N+1}$ theory \eqref{fig:FE_part_open} to arrive at frame 1:\footnote{To avoid cluttering, from now we will provide the R-charges only of the fields charged under the gauge symmetries. The R-charge of the singlets can be deducted from the superpotential. Also, we will not give the R-charge of antisymmetric operators of the $FE_N$ theories, we recall that in the ``standard" black $FE_N$ it is always $2-\tau$, conversly in the orange ``flipped" $FE_N$ it is $\tau$.}
\be\label{frame1} 
    \includegraphics[]{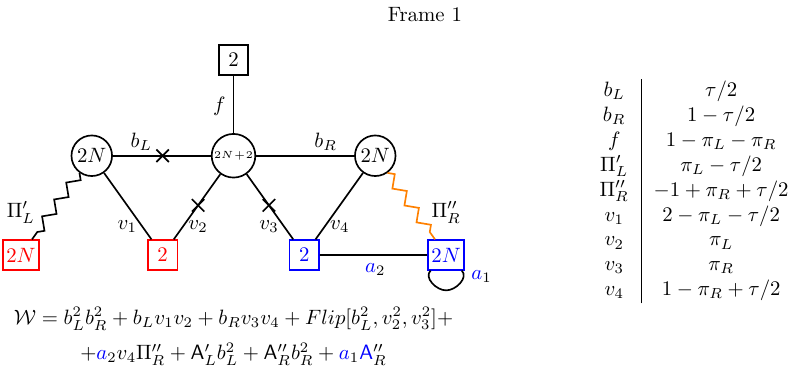}
\ee
Notice that the ${\color{blue}USp(2N+2)}$ antisymmetric $\color{blue}a$, with R-charge $2-\tau$, decomposes into the  ${\color{blue}USp(2N)}$  antisymmetric ${\color{blue}a}_1$, the ${\color{blue}USp(2)} \times {\color{blue}USp(2N)}$ bifundamental ${\color{blue}a}_2$ and a singlet (which disappears upon pairing with the flipper of $b_R^2$):

Let us write down the operator mapping between the initial frame \eqref{frame0} and frame one \eqref{frame1}, which can be obtained combining the discussion after \eqref{fig:FE_part_open} and the mapping \eqref{eq:flipflip_map}:
\begin{align*}
\renewcommand{\arraystretch}{1.1}
\resizebox{.95\hsize}{!}{
\begin{tabular}{c|c|c}
	Initial frame & Frame 1 & R-charge \\
	\hline
	$\Pi_L \Pi_R$ & $\{ \Pi'_L b_L v_3 \,,\, \Pi'_L b_L b_R \Pi''_R \,,\, v_2 v_3 \,,\, v_2 b_R \Pi''_R \}$ & $\pi_L + \pi_R$ \\
	$\Pi_L f$ & $\{ \Pi'_L b_L f \,,\, v_2 f \}$ & $1 - \pi_R$ \\
	$\Pi_R f$ & $\{ \Pi''_R b_R f \,,\, v_3 f \}$ & $1 - \pi_L$ \\
	${\color{red}\mathsf{A}}_L$ & $\{ {\color{red}\mathsf{A}}'_L \,,\, \Pi'_L v_1 \,,\, \CF[b_L^2] \}$ & $2-\tau$ \\
	${\color{blue}\mathsf{A}}_R$ & $\{ {\color{blue}a}_1 \,,\, {\color{blue}a}_2 \,,\, b_R^2 \}$ & $2-\tau$ \\
	$f^2 a^m \mathsf{A}^n$ & $f^2 b_L^{2m} b_R^{2n}$ & $2n+2-2(\pi_L + \pi_R) + (m-n)\tau$
\end{tabular}}
\end{align*}

\paragraph{Step 2}
In the second step we focus on the part of frame 1 highlighted in Figure \ref{fig:braid_proof_summary} and dualize the cental $USp(2N)$ node in \eqref{frame1}, using Intriligator-Pouliot duality. The quiver becomes:
\be\label{frame2}
    \includegraphics[]{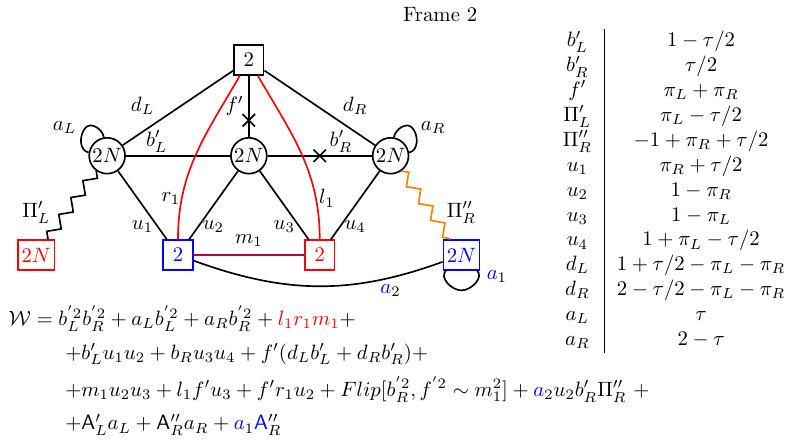}
\ee
Notice that 
\begin{itemize}
\item we added a term in the superpotential, $l_1 r_1 m_1$, highlighted in red. This term has R-charge $2$, and  it is generated, in a similar fashion as  superpotentials are generated in the confined frames of s-confining dualities. The superpotential term $l_1 r_1 m_1$ is the first term of a series that will build the cubic superpotential $l\, \Pi \, r$ going around the triangle in the final duality frame \eqref{framef}. 
\item since $f^{'2}$ and $m_1^2$ are degenerate, we claim that the operator $f^2$ in the previous frames maps to a singlet flipping a linear combination of $f^{'2}$ and $m_1^2$ (see \cite{Bajeot:2022kwt,Bajeot:2022lah} for similar subtleties arising when performing Seiberg dualities in quivers with degenerate holomorphic operators).
\end{itemize}
The mapping between frame  1 \eqref{frame1} and frame 2 \eqref{frame2} is:
\begin{align*}
\renewcommand{\arraystretch}{1.1}
\resizebox{.95\hsize}{!}{
\begin{tabular}{c|c|c}
	Frame 1 & Frame 2 & R-charge \\
	\hline
	$\{ \Pi'_L b_L v_3 \,,\, \Pi'_L b_L b_R \Pi''_R \,,\, v_2 v_3 \,,\, v_2 b_R \Pi''_R \}$ & $\{ \Pi'_L u_1 \,,\, \Pi'_L b'_L b'_R \Pi''_R \,,\, m_1 \,,\, u_4 \Pi''_R \}$ &$\pi_L + \pi_R$ \\
	$\{ \Pi'_L b_L f \,,\, v_2 f \}$ & $\{ \Pi'_L d_L \,,\, l_1 \}$ & $1 - \pi_R$ \\
	$\{ \Pi''_R b_R f \,,\, v_3 f \}$ & $\{ \Pi''_R d_R \,,\, r_1 \}$ & $1 - \pi_L$ \\
	$\{ {\color{red}\mathsf{A}}'_L \,,\, \Pi'_L v_1 \,,\, \CF[b_L^2] \}$ & $\{ {\color{red}\mathsf{A}}'_L \,,\, \Pi'_L b'_L u_3 \,,\, b_L^{'2} \}$ & $2-\tau$ \\
	$\{ {\color{blue}a}_1 \,,\, {\color{blue}a}_2 \,,\, b_R^2 \}$ & $\{ {\color{blue}a}_1 \,,\, {\color{blue}a}_2 \,,\, \CF[b_R^{'2}] \} $ & $2-\tau$ \\
	$f^2$ & $\CF[f^{'2} \sim m_1^2]$ & $2-2\pi_L-2\pi_R$ \\
	$f^2b_L^{2m}$ & $d_L^2 a_L^{m-1}$ & $2-2(\pi_L+\pi_R) + m \tau$ \\
	$f^2b_R^{2n}$ & $d_R^2 a_R^{n-1}$ & $2(1+n)-2(\pi_L+\pi_R) - n \tau$ \\
	$f^2b_L^{2m+2}b_R^{2n+2}$ & $d_L a_L^m b'_L b'_R a_R^n d_R$ & $2n+4-2(\pi_L+\pi_R)+(m-n)\tau$
\end{tabular}}
\end{align*}

\paragraph{Step 3}
In the third step we focus on the part of frame 2 highlighted in Figure \ref{fig:braid_proof_summary}.
We apply the improved Seiberg duality specialisation \eqref{fig:proof_defseib}.
The dualization produces also $2N \cdot 8$ flipping chiral fields, half of which become massive upon pairing with $f'$ and $u_3$.
\be\label{frame3}
    \includegraphics[]{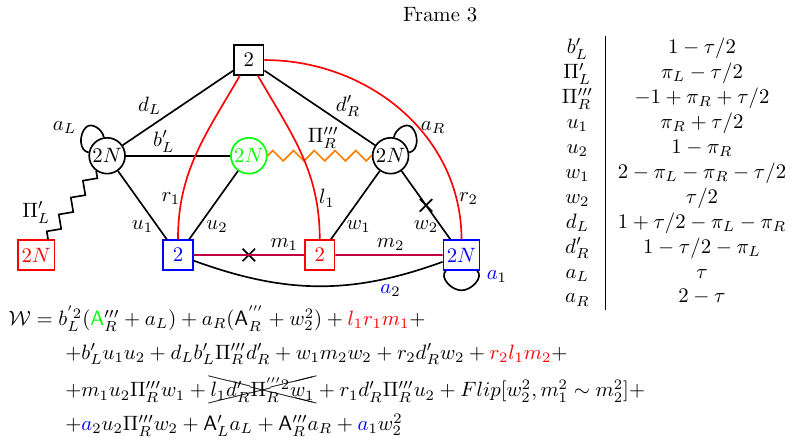}
\ee
Notice that the term $l_1 f_1 u_3$ in the superpotential in frame 2 \eqref{frame2} under the duality \eqref{fig:proof_defseib} using the map \eqref{mapSeiberg} becomes  the term $l_1 d'_R {\Pi'''}s^2_R w_1$ which we removed from the superpotential. This is because  the operator $(\Pi'''_R)^2$ is zero in the chiral ring (see Appendix \ref{app:FE_relations}). So the chiral ring stability criterion of \cite{Benvenuti:2017lle} forces us to remove the term $l_1 d'_R \Pi^{'''2}_R w_1$. This is counterbalanced by the appearance of the term $r_2 l_1 m_2$, which has R-charge $2$.\footnote{As a consistency check, let us consider what happens when we apply the improved Seiberg duality to the right node in  \eqref{frame3} (we must go back to \eqref{frame2}): the term $r_2 l_1 m_2$ is mapped to $u_4 \Pi_R^{''2} d_R l_1$, which is chiral ring unstable since $\Pi_R^{''2} \sim 0$, so it must be dropped, but the term $u_3 l_1 f'$ appears, and we precisely land on \eqref{frame3}.} Notice that without the red terms in the superpotential, that were added even if not directly produced from the dualizations, the field $l_1$ would be completely decoupled from the rest of the theory. \\
The mapping between frame 2 \eqref{frame2} and frame 3 \eqref{frame3} is:
\begin{align*}
\renewcommand{\arraystretch}{1.1}
\resizebox{.95\hsize}{!}{
\begin{tabular}{c|c|c}
	Frame 2 & Frame 3 & R-charge \\
	\hline
	$\{ \Pi'_L u_1 \,,\, \Pi'_L b'_L b'_R \Pi''_R \,,\, m_1 \,,\, u_4 \Pi''_R \}$ & $ \{ \Pi'_L u_1 \,,\, \Pi'_L b'_L \Pi'''_R w_2 \,,\, m_1 \,,\, m_2  \} $ & $\pi_L + \pi_R$ \\
	$\{ \Pi'_L d_L \,,\, l_1 \}$ & $\{ \Pi'_L d_L \,,\, l_1 \}$ & $1 - \pi_R$ \\
	$\{ \Pi''_R d_R \,,\, r_1 \}$ & $\{ r_2 \,,\, r_1 \}$ & $1 - \pi_L$ \\
	$\{ {\color{red}\mathsf{A}}'_L \,,\, \Pi'_L b'_L u_3 \,,\, b_L^{'2} \}$ & $\{ {\color{red}\mathsf{A}}'_L \,,\, \Pi'_L b'_L \Pi'''_R w_1 \,,\, b_L^{'2} \} $ & $2-\tau$ \\
	$\{ {\color{blue}a}_1 \,,\, {\color{blue}a}_2 \,,\, \CF[b_R^{'2}] \} $ & $\{ {\color{blue}a}_1 \,,\, {\color{blue}a}_2 \,,\, \CF[w_2^2] \}$ & $2-\tau$ \\
	$\CF[f^{'2} \sim m_1^2]$ & $\CF[m_1^2 \sim m_2^2]$ & $2-2\pi_L-2\pi_R$ \\	
	$d_L^2 a_L^{m-1}$ & $d_L^2 a_L^{m-1}$ & $2-2(\pi_L+\pi_R) + m \tau$ \\
	$d_R^2 a_R^{n-1}$ & $w_1^2 a_R^{n-1}$ & $2(n+1) - 2(\pi_L+\pi_R) - n \tau$ \\
	$d_L a_L^m b'_L b'_R a_R^n d_R$ & $d_L a_L^m b'_L {\color{green}\mathsf{A}}^{'''n}_R b'_L d_L$ & $2n+4-2(\pi_L+\pi_R)+(m-n)\tau$ 
\end{tabular}}
\end{align*}

\paragraph{Step 4}
In the fourth step, we focus on the part of frame 3 highlighted in Figure \ref{fig:braid_proof_summary}.
We first apply the deformed version of braid duality given in \eqref{fig:proof_defbraid}, then we apply flip-flip duality \eqref{fig:FE_flipflip} to the new orange $FE_N$. We reach:
\be\label{frame40} 
    \includegraphics[]{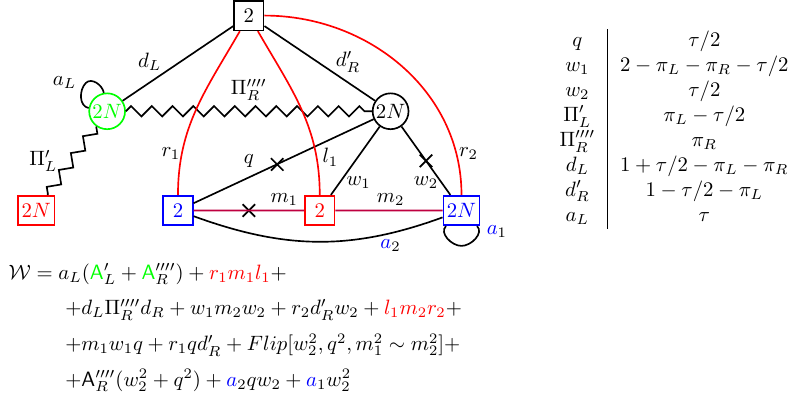}
\ee
At this point we can recombine the blue $USp(2N)$ and $USp(2)$
nodes into an $USp(2N+2)$ node  to reach frame 4:
\be\label{frame4} 
    \includegraphics[]{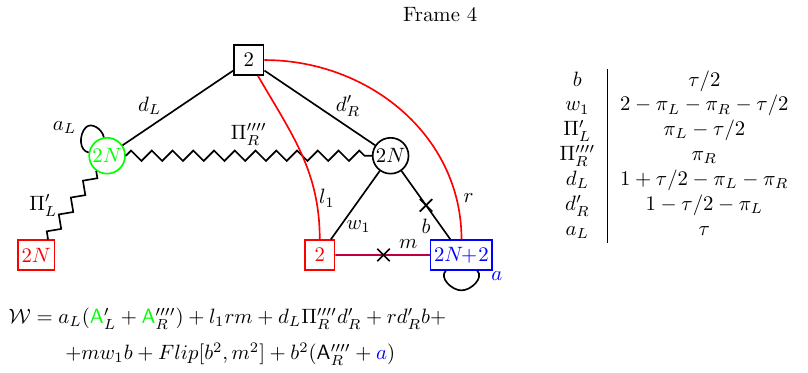}
\ee
where we combined $\{m_1, m_2\}$ into $m$, $\{r_1, r_2\}$ into $r$, $\{q, w_2\}$ into $b$ and $\{{\color{blue}a}_1, {\color{blue}a}_2, \mathcal{F}[w_2^2]\}$ into ${\color{blue}a}$.

The mapping between frame \eqref{frame3} and frame \eqref{frame4} is:
\begin{align*}
\renewcommand{\arraystretch}{1.1}
\resizebox{.95\hsize}{!}{
\begin{tabular}{c|c|c}
	Frame 3 & Frame 4 & R-charge \\
	\hline
	$ \{ \Pi'_L u_1 \,,\, \Pi'_L b'_L \Pi'''_R w_2 \,,\, m_1 \,,\, m_2  \} $ & $\{ \Pi'_L \Pi''''_R b \,,\, m \}$ & $\pi_L + \pi_R$ \\
	$\{ \Pi'_L d_L \,,\, l_1 \}$ & $\{ \Pi'_L d_L \,,\, l_1 \}$ & $1 - \pi_R$ \\
	$\{ r_1 \,,\, r_2 \}$ & $r$ & $1 - \pi_L$ \\
	$\{ {\color{red}\mathsf{A}}'_L \,,\, \Pi'_L b'_L \Pi'''_R w_1 \,,\, b_L^{'2} \} $ & ${\color{red}\mathsf{A}}'_L \,,\, \Pi'_L \Pi''''_R w_1 \,,\, \CF[b^2]$ & $2-\tau$ \\
	$\{ {\color{blue}a}_1 \,,\, {\color{blue}a}_2 \,,\, \CF[w_2^2] \}$ & $\color{blue}a$ & $2-\tau$ \\
	$\CF[m_1^2 \sim m_2^2]$ & $\CF[m^2]$ & $2-2\pi_L-2\pi_R$ \\	
	$d_L^2 a_L^{m-1}$ & $d_L^2 a_L^{m-1}$ & $2-2(\pi_L+\pi_R) + m \tau$ \\
	$w_1^2 a_R^{n-1}$ & $w_1^2 {\color{green}\mathsf{A}}^{n-1}$  & $2(n+1) - 2(\pi_L+\pi_R) - n \tau$ \\
	$d_L a_L^m b'_L {\color{green}\mathsf{A}}^{'''n}_R b'_L d_L$ & $ d_L^2 a_L^m {\color{green}\mathsf{A}}^{n+1} $ & $2n+4-2(\pi_L+\pi_R)+(m-n)\tau$ 
\end{tabular}}
\end{align*}
In the second column, the operator $\color{green}\mathsf{A}$ is given as: ${\color{green}\mathsf{A}} \sim {\color{green}\mathsf{A}}'_L \sim {\color{green}\mathsf{A}}''''_R$ where the relation is given by the F-term of $a_L$.

\paragraph{Final step}
In the final step we focus on the part of frame 4 highlighted in Figure \ref{fig:braid_proof_summary} and apply the star-triangle duality to confine the left $USp(2N)$.  We also produce the last red flipper field and obtain the theory:
\be\label{framef0} 
    \includegraphics[]{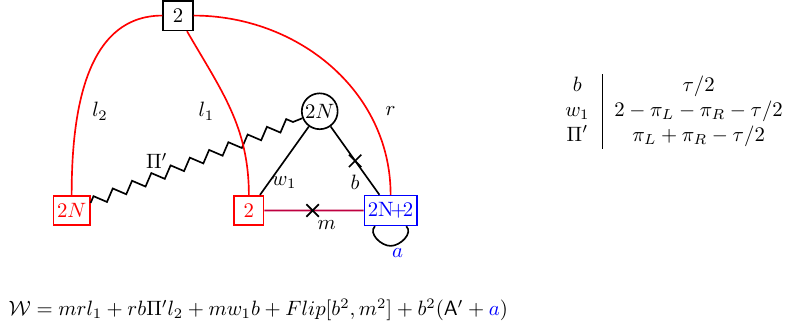}
\ee

Finally we repackage the result \eqref{framef0} using the recursive  definition of $FE_{N+2}$ \eqref{fig:FE_part_open}.
Notice that indeed the superpotential in \eqref{framef0} 
includes the terms appearing in eq. \eqref{fig:FE_part_open}, while the other terms can be easily repackaged in the $l \Pi r$ term and we obtain the triangle frame:
\be\label{framef}
    \includegraphics[]{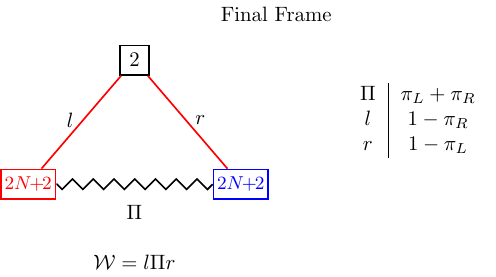}
\ee
This concludes our proof of the Braid star triangle duality, using only the Intriligator-Pouliot duality. 
The operator map between frame 4 \eqref{frame4} and the final frame \eqref{framef} is:
\begin{align*}
\renewcommand{\arraystretch}{1.1}
\begin{tabular}{c|c|c}
	Frame 4 & Final Frame & R-charge \\
	\hline
	$\{ \Pi'_L \Pi''''_R b \,,\, m \}$ & $ \Pi $ & $\pi_L + \pi_R$ \\
	$\{ \Pi'_L d_L \,,\, l_1 \}$ & $ l $ & $1 - \pi_R$ \\
	$r$ & $r$ & $1 - \pi_L$ \\
	${\color{red}\mathsf{A}}'_L \,,\, \Pi'_L \Pi''''_R w_1 \,,\, \CF[b^2]$ & $ {\color{red}\mathsf{A}} $ & $2-\tau$ \\
	$\color{blue}a$ & $ {\color{blue}\mathsf{A}} $ & $2-\tau$ \\
	$\CF[m^2]$ & $\mathsf{B}_{1,1}$ & $2-2\pi_L-2\pi_R$ \\	
	$d_L^2 a_L^m $ & $\mathsf{B}_{1,m+1}$ & $2 -2(\pi_L + \pi_R) + m\tau$ \\
    $w_1^2 {\mathsf{A}''''_R}^{n-1}$ & $2(n+1)-2(\pi_L+\pi_R) - n\tau$ \\
	$ d_L^2 a_L^m {\color{green}\mathsf{A}}^{n+1} $ & $ \mathsf{B}_{n+1,m+1} $ & $2n+4-2(\pi_L+\pi_R)+(m-n)\tau$ 
\end{tabular}
\end{align*}
Following the operator map step by step from frame zero \eqref{frame0} to the final frame \eqref{framef} one obtains precisely the operator map given in \eqref{braidmap}.

\section{$3d$ improved bifundamentals}\label{3dGB}    
\begin{figure}[h!]
    \centering
    \resizebox{\hsize}{!}{
    \includegraphics[scale=1]{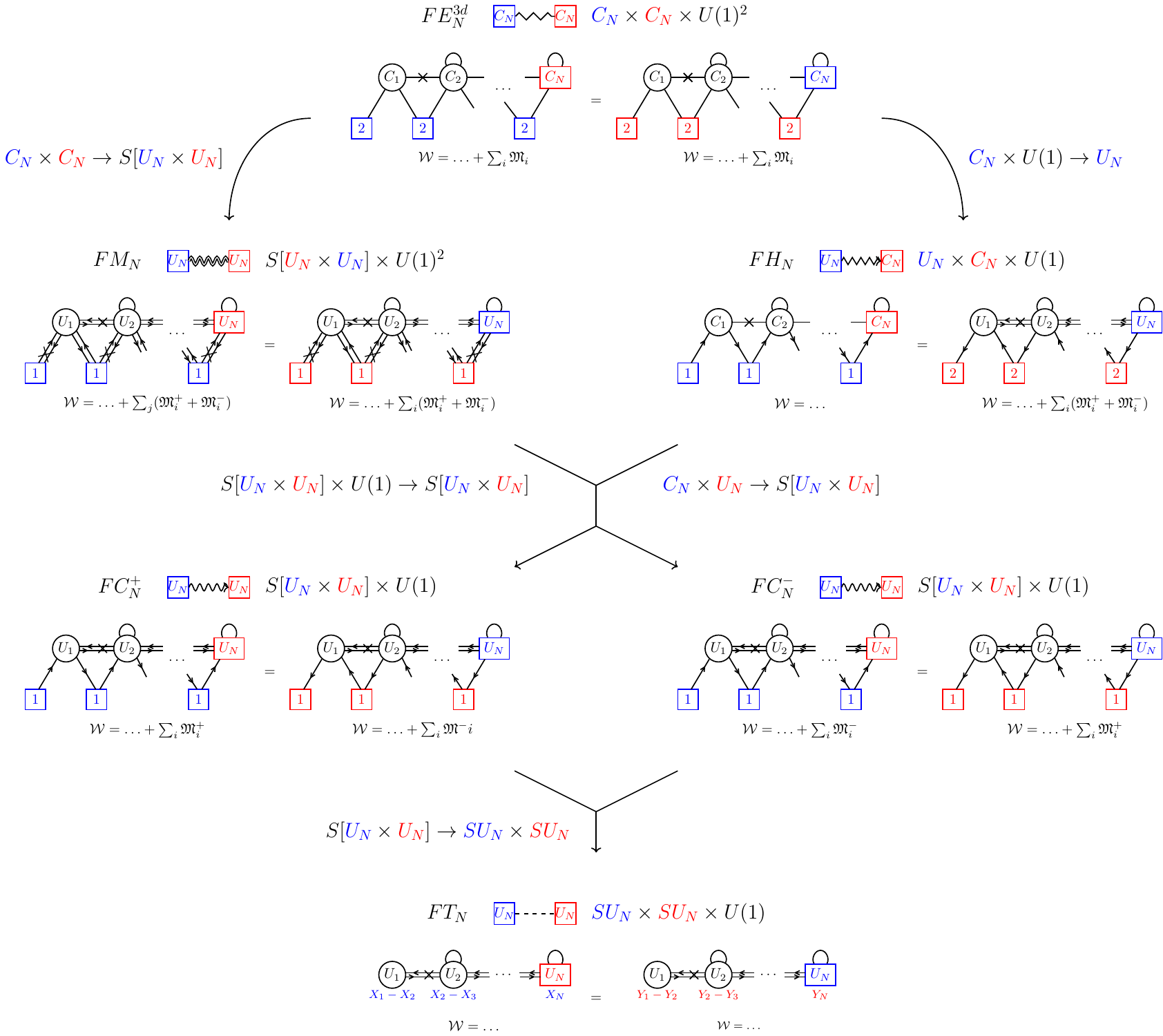}}
    \caption{Summary diagram of the improved bifundamental theories. For each theory we depict the two mirror-dual UV completions and the short notation, along with the corresponding global symmetry. To avoid cluttering, we just specify the part of the superpotential involving monopoles. The superpotential summarized with the dots contains: cubic coupling between adjoint chirals and bifundamental, the couplings associated to triangular loops and flipping terms. Arrows denote real mass deformations and they are labeled with the global symmetry breaking pattern. }
    \label{fig:impbif_summary}
\end{figure}

In this section we study the circle reduction of the $FE_N$ theory followed by various real mass deformations. With this procedure we obtain a set of $3d$ theories that we call \emph{improved bifundamentals}. The spectrum of all these theories, except the $FT_N$ theory, contains an operator in the bifundamental representation of two non-abelian global symmetries that can be either $SU(N)$ or $USp(2N)$. These theories \emph{improve} a standard bifundamental chiral multiplet in the sense that they carry an extra $U(1)$ global symmetry.
A summary diagram is given in Figure \ref{fig:impbif_summary}. On top we have the $FE^{3d}_N$ theory, which is simply the $3d$ reduction on a circle of the $FE_N$. Starting from the $FE^{3d}_N$ we partially break the global symmetries via real mass deformations, flowing to the various improved bifundamental theories, which we will describe in detail the following subsections. 

Interestingly, it was shown in \cite{BCP2,BCP3} that the $FM_N$ and $FC^\pm_N$ theories play an important role in the construction of mirror dualities for $3d$ $\mathcal{N}=2$ theories. In particular, it was shown that the $FM_N$ can be effectively associated to a $NS5$ brane, with $N-D3$ branes suspended on the two sides, inside Hanany-Witten brane setups composed of $NS5$ branes and rotated $D5$ branes\footnote{The rotated $D5$ branes are sometimes referred to as $D5'$ branes. The rotation angle is fixed so that it breaks only half SUSY, which might be broken further when considering generic angles. \cite{Elitzur:1997fh,Elitzur:1997hc}}. We therefore expect that all these theories will play a role in $3d$ $\mathcal{N}=2$ mirror dualities for $U(N)/USp(N)$ gauge theories and will admit an interpretation from the point of view of the brane construction of theories with four supercharges. 

\paragraph{Notation for $3d$ quivers} In the following sections we will deal with $3d$ theories with both $U(N)$ and $USp(2N)$ gauge/flavor groups. We therefore adopt the following notation. $U(N)/USp(2N)$ gauge or flavor nodes are labelled with $U_N / C_N$. Lines are $\CN=2$ chiral multiplets in the fundamental/antifundamental representation of the group where the arrow emerge/points to. Indeed for $USp(2N)$ group it is not necessary to distinguish between fundamental and antifundamental representations. Moreover, arches attached to $U(N)/USp(2N)$ nodes are traceless adjoint/antysimmetric chirals. To avoid cluttering, when we have double lines indicating fundamental/antifundamental chirals, we just write the name of one of the two chirals. The presence of a second chiral, distinguished by a tilde, is implied.

\subsection{$FE^{3d}_N$ theory: the $C_N \times C_N$ improved bifundamental}\label{sec:impbif}

The $FE^{3d}_N$ theory is a $3d$ $\CN=2$ SCFT  obtained as the circle reduction of the $4d$ $\CN=1$ $FE_N$ theory,
characterized by the global symmetry group:
\begin{align}\label{eq:FM_globalsym}
   {\color{red}USp(2N)}\times {\color{blue}USp(2N)} \times U(1)_\tau \times U(1)_\D \,.
\end{align}

We depict the $FE^{3d}_N$ theory with the same symbol as its $4d$ ancestor as in \eqref{FE_symbol} and the spectrum of gauge invariant operators is the same as that given in Table \ref{tab:FE_operators}.

As for the  $4d$ version, $FE_N^{3d}$ theory admits the two possible Lagrangian UV  completions  corresponding to  which $USp(2N)$ global symmetry is realized manifestly in the UV:
\begin{align}
    & \text{UV completion 1:} \qquad {\color{red}USp(2N)}\times {\color{blue}USp(2)}^N\times U(1)_\tau \times U(1)_\Delta \nn \\
    & \text{UV completion 2:} \qquad {\color{red}USp(2)}^N\times {\color{blue}USp(2N)} \times U(1)_\tau \times U(1)_\Delta \,.
\end{align}

The two $3d$ UV completion correspond to the same $4d$ quivers  in
\eqref{fig:FEmir}
with the addition of an  extra  superpotential term, linear in the fundamental monopoles of each gauge node. The effect of this superpotential is to ensure that the global symmetry of the $3d$ theories is the same as that of the $4d$ one. 
Also in this case we will refer to the self-duality as \emph{mirror}.
Under this self-mirror duality the two antisymmetric operators $\color{red}\mathsf{A}$ and $\color{blue}\mathsf{A}$ are exchanged, while the rest of the operators are trivially mapped to themselves.

The $S^3_b$ partition function of this theory is given as\footnote{The $S^3_b$ partition function can be obtained by performing a circle reduction limit of the $4d$ superconformal index. We give more details about the limit in Appendix \ref{app:3dlim}, see also \cite{Aharony:2013dha,Aharony:2013kma,Benini:2017dud}.}:
\begin{align}\label{eq:FE3d_parfun}
	Z_{FE^{3d},1}^{(N)} & (\vec{X},\vec{Y},\tau,\D) = s_b \big( -i\tfrac{Q}{2} + 2\D \big) \prod_{j=1}^{N} s_b \big( i\tfrac{Q}{2} - \D \pm Y_N \pm X_j \big) \times \nn \\
    & \times s_b \big( \tfrac{iQ}{2} -\tau \big)^{N-1} \prod_{j<k}^N s_b \big( \tfrac{iQ}{2} - \tau \pm X_j \pm X_k \big) \int d\vec{Z}_{N\!-\!1} \D_{USp(2N-2)} (\vec{Z}) \times \nn \\
	& \times \prod_{j=1}^{N-1} \left[ \, \prod_{k=1}^{N} 
    s_b(\tfrac{iQ}{2} - \tfrac{\tau}{2} \pm Z_j \pm X_k) s_b \big(-\tfrac{iQ}{2} + \tfrac{\tau}{2} + \D \pm Z_j \pm Y_N) \big) \right] \times \nn \\
	& \times Z_{FE^{3d},1}^{(N-1)} \big( \vec{Z},\{ Y_1,\ldots,Y_{N-1}\},\tau, \D - \tfrac{\tau}{2} \big)  \,,
\end{align}
with the basis of the recursion given by:
\begin{align}
	Z_{FE^{3d},1}^{(1)} (X,Y,\tau,\D) = s_b(-i\tfrac{Q}{2} + 2\D) s_b(i\tfrac{Q}{2} - \D \pm X \pm Y) \,.
\end{align}
Our convention for the $S^3_b$ partition function is given in Appendix \ref{app:conventions}. As in the $4d$ case, in the first UV completion the first entry of the partition function is associated to the manifest $USp(2N)$ symmetry and the second to the emergent $USp(2N)$ symmetry. 

The second UV completion is not actually independent and its partition function can be obtained  from \eqref{eq:FE3d_parfun} by swapping the two sets $\vec{X} \leftrightarrow \vec{Y}$.
\begin{align}\label{eq:FE3d_mirparfun}
    Z^{(N)}_{FE,2} (\vec{X}, \vec{Y}, \tau, \D ) = Z^{(N)}_{FE,1} (\vec{Y}, \vec{X}, \tau, \D) \,,
\end{align}
where now the first entry of the partition function $Z^{(N)}_{FE,2}$ is associated to the emergent $USp(N)$ symmetry while  the second to the manifest $USp(N)$ symmetry.

Since the two UV completions are IR dual the two apparently different partition functions match as a function of the real mass parameters. We can therefore define the partition function of the $FE^{3d}_N$ SCFT simply as:

\begin{align}\label{eq:FE3d_mir}
    Z^{(N)}_{FE} (\vec{X}, \vec{Y}, \tau, \D) \overset{\text{def}}{=} Z^{(N)}_{FE,1} (\vec{X}, \vec{Y}, \tau, \D ) = Z^{(N)}_{FE,2} (\vec{X}, \vec{Y}, \tau, \D) 
    \,.
\end{align}

The $FE^{3d}_N$ inherits all the properties of the $4d$ $FE_N$ ancestor theory, therefore we will not discuss them in detail and we refer to Section \ref{sec:FE}.  

\subsection{$FM_N$ theory: the $A_{N-1} \times A_{N-1}$ improved hyper}

Starting from the $FE_N^{3d}$ theory we can perform a real mass deformation with the effect of breaking the ${\color{red}USp(2N)}\times {\color{blue}USp(2N)}$ global symmetry to $S[{\color{red}U(N)}\times {\color{blue}U(N)}]$ (see Appendix \ref{app:3dlim}) to reach the $FM_N$ theory. This theory was first introduced in \cite{Pasquetti:2019tix}\footnote{See also \cite{BCP2} for applications of the $FM_N$ SCFT in the construction of mirror dualities with four supercharges.}, some of its properties and interesting deformations were  discussed in \cite{BCP2}. Notice that all the properties that we will discuss in this subsection can be derived starting from those of the $FE^{3d}_N$ theory and performing the suitable real mass deformation described above.

The $FM_N$ theory is a $3d$ $\CN=2$ SCFT characterized by the global symmetry group:
\begin{align}\label{eq:FM_globalsym}
    S[{\color{red}U(N)}\times {\color{blue}U(N)}] \times U(1)_\tau \times U(1)_\D \,.
\end{align}
Thus we depict the $FM_N$ SCFT in short as:
\be\label{eq:FM_symbol}
    \includegraphics[]{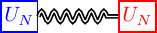}
\ee
to recall that the theory is characterized by two $U(N)$ global symmetries and also that the spectrum contains a pair of bifundamental operators, as we will discuss in detail shortly. 

\begin{figure}[]
\centering
\resizebox{.8\hsize}{!}{
    \includegraphics[]{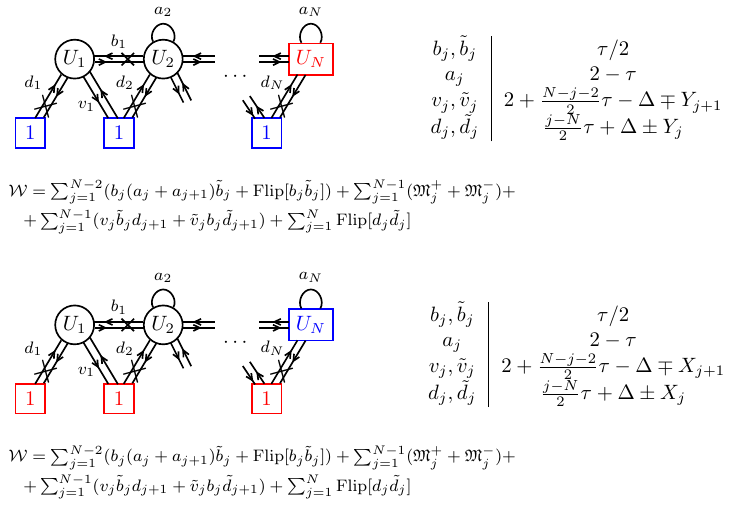}
}
\caption{The two UV Lagrangian completions of the $FM_N$ SCFT. On the right we specify the R-charges of all the fields composing the two theories. These are given as a trial value mixed with the other abelian global symmetries
$\tau,\Delta$.
For convenience we included also the ${\color{blue}U(1)}^N$ and ${\color{red}U(1)}^N$ charges which however do  not mix with the R-charge because of the IR-enhancement.}
\label{fig:FM_quiver}
\end{figure}

The $FM_N$ SCFT admits two possible Lagrangian UV completions that are depicted in Figure \ref{fig:FM_quiver}.

These two theories do not realize manifestly the full global symmetry \eqref{eq:FM_globalsym}, instead they realize the full Cartan subgroup of one of the two $U(N)$:
\begin{align}
    & \text{UV completion 1:} \qquad S[{\color{red}U(N)}\times {\color{blue}U(1)}^N] \times U(1)_\tau \times U(1)_\D \nn \\
    & \text{UV completion 2:} \qquad S[{\color{red}U(1)}^N\times {\color{blue}U(N)}] \times U(1)_\tau \times U(1)_\D
\end{align}
Indeed, along at the IR fixed point the UV global symmetries enhance to give precisely the IR global symmetries in \eqref{eq:FM_globalsym}. 

Notice that the two UV completions in Figure \ref{fig:FM_quiver} are related by a simple transformation swapping the two set of parameters $\vec{X} \leftrightarrow \vec{Y}$ and thus swapping the emergent and manifest $U(N)$ global symmetries. In some sense, we can view the two possible UV completions as a choice of which $U(N)$ global symmetry is realized manifestly in the UV, without changing the Lagrangian description. We will come back to this property later.

\paragraph{Gauge invariant operators}
The spectrum of gauge invariant operators of the $FM_N$ SCFT is listed in Table \ref{tab:FM_operators}. 
\begin{table}[]
\renewcommand{\arraystretch}{1.2}
\centering
\begin{tabular}{|c|cc|c|}\hline
{} & ${\color{red}U(N)}$ & ${\color{blue}U(N)}$ & R charge\\ \hline
${\color{red}\mathsf{A}}$ & ${\bf N^2-1}$ & $\bf 1$  & $2-\tau$ \\
${\color{blue}\mathsf{A}}$ & $\bf1$ & ${\bf N^2-1}$  & $2-\tau$ \\
$\Pi$ & $\bf \bar{N}$ & $\bf N$  & $\D$ \\
$\tilde{\Pi}$ & $\bf N$ & $\bf \bar{N}$  & $\D$ \\
$\mathsf{B}_{n, m}$ & $\bf1$ & $\bf1$  & $2n-2\D+(m-n)\tau$ \\
 \hline
\end{tabular}
\caption{List of all the gauge invariant operator that compose the spectrum of the $FM_N$ SCFT. The R-charge is given as a trial value mixed with the other two abelian symmetries of the theory, $U(1)_\tau$ and $U(1)_\D$, whose mixing values are given by the two real variables $\tau$ and $\D$.}
\label{tab:FM_operators}
\end{table}
We now discuss how these operators are constructed in each UV completion given in Figure \ref{fig:FM_quiver}.
\begin{itemize}
	\item The operator $\color{red}\mathsf{A}$ in the traceless adjoint representation of $\color{red}U(N)$.
    \begin{itemize}
        \item In the UV completion 1 it is realized as the gauge singlet $a_N$.
        \item In the UV completion 2 it is realized as mesonic operators collected into a $N \times N$ matrix. On the diagonal we place the singlets flipping the bifundamentals $\CF[b_j \tilde{b}_j]$ such that the trace of the resulting matrix is zero. Outside the diagonal we place the operators $d_i b_i \ldots b_{j-1} v_j$ and $\tilde{v}_j \tilde{b}_{j-1} \ldots \tilde{b}_i \tilde{d}_i$, for $i<j$, that are mesons in the bifundamental of a pair of $\color{red}U(1)$ symmetries
    \end{itemize}

    \item The operator $\color{blue}\mathsf{A}$ is in the traceless adjoint representation of $\color{blue}U(N)$. Its construction in the first UV completions is analogous to that of the operator $\color{red}\mathsf{A}$ in the second UV completion, and viceversa.
	
	\item The pair of bifundamentals $\Pi$ and $\tilde{\Pi}$ are constructed in the UV completion 1 by collecting $N$ mesons in the bifundamental of ${\color{blue}U(1)} \times {\color{red}U(N)}$. In particular $\Pi$ is constructed by collecting all the mesons: $d_i b_i b_{i+1} \ldots b_{N-1}$, for $i=1,\ldots,N$, while $\tilde{\Pi}$  collecting: $\tilde{b}_{N-1} \ldots \tilde{b}_i \tilde{d}_i$. In the UV completion 2 they are constructed analogously.
	
	\item The matrix $\mathsf{B}_{n,m}$ is a collection of singlets under the non-abelian global symmetries and it is constructed in the same way in the two UV completions. For $n=1$ they are simply the flipping fields: $\mathsf{B}_{1,m} = \CF[d_{N+1-m} \tilde{d}_{N+1-m}]$, while for $n \geq 2$ they are given as: $\mathsf{B}_{n,m} = v_{N-m} a_{N-m}^{n-2} \tilde{v}_{N-m}$. 
	

 \end{itemize}
As for the $4d$ $FE_N$ theory discussed in Section \ref{sec:FE}, we do not attempt to analyze in full detail the relations obeyed by the chiral ring generators, but we comment on the composite operators of the form $\Pi \tilde{\Pi}$, which decompose in irreducible ${\color{red}U(N)} \times {\color{blue}U(N)}$ representations as
\be  
(adj, adj) \oplus (adj, 1)  \oplus (1, adj)  \oplus (1,1) \,.
\ee     
We find that  $(adj, adj)$ is non-trivial in the chiral ring, while $(adj,1)$, $(1,adj)$ and $(1,1)$ are zero in the chiral ring. The vanishing of these $3$ irreps can be shown in a way analog to the $FE_N$ case, using the classical F-terms relations. \\
We also look at index contributions of the composite operator of the form $\Pi \Pi$ (similarly $\tilde{\Pi}\tilde{\Pi}$), which decompose as:
\be
    (sym, \overline{sym}) \oplus (antisymm, \overline{antisymm})  
\ee
We find that both the combinations are non-trivial in the chiral ring. \\
Finally, we observe that the traces of the two adjoint operators $\color{red}\mathsf{A}$ and $\color{blue}\mathsf{A}$ are related as: $\Tr{\color{red}\mathsf{A}}^j \sim 
 \Tr{\color{blue}\mathsf{A}}^j$, for $j \geq 2$. See \cite{BenvenutiPedde:2024} for the full Hilbert Series of the $FM_N$ theories.

\paragraph{Partition function and self-mirror property}
From the UV Lagrangian completions in Figure \ref{fig:FM_quiver} we can write the partition function for the $FM_N$ theory. We turn on the following real mass parameters:
\begin{align}
    & \vec{X} \quad \text{for} \quad {\color{red}U(N)} \nn \,, \\
    & \vec{Y} \quad \text{for} \quad {\color{blue}U(N)} \nn \,, \\
    & \D \quad \text{for} \quad U(1)_\D \nn \,, \\
    & \tau \quad \text{for} \quad U(1)_\tau \,.
\end{align}
For the first UV completion we have:
\begin{align}\label{eq:FM_parfun}
    Z^{(N)}_{FM,1} & (\vec{X}, \vec{Y}, \tau, \D) = s_b \big( -i\tfrac{Q}{2} + 2\D \big) \prod_{j=1}^{N} s_b \big( i\tfrac{Q}{2} - \D \pm (Y_N - X_j) \big) \times \nn \\
	& \times s_b \big( -\tfrac{iQ}{2} + \tau \big)^{N} \prod_{j<k}^{N} s_b \big( -\tfrac{iQ}{2} + \tau \pm (X_j - X_k) \big) \int d\vec{Z}_{N\!-\!1} \D_{U(N\!-\!1)} \big(\vec{Z} \big) \times \nn \\
	& \times \prod_{j=1}^{N-1} \left[ \prod_{k=1}^N s_b \big( \tfrac{iQ}{2} - \tfrac{\tau}{2} \pm (Z_j - X_k) \big) 
	s_b \big( -\tfrac{iQ}{2} + \tfrac{\tau}{2} + \D \pm (Z_j - Y_N) \big) \right] \times\nn \\
	& \times Z_{FM,1}^{(N-1)} \big( \vec{Z},\{ Y_1,\ldots,Y_{N-1} \},\tau, \D - \tfrac{\tau}{2} \big) \,,
\end{align}
where the basis of the recursion is:
\begin{align}\label{eq:FM_parfun_N=1}
    Z_{FM,1}^{(1)}(X,Y,\tau,\D) = s_b \big( -\tfrac{iQ}{2} + 2\D \big) s_b \big( \tfrac{iQ}{2} - \D \pm (X - Y) \big) \,.
\end{align}
Notice that the manifest $U(N)$ parameter $\vec X$ sits in the first entry of the partition function. Moreover, from the above expressions, one can observe that by performing shifts of the gauge parameters $\vec{Z}$ it is possible to set $\sum_{j=1}^N (X_j + Y_j) = 0$. This consist in implementing the constraint that the faithful global symmetry is $S[{\color{red}U(N)} \times {\color{blue}U(N)}]$. However we prefer to give the partition function ``off-shell" where the constraint is not implemented. This is useful when we consider bigger theories containig the $FM_N$ as a building block, where one or both the global $U(N)$ symmetries are gauged.

As already mentioned, the second UV completion is not actually independent and its partition function can be obtained from \eqref{eq:FM_parfun} by swapping the two sets $\vec{X} \leftrightarrow \vec{Y}$
\begin{align}
    Z^{(N)}_{FM,2} (\vec{X}, \vec{Y}, \tau, \D ) = Z^{(N)}_{FM,1} (\vec{Y}, \vec{X}, \tau, \D) \,,
\end{align}
so the first entry of the partition function  $ Z^{(N)}_{FM,2}$ is associated to the emergent $U(N)$ parameter while  the second to the manifest $U(N)$ one.

Indeed, since the two UV completions are IR dual, i.e.~they flow to the same IR SCFT, the two apparently different partition functions must match as a function of the real mass parameters. We can therefore define the partition function of the $FM_N$ SCFT simply as:
\begin{align}\label{eq:FM_parfunmir}
    Z^{(N)}_{FM} (\vec{X}, \vec{Y}, \tau, \D) \overset{\text{def}}{=} Z^{(N)}_{FM,1} (\vec{X}, \vec{Y}, \tau, \D ) = Z^{(N)}_{FM,2} (\vec{X}, \vec{Y}, \tau, \D) 
    \,.
\end{align}
We will refer to this exact self-duality as \emph{mirror} and we interpret it as the fact that the $FM_N$ SCFT do not admit any (knwon) UV Lagrangian completion where both the $\color{red}U(N)$ and $\color{blue}U(N)$ symmetries are realized manifestly. One has to decide which of the two is manifest in the UV and which one is not. 

Under the self-mirror duality the two adjoint operators $\color{red}\mathsf{A}$ and $\color{blue}\mathsf{A}$ are exchanged, while the rest of the operators are trivially mapped to themselves.

\paragraph{Interesting deformations}
\begin{figure}
	\centering
	\includegraphics[]{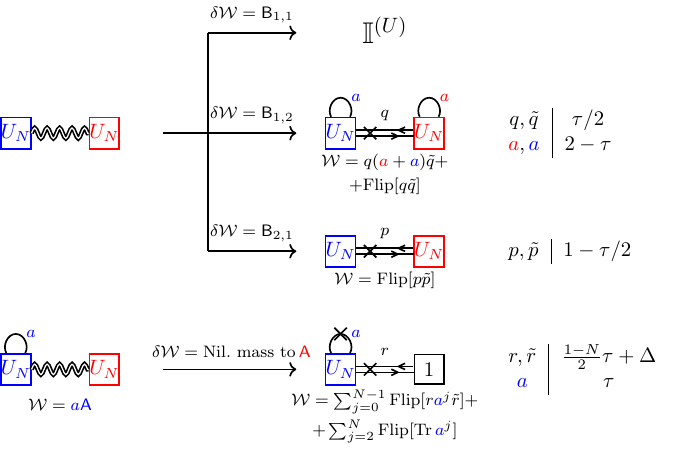}
	\caption{Summary of the interesting deformations of the $FM_N$ theory.}
	\label{fig:FM_deformations}
\end{figure}
Some interesting deformations are depicted in Figure \ref{fig:FM_deformations}.

The first deformation in Figure \ref{fig:FM_deformations} consist in turning on linearly the $\mathsf{B}_{1,1}$ singlet in the superpotential. This deformation breaks completely the $U(1)_\D$ symmetry so that the R-charge of the $\Pi, \tilde{\Pi}$ operators is set to zero indicating that they have acquired a VEV. Such VEVs break the $S[{\color{red}U(N)}\times{\color{blue}U(N)}]$ global symmetry to the diagonal $U(N)$ subgroup so that the $FM_N$ theory behaves as in Identity-wall.
From the point of view of the partition function we have:
\begin{align}
    Z_{FM}^{(N)} (\vec{X}, \vec{Y}, \tau, \D = 0) = \frac{1}{\D_N(\vec{X},\vec{Y},\tau)} \sum_{\s \in S_N} \prod_{j=1}^N \d \big( X_j - Y_{\s(j)} \big) \overset{\text{def}}{=} {}_{\vec{X}} \mathbb{I}^{(U)}_{\vec{Y}} (\tau) \,,
\end{align}
where the sum runs over the elements $\s$ of the $U(N)$ Weyl group $S_N$.

The second and third deformations in Figure \ref{fig:FM_deformations} show that the $FM_N$ theory reduces to a standard bifundamental hypermultiplet when we introduce a superpotential linear in the $\mathsf{B}_{1,2}$ or $\mathsf{B}_{2,1}$ singlets. This property translates to the following partition function identities.
\begin{align}
    Z_{FM}^{(N)} (\vec{X}, \vec{Y}, \tau, \D = \tfrac{\tau}{2} ) = & \prod_{j<k}^N s_b \big( -\tfrac{iQ}{2} + \tau \pm (X_j - X_k) \big) s_b \big( -\tfrac{iQ}{2} - \tau \pm (Y_j - Y_k) \big) \times \nn \\
    & \times s_b \big( -\tfrac{iQ}{2} - \tau \big)^{2N-1} \prod_{j,k=1}^N s_b \big( \tfrac{iQ}{2} - \tfrac{\tau}{2} \pm ( X_j - Y_j ) \big) \,.
\end{align}
\begin{align}
    Z_{FM}^{(N)} (\vec{X}, \vec{Y}, \tau, \D = \tfrac{iQ - \tau}{2} ) = s_b \big( \tfrac{iQ}{2} - \tau \big) \prod_{j,k=1}^N s_b \big( \tfrac{\tau}{2} \pm ( X_j - Y_j ) \big) \,,
\end{align}

The last deformation depicted in Figure \ref{fig:FM_deformations} is obtained by giving a maximal nilpotent mass to one of the two adjoint operators, say to $\color{red}\mathsf{A}$. Under this deformation we obtain a fundamental hypermultiplet coupled to singlets. This property translates to:
\begin{align}
    Z_{FM}^{(N)} & (\vec{X}, \vec{Y} = \{ \frac{N-1}{2} \tau + V, \ldots, \frac{1-N}{2} \tau + V \}, \tau, \D) = \nn \\ 
    & = \prod_{j=2}^N s_b \big( -\tfrac{iQ}{2} + j \tau ) \prod_{j=1}^N s_b \big( \tfrac{iQ}{2} - \tfrac{1-N}{2} \tau - \D \pm (X_j - Y_j) \big) s_b( -\tfrac{iQ}{2} - (N-j)\tau + 2\D \big) \,.
\end{align}




\subsection{$FH_N$ theory: the $C_N \times A_{N-1}$ improved bifundamental}
\begin{figure}[h]
	\centering
    \resizebox{.8\hsize}{!}{
	\includegraphics[]{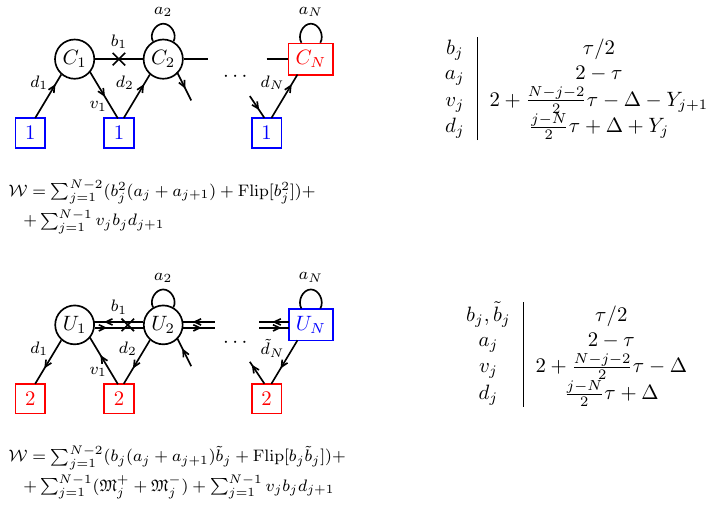}}
	\caption{The two UV completions of the $FH_N$ SCFT. On the right we specify the R-charges of the fields composing the theories. These are given as a trial value mixed with all the other abelian symmetries of the theory. In particular $\D$ is associated to the $U(1)_\D \subset {\color{blue}U(N)}$ group. The set $\vec{Y}$ satisfies $\sum_{j=1}^N Y_j = 0$ in both the two UV completions.
    For convenience we included also the ${\color{blue}U(1)}^N$ charges  which however do  not mix with the R-charge because of the IR-enhancement to ${\color{blue}U(N)}$. }
	\label{fig:FH_quiver}
\end{figure}
Starting from the $FE^{3d}_N$ theory we can perform a real mass deformation with the effect of breaking the ${\color{blue} USp(2N)} \times U(1)_\D$ global symmetry down to $\color{blue}U(N)$ (see Appendix \ref{app:3dlim}). The result of the limit is the $FH_N$ theory. \\
The $FH_N$ theory is a $3d$ $\CN=2$ SCFT whose global symmetry is:
\begin{align}\label{eq:FH_globalsymm}
	{\color{red}USp(2N)} \times {\color{blue}U(N)} \times U(1)_\tau \,.
\end{align}
We depict the $FH_N$ SCFT in short as:
\be\label{fig:FH_symbol}
    \includegraphics[]{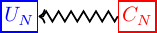}
\ee
to recall that the theory is characterized by a ${\color{red}USp(2N)} \times {\color{blue}U(N)}$ non-abelian global symmetry and that also the theory possesses an operator in the $\bf 2N \times \bar{N}$ representation of the aforementioned symmetry, as we will discuss in detail briefly. 

The $FH_N$ SCFT admits two possible Lagrangian UV completions that are depicted in Figure \ref{fig:FH_quiver}. The two completions however do not realize manifestly the full global symmetry group given in \eqref{eq:FH_globalsymm} but instead they realize the following UV global symmetry groups:
\begin{align}
    & \text{UV completion 1:} \qquad {\color{red}USp(2N)} \times {\color{blue}U(1)^N} \times U(1)_\tau \,, \nn \\
    & \text{UV completion 2:} \qquad {\color{red}USp(2)^N} \times {\color{blue}U(N)} \times U(1)_\tau \,.
\end{align}
Indeed, along the RG flow the UV global symmetries enhance to give precisely the group in \eqref{eq:FH_globalsymm}. 

\paragraph{Gauge invariant operators} The spectrum of gauge invariant operators of the $FH_N$ SCFT is listed in Table \ref{tab:FH_operators}.
\begin{table}[]
\renewcommand{\arraystretch}{1.2}
\centering
\begin{tabular}{|c|cc|c|}\hline
{} & ${\color{red}USp(2N)}$ & ${\color{blue}U(N)}$ & R charge\\ \hline
${\color{red}\mathsf{A}}$ & ${\bf N(2N-1)-1}$ & $\bf 1$  & $2-\tau$ \\
${\color{blue}\mathsf{A}}$ & $\bf 1$ & ${\bf N^2-1}$  & $2 - \tau$ \\
$\Pi$ & $\bf 2N$ & $\bf \bar{N}$  & $\D$ \\
\hline
\end{tabular}
\caption{List of all the gauge invariant operator that compose the spectrum of the $FH_N$ SCFT. The R-charge of the operators is given as a trial value mixed with $\tau$ and $\D$. $\tau$ is the mixing parameter for the $U(1)_{\tau}$ symmetry while $\D$ is that of the $U(1)$ centre of $ {\color{blue}U(N)}$.}
\label{tab:FH_operators}
\end{table}
We now discuss how these operators are constructed in each UV completion given in Figure \ref{fig:FH_quiver}.
\begin{itemize}
	\item The operator $\color{red} \mathsf{A}$ in the
 traceless antisymmetric  representation of $\color{red}USp(2N)$
 is simply the singlet $a_N$ in the first UV completion. In the second UV completion this operator is obtained by collecting all the mesons that are charged under pairs of  $\color{red}USp(2)$ symmetries: $v_i \tilde{b}_{i-1} \ldots \tilde{b}_j d_j$, for $i > j$, together with the singlets $\CF[b_i \tilde{b}_i]$. 
	
	\item The operator $\color{blue} \mathsf{A}$ in the traceless adjoint representation of $\color{blue}U(N)$ is simply the singlet $a_N$ in the second UV completion. In the first UV completion is instead given by collecting: the singlets $\CF[b_i^2]$, mesons charged under pairs of $\color{blue}U(1)$: $d_i b_i \ldots b_{j-1} v_j$, and all the monopoles charged under a sequence of adjacent gauge nodes.
    For example for $N=3$ we have:\footnote{For $USp(2N)$ monopoles the superscript $1$ means that we turn on the lowest magnetic flux $(1,0,\ldots,0)$. }
    \begin{align}
        \begin{pmatrix}
            \mathcal{F}[b_1^2] & d_1  v_1 & d_1 b_1 v_2\\
            \mathfrak{M}^{1,0} & \mathcal{F}[b_2^2] & d_2 v_2\\
            \mathfrak{M}^{1,1}&\mathfrak{M}^{0,1}&-(\mathcal{F}[b_1^2] +\mathcal{F}[b_2^2]) 
        \end{pmatrix}
    \end{align}

	\item The operator $\Pi$ is obtained in both theories as a collection of mesons charged under the manifest global symmetry and one of the symmetries in the saw-like structure. In the first UV completion we collect the operators: $d_i b_i \ldots b_{N-1}$. In the second UV completion we instead collect: $\tilde{b}_{N-1} \ldots \tilde{b}_i d_i$.
	
\end{itemize}

\paragraph{Partition function and mirror duality}
From the two UV completions in Figure \ref{fig:FH_quiver} we can write the $S^3_b$ partition function of the $FH_N$ theory. We turn on the following real mass parameters:
\begin{align}
    & \vec{X} \quad \text{for} \quad {\color{red}USp(2N)} \nn \,, \\
    & \vec{Y} \quad \text{for} \quad SU(N) \subset {\color{blue}U(N)} \nn \,, \\
    & \D \quad \text{for} \quad U(1) \subset {\color{blue}U(N)} \nn \,, \\
    & \tau \quad \text{for} \quad U(1)_\tau \,,
\end{align}
since $\vec{Y}$ is a $SU(N)$ parameter we can impose: $\sum_{j=1}^N Y_j = 0$.

For the first UV completion we write:
\begin{align}
	Z_{FH,1}^{(N)} &(\vec{X},\vec{Y},\tau,\D) = s_b \big( \tfrac{iQ}{2} - \tau \big)^{N} \prod_{j<k}^{N} s_b \big( -\tfrac{iQ}{2} + \tau \pm X_j \pm X_k \big) \times \nn \\
	& \times \prod_{j=1}^{N} s_b \big( i\tfrac{Q}{2} - \D \pm X_j + Y_N \big) \int d\vec{Z}_{N\!-\!1} \D_{USp(2N\!-\!2)} \big( \vec{Z} \, \big) \times \nn \\
    & \times \prod_{j=1}^{N-1} \left[ \, \prod_{k=1}^N s_b \big( \tfrac{iQ}{2} - \tfrac{\tau}{2} \pm Z_j \pm X_k \big) s_b \big( -\tfrac{iQ}{2} + \tfrac{\tau}{2} + \D \pm Z_j - Y_N \big) \right] \times \nn \\
	& \times Z_{FH,1}^{(N-1)} \big( \vec{Z},\{ Y_1,\ldots,Y_{N-1}\},\tau, \D - \tfrac{\tau}{2} \big) \,,
\end{align}
Notice that the manifest $USp(2N)$ parameter $\vec X$  sits in the first entry of the partition function.

For the second UV completion we instead have:
\begin{align}
	Z_{FH,2}^{(N)} & ( \vec{X},\vec{Y},\tau,\D) = s_b \big( -\tfrac{iQ}{2} + \tau \big)^{N} \prod_{j<k}^{N} s_b \big( -\tfrac{iQ}{2} + \tau \pm( Y_j - Y_k) \big) \times \nn \\
	& \times \prod_{j=1}^{N} s_b \big( i\tfrac{Q}{2} - \D + Y_j \pm X_N \big) \int d\vec{Z}_{N\!-\!1} \D_{U(N\!-\!1)} \big( \vec{Z} \, \big) \times \nn \\
    & \times \prod_{j=1}^{N-1} \left[ \, \prod_{k=1}^N s_b \big( \tfrac{iQ}{2} - \tfrac{\tau}{2} \pm (Z_j - Y_k) \big) s_b \big( -\tfrac{iQ}{2} + \tfrac{\tau}{2} + \D - Z_j \pm X_N \big) \right] \times \nn \\
	& \times Z_{FH,2}^{(N-1)} \big( \{ X_1,\ldots,X_{N-1}\}, \vec{Z},\tau, \D - \tfrac{\tau}{2} \big) \,,
\end{align}
where now the manifest  $U(N)$ parameter $\vec Y$ sits in the second entry of the partition function.
The two UV completions have  the same basis of the recursion:
\begin{align}
	Z_{FH,1}^{(1)} (X,Y,\tau,\D) = Z_{FH,2}^{(1)} (X,Y,\tau,\D) = s_b \big( \tfrac{iQ}{2} - \D \pm X + Y \big) \,.
\end{align}
The expressions for the partition functions are given in an off-shell parameterization, where the constraint $\sum_{j=1}^N Y_j = 0$ can be imposed afterwards.

Both the two UV theories flow to the same IR SCFT, therefore their partition function must be equal. This means that we can define in general:
\begin{align}\label{eq:FH_parfunmir}
	Z_{FH}^{(N)}(\vec{X},\vec{Y},\tau,\D) \stackrel{\text{def}}{=} Z_{FH,1}^{(N)}(\vec{X},\vec{Y},\tau,\D) = Z_{FH,2}^{(N)}(\vec{X},\vec{Y},\tau,\D) \,,
\end{align}
introducing $Z_{FH}^{(N)}$ which is a quantity that do not refer to any specific choice of UV completion with the first entry reserved for the  $USp(2N)$ parameter and
and the second entry  for the $U(N)$ parameter.

We can also consider a variant of the $FH_N$ theory
differing from the $FH_N$ theory only in the fact that 
 the bifundamental $\Pi$ operator in Table \ref{tab:FH_operators} is in the $\bf 2N \times N$ 
(rather than in the $\bf 2N \times \bar N$)
   representation of ${\color{red}USp(2N)} \times {\color{blue}U(N)}$. The partition function of this variant   can be obtained a simple sign-flip as $Z_{FH}^{(N)}( \vec{X}, -\vec{Y}, \tau, \D)$\footnote{Notice that instead $Z_{FH}^{(N)}(-\vec{X},\vec{Y},\tau,\D)=Z_{FH}^{(N)}(\vec{X},\vec{Y},\tau,\D) $ since $\vec{X}$ are $USp(2N)$ cartans.}, which means that in the two UV completions in Figure \ref{fig:FH_quiver} we are reversing all the arrows in the saw structure. Graphycally we distinguish these two situations by an arrow-flip in the short notation \eqref{fig:FH_symbol}.
\be
\begin{tikzpicture}[thick,node distance=3cm,gauge/.style={circle,draw,minimum size=5mm},flavor/.style={rectangle,draw,minimum size=5mm}]
	
\begin{scope}
	\path  (1,0) node[flavor,red] (x) {$\!C_N\!$}  -- (3,0) node[flavor,blue] (y) {$\!U_N\!$} -- (4,0) node[right] {associated to: \qquad $Z_{FH}^{(N)}(\vec{X},\vec{Y},\tau,\D)$};
	
	\wigMB (x) -- (y);
\end{scope}

\begin{scope}[shift={(0,-1.5)}]
	\path (1,0) node[flavor,red] (x) {$\!C_N\!$}  -- (3,0) node[flavor,blue] (y) {$\!U_N\!$} -- (4,0) node[right] {associated to: \qquad $Z_{FH}^{(N)}(\vec{X},-\vec{Y},\tau,\D)$};
	
	\wigMB (y) -- (x);
\end{scope}
\end{tikzpicture}
\label{FHsf}
\ee

\paragraph{Interesting deformations}
We now discuss two interesting deformations of the $FH_N$ theory. As for the $FM_N$ theory, these can be derived by performing a real mass deformation starting from the properties of the $FE^{3d}_N$ theory (see Appendix \ref{app:3dlim}).

The first deformation consists in performing a nilpotent mass for the adjoint operator $\color{blue}\mathsf{A}$
which has the effect of deforming the $FH_N$ improved bifundamental  in an ${\color{red}USp(2N)}\times {\color{blue}U(1)}$ bifundamental.
At the level of partiton function we have:
\begin{align}\label{eq:FHnilmass_UN}
    Z_{FH}^{(N)} ( & \vec{X} = \{ \tfrac{N-1}{2}\tau + V, \ldots, \tfrac{1-N}{2}\tau + V \}, \vec{Y}, \tau, \D ) = \nn \\
    & = \prod_{j=2}^N s_b \big( -\tfrac{iQ}{2} + j\tau \big) 
    \prod_{j=1}^N s_b \big( \tfrac{iQ}{2} - \tfrac{1-N}{2}\tau - \D \pm Y_j + V \big) \,.
\end{align}

The second deformation consist in performing a nilpotent mass for the antisymmetric  operator $\color{red}\mathsf{A}$
which has the effect of deforming the $FH_N$ improved bifundamental  in an ${\color{red}USp(2)}\times {\color{blue}U(N)}$ bifundamental.
At the level of partiton function we have:
\begin{align}\label{eq:FHnilmass_USpN}
    Z_{FH}^{(N)} ( & \vec{X} , \vec{Y} = \{ \tfrac{N-1}{2}\tau + V, \ldots, \tfrac{1-N}{2}\tau + V \}, \tau, \D ) = \nn \\
    & = \prod_{j=2}^N s_b \big( -\tfrac{iQ}{2} + j\tau \big) 
    \prod_{j=1}^N s_b \big( \tfrac{iQ}{2} - \tfrac{1-N}{2}\tau - \D + X_j \pm V \big) \,.
\end{align}

\subsection{The $FC_N^\pm$ theory: the $A_{N-1} \times A_{N-1}$ improved chiral}\label{sec:FC}
\begin{figure}[t]
	\centering
    \resizebox{.85\hsize}{!}{
	\includegraphics[]{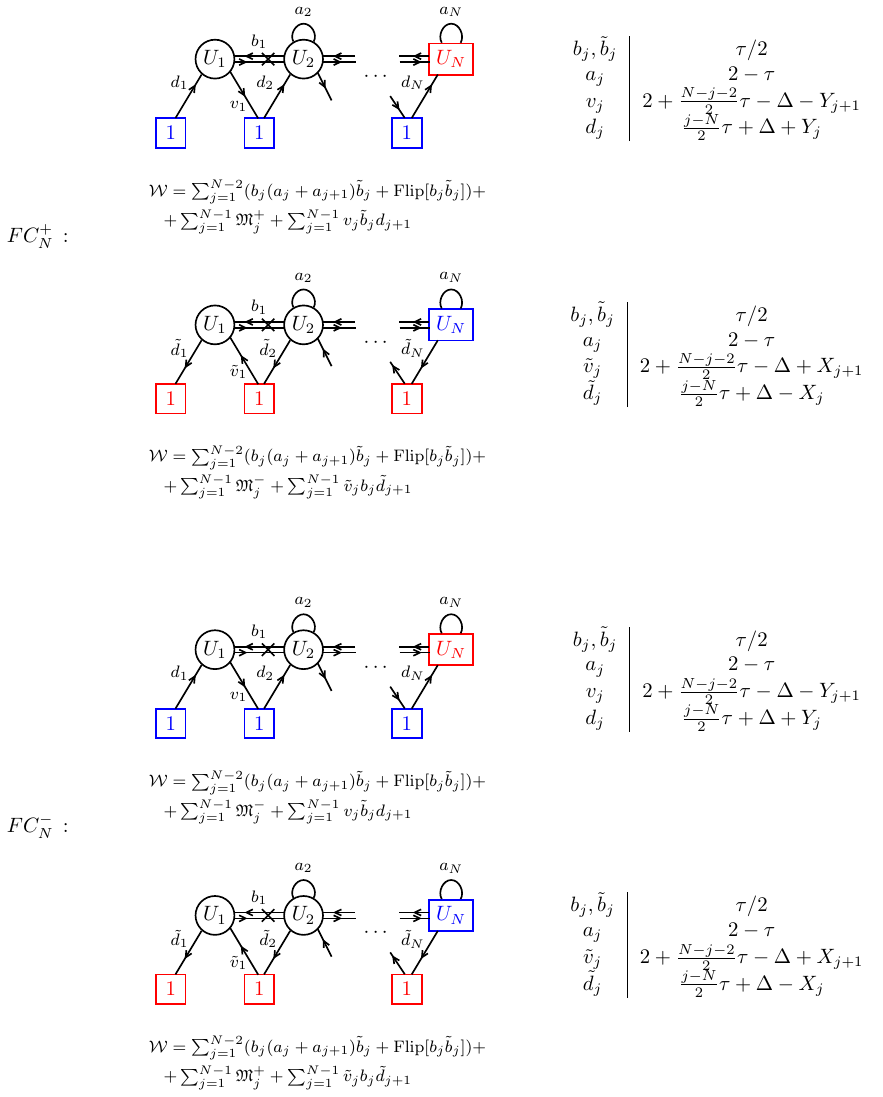}
    }
	\caption{UV completions of the $FC^{\pm}_N$ theory. The first and third quivers  are the first UV completions of the $FC^+$ and $FC^-$ theories respectively. The second and fourth quivers  are the second completions of the $FC^+$ and $FC^-$ theories. Notice that in both the first UV completions  the $d_N$ arrow points  towards the manifest $\color{red}U(N)$ global symmetry and the superpotential contains respectively the positively/negatively charged monopoles. In the second UV completions 
 the $\tilde d_N$ arrow emerges from the manifest
 we flip both the arrows and the sign of the monopoles turned on in the superpotentials. On the right we give the list of R-charges of the field of the theories, these are specified by a trial value mixed with the other abelian symmetries. In particular, the parameter $\D$ is associated to $U(1)_\D \subset S[{\color{red}U(N)} \times {\color{blue}U(N)}]$. 
 }
	\label{fig:FC_quiver}
\end{figure}
Starting from both the $FM_N$ and the $FH_N$ theories we can perform a real mass deformation and flow to a new improved bifundamental  called $FC_N^\pm$ (see Appendix \ref{app:3dlim}). If we start from the $FM_N$ we break the $S[{\color{red}U(N)} \times {\color{blue}U(N)}] \times U(1)_\D$ symmetry down to simply $S[{\color{red}U(N)} \times {\color{blue}U(N)}]$. While starting from the $FH_N$ we break the global ${\color{red}USp(2N)} \times {\color{blue}U(N)}$ global symmetry down to $S[{\color{red}U(N)} \times {\color{blue}U(N)}]$. \\
Independently of the theory we started from we land on a $3d$ $\mathcal{N}=2$ theory called $FC_N^\pm$, that is characterized by the following global symmetry:
\begin{align}\label{eq:FC_globalsymm}
	S[{\color{red}U(N)} \times {\color{blue}U(N)}] \times U(1)_\tau \,.
\end{align}
We depict the theory in short as:
\be\label{fig:FC_symbol}
    \includegraphics[]{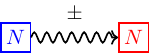}
\ee
to recall that it is characterized by two $U(N)$ global symmetries and that the spectrum contains a bifundamental operator in the $\bf N \times \bar{N}$ of ${\color{blue} U(N)} \times {\color{red}U(N)}$, as specified by the arrow. Also, we label the theory with a $\pm$ sign to distinguish between the $FC_N^+$ and $FC_N^-$ theories. 

Both the $FC_N^+$ and $FC_N^-$ SCFTs admit two possible UV Lagrangian completions that are depicted in Figure \ref{fig:FC_quiver}\footnote{If the arrows in the saw point to the manifest global symmetry then the sign of the $FC_N^\pm$ theory is the same as the sign of the monopoles turned on in the superpotential as in  the first UV completions. Conversely, if the arrows in the saw point to the emergent symmetry then the sign of the corresponding $FC_N^\pm$ theory is the opposite as that of the monopoles in the superpotential as in the second UV completions.}. All the UV completions do not posses the full global symmetry in \eqref{eq:FC_globalsymm} but they realize manifestly only one $U(N)$ global symmetry while it is visible only the Cartan subgroup of the other.

Notice that starting from this point of the draft, in addition to the data, such as matter content and superpotential, which we usually report in the Figures,  as in \ref{fig:FC_quiver}, the definition of the UV completions of the $3d$ theories, theories can include a collection of mixed background CS terms which we give explicitly when we write  the partition functions.

\paragraph{Gauge invariant operators}
The spectrum of gauge invariant operators of the $FC^{\pm}_N$ SCFT is given in Table \ref{tab:FC_operators}.
\begin{table}[]
\renewcommand{\arraystretch}{1.2}
\centering
\begin{tabular}{|c|cc|c|}\hline
{} & ${\color{red}U(N)}$ & ${\color{blue}U(N)}$ & R charge\\ \hline
${\color{red}\mathsf{A}}$ & ${\bf N^2-1}$ & $\bf 1$  & $2-\tau$ \\
${\color{blue}\mathsf{A}}$ & $\bf1$ & ${\bf N^2-1}$  & $2-\tau$ \\
$\Pi$ & $\bf \bar{N}$ & $\bf N$  & $\Delta$ \\
 \hline
\end{tabular}
\caption{List of all the gauge invariant operator that compose the spectrum of the $FC^{\pm}_N$ SCFT. The R-charge is given as a trial value mixed with the other two abelian symmetries of the theory. $U(1)_\D$ is the only abelian subgroup of the $S[{\color{red}U(N)}\times{\color{blue}U(N)}]$ global symmetry group and its associated mixing parameter is $\D$.}
\label{tab:FC_operators}
\end{table}
These operators are constructed from the UV completions in Figure \ref{fig:FC_quiver} as follows.

\begin{itemize}
    \item The operator $\color{red}\mathsf{A}$ is simply the gauge sinlgets $a_N$ in the first UV completion of both the $FC^+_N$ and $FC^-_N$ theories. In both the second UV completions it is construced by collecting together monopole and mesonic operators in a $N \times N$ traceless matrix. For simplicity let us give an example when $N=3$, meaning that the UV completions are quiver theories with two gauge nodes. For the second UV completion of the $FC^+_N$ theory we construct:
    \begin{align}
        {\color{red}\mathsf{A}} = \begin{pmatrix}
            \CF[b_1 \tilde{b}_1] & \M^{+,0} & \M^{+,+} \\
            \tilde{d}_1 \tilde{v}_1 & \CF[b_2 \tilde{b}_2] & \M^{0,+} \\
            \tilde{d}_1 \tilde{b}_1 \tilde{v}_2 & \tilde{d}_2 \tilde{v}_2 & -(\CF[b_1\tilde{b}_1] + \CF[b_2 \tilde{b}_2])
        \end{pmatrix}
    \end{align}
    While for the second UV completion of the $FC^-_N$ theory we construct:
    \begin{align}
        {\color{red}\mathsf{A}} = \begin{pmatrix}
            \CF[b_1 \tilde{b}_1] & \M^{-,0} & \M^{-,-} \\
            \tilde{d}_1 \tilde{v}_1 & \CF[b_2 \tilde{b}_2] & \M^{0,-} \\
            \tilde{d}_1 \tilde{b}_1 \tilde{v}_2 & \tilde{d}_2 \tilde{v}_2 & -(\CF[b_1\tilde{b}_1] + \CF[b_2 \tilde{b}_2])
        \end{pmatrix}
    \end{align}

    \item The $\color{blue}\mathsf{A}$ operator is constructed analogously as the $\color{red}\mathsf{A}$ operator described above. It is realized by the gauge singlet $a_N$ in the second UV completion of both the $FC^+_N$ and $FC^-_N$, while it consist of a collection of monopoles and mesons in the first UV completions.

    \item The operator $\Pi$ in the first UV completions is constructed by collecting mesons of the form: $d_i b_i \ldots b_{N-1}$ for each $i = 1,\ldots,N$ to form the ${\bf N} \times {\bf \bar N}$ of ${\color{blue}U(N)} \times {\color{red}U(N)}$.
    In the second UV completions  (notice the sign of the arrows is flipped) we collect mesons of the form: $\tilde d_i \tilde b_i \ldots \tilde b_{N-1}$ for each $i = 1,\ldots,N$ to form again the ${\bf N} \times {\bf \bar N}$ of ${\color{blue}U(N)} \times {\color{red}U(N)}$.
    
\end{itemize}

\paragraph{Partition function and mirror duality}
From the the UV Lagrangian completions in Figure \ref{fig:FC_quiver} we can write down the $S^3_b$ partition function of the $FC^\pm_N$ theories. We turn on the following real masses:
\begin{align}
    & \vec{X} \quad \text{for} \quad {\color{red}SU(N)} \nn \,, \\
    & \vec{Y} \quad \text{for} \quad {\color{blue}SU(N)} \nn \,, \\
    & \D \quad \text{for} \quad U(1) \subset S[{\color{red}U(N)} \times {\color{blue}U(N)}] \nn \,, \\
    & \tau \quad \text{for} \quad U(1)_\tau \,,
\end{align}
since $\vec{X}$ and $\vec{Y}$ are $SU(N)$ variables we impose: $\sum_{j=1}^N X_j = \sum_{j=1}^N Y_j = 0$. \\
The first UV completion of the $FC_N^+$ theory has:\footnote{Notice that the $j$-th gauge node has FI parameter set to: $(\tau - Y_{j+1} + Y_j)/2$. This is consistent with the fact that the positive monopole of the $j$-th node is in the superpotential and therefore it has R-charge 2. Before turning on the monopole in the superpotential its R-charge and charge under the abelian global symmetries can be computed to be:
\begin{align}
    (1-j)\tau + \tfrac{1}{2}[ 4j(\tfrac{\tau}{2}-1) + (\D + \tfrac{j-N}{2}\tau + Y_j -1) + (1 - \D +\tfrac{N-j-2}{2}\tau - Y_{j+1}) ] +\xi_j = 2 - \tfrac{\tau + Y_j - Y_{j+1}}{2} +\xi_j\,.
\end{align}
Turning on the monopole in the superpotential freezes the FI $\xi_j=(\tau - Y_{j+1} + Y_j)/2$. 
Notice also that, by using $N$ times  the recursive definition  of the $FC^+_N$ theory, we reconstruct a mixed background CS contributing to the partition function as: $\exp \big[ i\pi (1-N)\frac{\tau}{2} \sum_{j=1}^N Y_j \big]$.}
\begin{align}\label{eq:FC+_parfun}
	Z^{(N)}_{FC^+,1} & (\vec{X}, \vec{Y}, \tau, \D) = e^{\pi i [ Y_N (\sum_{j=1}^N X_j + (1-N)\frac{\tau}{2}) - \frac{\tau}{2} \sum_{j=1}^{N-1} Y_j ]} 
	\prod_{j=1}^{N} s_b \big( \tfrac{iQ}{2} - \D - Y_N + X_j \big) \times \nn \\
	& \times s_b \big( -\tfrac{iQ}{2} + \tau \big)^{N} \prod_{j<k}^{N} s_b \big( -\tfrac{iQ}{2} + \tau \pm (X_j - X_k) \big) \times \nn \\	
    & \times \int d\vec{Z}_{N\!-\!1} \D_{U(N\!-\!1)} \big( \vec{Z} \big) e^{\pi i(\tau - Y_N)\sum_{j=1}^{N-1}Z_j} \times \nn \\
	& \times \prod_{j=1}^{N-1} \left[ \, \prod_{k=1}^N s_b \big( \tfrac{iQ}{2} - \tfrac{\tau}{2} \pm (Z_j - X_k) \big) 
	s_b \big( -\tfrac{iQ}{2} + \tfrac{\tau}{2} + \D - Z_j + Y_N \big) \right] \times \nn \\
	& \times Z_{FC^+,1}^{(N-1)} (\vec{Z},\{ Y_1,\ldots,Y_{N-1} \},\tau, \D - \tfrac{\tau}{2}) \,,
\end{align}
with the basis of the recursion given by a single chiral bifundamental times a background mixed CS interaction:
\begin{align}\label{eq:FC+_N=1parfun}
	Z_{FC^+,1}^{(1)} (X,Y,\tau,\D) = e^{i \pi X Y} s_b \big( \tfrac{iQ}{2} - \D + X - Y \big) \,.
\end{align}

The partition function of the second UV completion can be obtained by a redefinition of the parameters starting from \eqref{eq:FC+_parfun}.
\begin{align}
    Z_{FC^+,2}^{(N)} (\vec{X},\vec{Y},\tau, \D) = Z_{FC^+,1}^{(N)} ( -\vec{Y}, -\vec{X}, \tau, \D ) \,.
\end{align}
Both the UV theories flow to the same IR fixed point, therefore their partition function match. We can then define:
\begin{align}\label{eq:FC+_parfunmir}
    Z_{FC^+}^{(N)} (\vec{X}, \vec{Y}, \tau, \D ) \overset{\text{def}}{=} Z_{FC^+,1}^{(N)} (\vec{X}, \vec{Y}, \tau, \D ) = Z_{FC^+,2}^{(N)} (\vec{X}, \vec{Y}, \tau, \D )
\end{align}
introducing the quantity $Z_{FC^+}^{(N)}$ which do not refer to any specific choice of UV completion. Notice that the $\vec{X}$ and $\vec{Y}$ entries in the partition function defined in \eqref{eq:FC+_parfunmir} require a precise ordering. The rule is that given the compact description in \eqref{fig:FC_symbol} the arrow emerges from the $U(N)$ symmetry parameterized by the second entry of the partition function and points towards the $U(N)$ parameterized by the first one. This means that the first entry corresponds to the $U(N)$ global symmetry which is manifest in  $ Z^{(N)}_{FC^+,1}$ 
and emergent in $ Z^{(N)}_{FC^+,2}$  while the second entry corresponds to the $U(N)$ symmetry which is emergent in in  $ Z^{(N)}_{FC^+,1}$  and manifest in $ Z^{(N)}_{FC^+,2}$.

Similarly for the $FC_N^-$ theory we write the partition function of its first UV completion\footnote{The FI parameter of the $j$-th gauge node is $(-\tau + Y_{j+1} - Y_j)/2$, which is frozen to this value by the presence of the monopole superpotential.
Notice also that, by using $N$ times  the recursive definition  of the $FC^-_N$ theory, we reconstruct a mixed background CS contributing to the partition function as: $\exp \big[ -i\pi (1-N)\tau/2 \sum_{j=1}^N Y_j \big]$.}:
\begin{align}\label{eq:FC-_parfun}
	Z_{FC^-,1}^{(N)} & (\vec{X}, \vec{Y}, \tau, \D) = e^{\pi i Y_N (-\sum_{j=1}^N X_j - (1-N)\frac{\tau}{2}) + \pi i \frac{\tau}{2} \sum_{j=1}^{N-1} Y_j} \prod_{j=1}^{N} s_b \big( i\tfrac{Q}{2} - \D - Y_N + X_j \big) \times \nn \\
	& \times s_b \big( -\tfrac{iQ}{2} + \tau \big)^{N} \prod_{j<k}^{N} s_b \big( -i\tfrac{Q}{2} + \tau \pm (X_j - X_k) \big) \times \nn \\
	& \times \int d\vec{Z}_{N\!-\!1} \D_{U(N\!-\!1)} \big( \vec{Z} \big) e^{\pi i(-\tau + Y_N)\sum_{j=1}^{N-1}Z_j} \times \nn \\
	& \times \prod_{j=1}^{N-1} \left[ \, \prod_{k=1}^N s_b \big( \tfrac{iQ}{2} - \tfrac{\tau}{2} \pm (Z_j - X_k) \big) 
	s_b \big( -\tfrac{iQ}{2} + \tfrac{\tau}{2} + \D - Z_j + Y_N \big) \right] \times \nn \\
	& \times Z_{FC^-,1}^{(N-1)}(\vec{Z},\{ Y_1,\ldots,Y_{N-1} \},\tau, \D - \tfrac{\tau}{2}) \,,
\end{align}
with the basis of the recursion given by a single chiral bifundamental times a background CS interaction:
\begin{align}\label{eq:FC-_N=1parfun}
	Z_{FC^-,1}^{(1)} (X,Y,\tau,\D) = e^{-\pi i X Y}s_b \big( \tfrac{iQ}{2} - \D + X - Y \big) \,.
\end{align}
Where again $\vec{X},\vec{Y}$ are the real masses of ${\color{red} SU(N)} \subset {\color{red} U(N)}$ and ${\color{blue} SU(N)} \subset {\color{blue}U(N)}$ symmetries respectively so one can impose the constraint $\sum_{j=1}^N X_j = \sum_{j=1}^N Y_j = 0$ in all the provided partition functions. Also, we recall that $\D$ is the real mass associated to the $U(1)$ centre of the global symmetry $S[ {\color{red}U(N)} \times {\color{blue}U(N)} ]$. \\
The partition function of the second UV completion can be obtained as:
\begin{align}
    Z_{FC^-,2}^{(N)} (\vec{X}, \vec{Y}, \tau, \D) = Z_{FC^-,1}^{(N)} ( -\vec{Y}, -\vec{X}, \tau , \D) \,.
\end{align}
We then define:
\begin{align}\label{eq:FC-_parfunmir}
    Z_{FC^-} (\vec{X},\vec{Y},\tau,\D) \overset{\text{def}}{=} Z_{FC^-,1} (\vec{X},\vec{Y},\tau,\D) = Z_{FC^-,2} (\vec{X},\vec{Y},\tau,\D)\,,
\end{align}
where as above we introduced the quantity $Z_{FC^-}^{(N)}$ which do not refer to any specific choice of UV completion following the same rule  that given the compact description  in \eqref{fig:FC_symbol} the arrow emerges from the $U(N)$ symmetry parameterized by the second entry of the partition function and points towards the $U(N)$ parameterized by the first one. 

This means that the first entry corresponds to the $U(N)$ global symmetry which is manifest in  $ Z^{(N)}_{FC^-,1}$ 
and emergent in $ Z^{(N)}_{FC^-,2}$  while the second entry corresponds to the $U(N)$ symmetry which is emergent in in  $ Z^{(N)}_{FC^-,1}$  and manifest in $ Z^{(N)}_{FC^-,2}$.

\paragraph{Interesting deformations}
We now discuss two interesting deformations of the $FC^\pm_N$ theory. As for the other improved bifundamentals, these can be derived by performing a suitable real mass deformation starting from the parent theories (see Appendix \ref{app:3dlim}).

The first deformation consist in performing a nilpotent mass for the adjoint operator $\color{red}\mathsf{A}$. Under this deformation the $FC^\pm_N$ theory becomes simply an $\color{blue}U(N)$ antifundamental chiral\footnote{Similarly, performing a similar deformation by giving a nilpotent mass for the operator $\color{red}\mathsf{A}$ provides a fundamental of $\color{blue}U(N)$.}
\begin{align}\label{fig:FCtochiral}
    \includegraphics[]{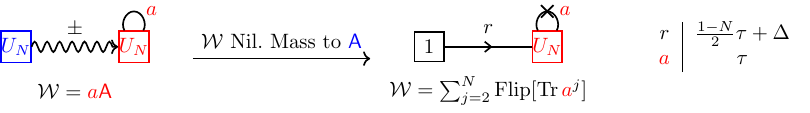}
\end{align}
Which translates to the following partition function identity:
\begin{align}
    Z_{FC^{\pm}}^{(N)} & (\vec{X}, \vec{Y} = \{ \tfrac{N-1}{2}\tau + V , \ldots, \frac{1-N}{2}\tau + V \} , \tau , \D ) = \nn \\
    = & e^{ \pm \tfrac{i \pi }{2} [ (N-1)\tau \big( \sum_{j=1}^N X_j - N V \big) + N V \sum_{j=1}^N X_j ] } \times \nn \\
    & \times \prod_{j=2}^N s_b \big( -\tfrac{iQ}{2} + j \tau \big) \prod_{j=1}^N s_b \big( \tfrac{iQ}{2} - \tfrac{1-N}{2}\tau - \D + X_j - V \big) \,.
\end{align}
Notice that the limits for the $FC^+_N$ and $FC^-_N$ differ only for a background term.

The second deformation relates a $FC_N^\pm$ theory with an identity operator.
\begin{align}\label{fig:FCtodelta}
    \includegraphics[]{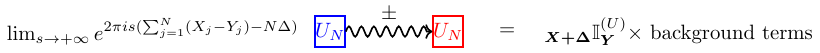}
\end{align}
On the l.h.s.~we have an $FC^\pm_N$ theory coupled to a background BF coupling of the form:
\begin{align}
    \exp \big[ 2\pi i s(\sum_{j=1}^N (X_j - Y_j) - N\D)
    \big]
\end{align}
which is divergent as $s \to + \infty$. Under this limit the $FC_N^\pm$ behaves as an identity operators times contact terms that can be read from the corresponding partition function identity provided below:
\begin{align}
    _{\vec{X}+\D} \mathbb{I}^{(U)}_{\vec{Y}} (\tau) 
    e^{ \pm \frac{i \pi}{2} \big[ 
    \sum_{j=1}^N (X_j^2 + Y_j^2) 
    + \tfrac{1}{2}N \D^2 
    + N (N-1) \tau \D
    - \tfrac{N(N^2-1)}{6} \tau^2
    \big] } = \nn \\ 
    = \lim_{s \to +\infty} e^{ 2\pi i s \big( \sum_{j=1}^N (X_j - Y_j) - N\D \big) } Z_{FC^\pm }^{(N)} (\vec{X}, \vec{Y}, \tau ,\D ) \,.
\end{align}
We provide a proof of this relation at the end of Section \ref{sec:chiralbasicdual}.

\subsection{The $FT_N$ theory}
\begin{figure}[h]
	\centering
	\includegraphics[]{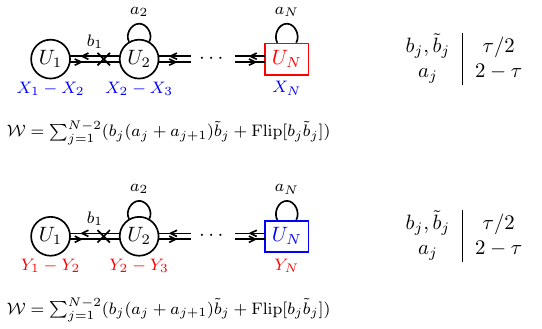}
	\caption{UV completions of the $FT_N$ theory. In the figure we explicitly wrote the FI parameters associated to each gauge node (BF couplings for flavors nodes).
    On the right we give the R-charges of the fields composing the theories given as a trial value mixed with the $U(1)_\tau$ symmetry.}
	\label{fig:FT_quiver}
\end{figure}
Starting from any of the previously introduced theories we can perform a series of suitable real mass deformation and land on the $FT_N$ SCFT (see Appendix \ref{app:3dlim}), which is characterized by the following global symmetry group.
\begin{align}
    {\color{red}SU(N)} \times {\color{blue}SU(N)} \times U(1)_\tau \,.
\end{align}
We depict the theory in short as:
\begin{align}
    \includegraphics[]{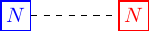}
\end{align}
to recall that the theory possesses two $U(N)$ global symmetries. 

The $FT_N$ theory coincides with the $T[SU(N)]$ theory of \cite{Gaiotto:2008ak} with the  addition of an adjoint chiral multiplet flipping the Higgs branch moment map. 

The $FT_N$ SCFT admits two possible UV completions that are depicted in Figure \ref{fig:FT_quiver}. In each completion only one of the two global $SU(N)$ symmetries is realized manifestly as the flavor symmetry rotating the last bifundamental. The second $SU(N)$ global symmetry instead enhances from the $N-1$ topological symmetries due to all the nodes being balanced \cite{Gaiotto:2008ak}. From the $\mathcal{N}=4$ perspective the $U(1)_\tau$ flavor symmetry is the commutant inside the $SU(2) \times SU(2)$ R-symmetry.

\paragraph{Gauge invariant operators}
The gauge invariant operators of the $FT_N$ theory are simply given by two $SU(N)$ adjoint operators, as listed in Table \ref{tab:FT_operators}. 
\begin{table}[]
\renewcommand{\arraystretch}{1.2}
\centering
\begin{tabular}{|c|cc|c|}\hline
{} & ${\color{red}SU(N)}$ & ${\color{blue}SU(N)}$ & R charge\\ \hline
${\color{red}\mathsf{A}}$ & ${\bf N^2-1}$ & $\bf 1$  & $2-\tau$ \\
${\color{blue}\mathsf{A}}$ & $\bf1$ & ${\bf N^2-1}$  & $2-\tau$ \\
 \hline
\end{tabular}
\caption{List of all the gauge invariant operators that compose the spectrum of the $FT^{\pm}_N$ SCFT. The R-charge is given as a trial value mixed with the other abelian symmetry of the theory, $U(1)_\tau$.}
\label{tab:FT_operators}
\end{table}
These are constructed from the UV completions in Figure \ref{fig:FT_quiver} as follows.
\begin{itemize}
    \item The $\color{red}\mathsf{A}$ operator is simply the gauge singlet $a_N$ in the first UV completion. In the second completion is instead obtained by collecting singlets and monopole operators in an $N \times N$ matrix. The $N-1$ singlets $\CF[b_i \tilde{b}_i]$ compose the diagonal of the matrix. The off-diagonal entries are filled with monopoles that are positively/negatively charged under strings of consequent gauge nodes.

    \item The $\color{blue}\mathsf{A}$ operator is construceted analogously as above. This time it is obtained as a collection of monopoles in the first UV completion, while it is simply the singlet $a_N$ in the second completion.
\end{itemize}

\paragraph{Partition function and mirror duality}
From the UV Lagrangian completions in Figure \ref{fig:FT_quiver} we can write down the $S^3_b$ partition function. We turn on the following real masses:
\begin{align}
    & \vec{X} \quad \text{for} \quad {\color{red}SU(N)} \nn \,, \\
    & \vec{Y} \quad \text{for} \quad {\color{blue}SU(N)} \nn \,, \\
    & \tau \quad \text{for} \quad U(1)_\tau \,,
\end{align}
with the constraints: $\sum_{j=1}^N X_j = \sum_{j=1}^N Y_j = 0$. For the first UV completion we have:
\begin{align}\label{eq:FT_parfun}
	Z_{FT,1}^{(N)} (\vec{X},\vec{Y},\tau) = & e^{2\pi i Y_N \sum_{j=1}^N X_j} 
	s_b \big( -\tfrac{iQ}{2} + \tau)^{N} \prod_{j<k}^{N} s_b \big( -\tfrac{iQ}{2} + \tau \pm (X_j - X_k) ) \times \nn \\
	& \times \int d\vec{Z}_{N-1} \D_{N-1}(\vec{Z}) e^{-2\pi i Y_N \sum_{j=1}^{N-1} Z_j}
	Z_{FT,1}^{(N)} (\vec{Z},\{ Y_1,\ldots,Y_{N-1}\},\tau) \,,
\end{align}
with the basis of the recursion given by a mixed CS interaction:
\begin{align}\label{eq:FT_parfun_N=1}
	Z_{FT,1}^{(1)}(X,Y,\tau) = e^{2i\pi XY} \,.
\end{align}
As usual, in the first UV completion the first entry of the partition function is associated to the manifest $U(N)$ symmetry and the second to the emergent $U(N)$ symmetry. The definitions provided are ``off-shell", one need to impose afterwards that $\sum_{j=1}^N X_j = \sum_{j=1}^N Y_j = 0$.

The partition function of the second UV completion can be obtained from \eqref{eq:FT_parfun} by simply swapping $\vec{X} \leftrightarrow \vec{Y}$.
\begin{align}
    Z_{FT,2}^{(N)} (\vec{X}, \vec{Y}, \tau) = Z_{FT,1}^{(N)} (\vec{Y}, \vec{X}, \tau) \,,
\end{align}
where now  the first entry of the partition function is associated to the emergent $U(N)$ symmetry and the second to the manifest $U(N)$ symmetry.

We can then define:
\begin{align}\label{eq:FT_parfunmir}
    Z_{FT}^{(N)} (\vec{X}, \vec{Y}, \tau) \overset{\text{def}}{=} Z_{FT,1}^{(N)} (\vec{X}, \vec{Y}, \tau) = Z_{FT,2}^{(N)} (\vec{X}, \vec{Y}, \tau) \,.
\end{align}

Notice that we provided the partition function expressions without imposing the constraints $\sum_{j=1}^N X_j = \sum_{j=1}^N Y_j = 0$. One may impose them afterwards.

\subsection{Fusion to Identity}\label{sec:idwalls}

In this section we study the fusion  to identity property of the improved bifundamentals.

All the results in this subsection can be  obtained starting from the fusion to identity property of the $4d$ $FE_N$ theory given in \eqref{fig:FEdelta}
and performing the $3d$ circle reduction combined with a  suitable real mass deformations, following Appendix \ref{app:3dlim}.
Alternatively we can prove each fusion to identity as in \cite{Bottini:2021vms}
starting from the UV Lagrangian
description and dualizing sequentially
the nodes using an appropriate Seiberg-like duality (Aharony duality \cite{Aharony:1997gp} or its monopole deformations \cite{Benini:2017dud}).

\paragraph{\boldmath{$FE^{3d}_N$}}
Two $FE^{3d}_N$ theories fuse to an identity wall when they are glued with an extra antisymmetric $a$, coupled to the antisymmetric $\mathsf{A}_{L/R}$ of the two theories and a linear superpotential term for the fundamental monopole $\M$ of the new gauge node.
\begin{align}
\resizebox{.95\hsize}{!}{ 
    \includegraphics[]{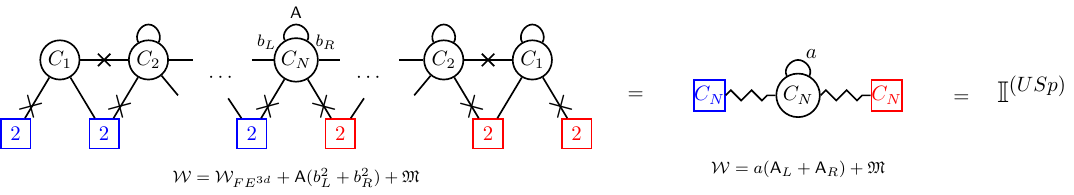}
}
\label{fig:FE3d_idwall1}
\end{align}
On the l.h.s.~we depicted the UV perspective, where the two theories are glued through the gauging of the manifest $USp(2N)$ symmetries. The F-term for the $a$ field imply that $\mathsf{A}_L = \mathsf{A}_R$. The only antisymmetric field left, which we rename $\mathsf{A}$, is then coupled to the bifundamentals on its sides. In addition, we encoded the superpotential terms of the UV completions in the $FE^{3d}_N$ theories in the $\CW_{FE^{3d}}$ term.

Due to the self-mirror property for the $FE^{3d}$ theory \eqref{eq:FE3d_mir}, which exchanges the manifest and emergent $USp(2N)$ symmetries, the fusion to identity property holds also when one, or both, of the gauged symmetries are emergent.

Starting from the UV picture we can write down the corresponding partition function identity, which is:
\begin{align}
    \int & d\vec{Z}_{N} \D_{USp(2N)} (\vec{Z},\tau) Z_{FE^{3d},1}^{(N)}(\vec{Z},\vec{X},\tau,\D) Z_{FE^{3d},1}^{(N)}( \vec{Z},\vec{Y},\tau,-\D) = \nn \\
    & = \int d\vec{Z}_{N} \D_{USp(2N)} (\vec{Z},\tau) Z_{FE^{3d}}^{(N)}(\vec{X},\vec{Z},\tau,\D) Z_{FE^{3d}}^{(N)}( \vec{Z},\vec{Y},\tau,-\D) = \nn \\
    & = \frac{1}{\D_{USp(2N)}(\vec{X},\tau)} \sum_{l=0,1} \sum_{\s \in S_N} \prod_{j=1}^N \d \big( X_j - (-1)^l Y_{\s(j)} \big) = {}_{\vec{X}} \mathbb{I}^{(USp)}_{\vec{Y}} (\tau)\,.
\end{align}

\paragraph{\boldmath{$FM_N$}}
Two $FM_N$ theories fuse to an identity wall when they are glued with an extra adjoint $a$, coupled to the adjoint $\mathsf{A}_{L/R}$ of the two theories and a linear superpotential term for the $\M^\pm$ monopoles of the new gauge node.
\begin{align}\label{fig:FM_idwall}
\resizebox{.95\hsize}{!}{ 
    \includegraphics[]{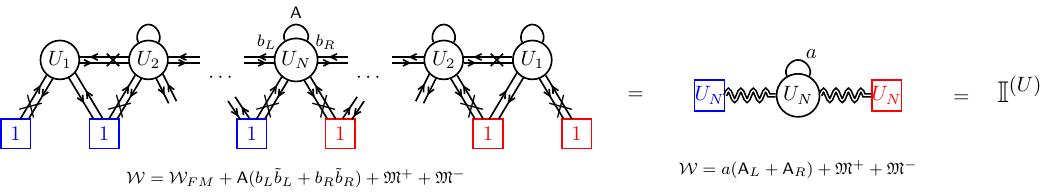}
}
\end{align}
On the l.h.s.~we depicted the UV perspective where the gluing is performed via the gauging of two manifest global symmetries. The F-terms for $a$ imply the relation $\mathsf{A}_L =\mathsf{A}_R$. The remaining adjoint chiral, which we relabel simply as $\mathsf{A}$, is then coupled to the bifundamentals on its sides. Also, we encoded all the superpotential terms of the UV completion of the $FM_N$ theories in $\cW_{FM}$. 

Due to the self-mirror property of the $FM_N$ theory \eqref{eq:FM_parfun}, which exchanges the manifest and emergent $U(N)$ symmetries, the fusion to identity property holds not only when we glue manifest symmetries but also when one, or both, the gauged symmetries are taken to be emergent.

As a partition function identity this property translates to:
\begin{align}\label{eq:FM_idwall}
	\int & d\vec{Z}_N \D_{U(N)}(\vec{Z},\tau) Z_{FM,1}^{(N)}(\vec{Z},\vec{X},\tau,\D) Z_{FM,1}^{(N)}(\vec{Z},\vec{Y},\tau,-\D) 
	= \nn \\
    & = \int d\vec{Z}_N \D_{U(N)}(\vec{Z},\tau) Z_{FM}^{(N)}(\vec{X},\vec{Z},\tau,\D) Z_{FM}^{(N)}(\vec{Z},\vec{Y},\tau,-\D) 
	= \nn \\
    & = \frac{1}{\D_{U(N)}(\vec{X},t)} \prod_{\sigma \in S_N} \prod_{j=1}^{N} \delta \big( X_j - Y_{\sigma(j)} \big) 
    ={}_{\vec{X}}\mathbb{I}^{(U)}_{\vec{Y}}(\tau) \,,
\end{align}

\paragraph{\boldmath{$FH_N$}} 
There are two possibile ways to fuse two $FH_N$ theories to an identity wall. We can either glue the theories thorugh identification and gauging of their $USp(2N)$ or $U(N)$ symmetries. In both cases we introduce an extra field $a$ coupling to the adjoint/antisymmetric operators of the two $FH_N$ theories. In the case of the $U(N)$ gluing we also turn on a superpotential linear in the $\M^\pm$ monopoles. 
\begin{align}\label{fig:FE3d_idwall}
\resizebox{.95\hsize}{!}{ 
    \includegraphics[]{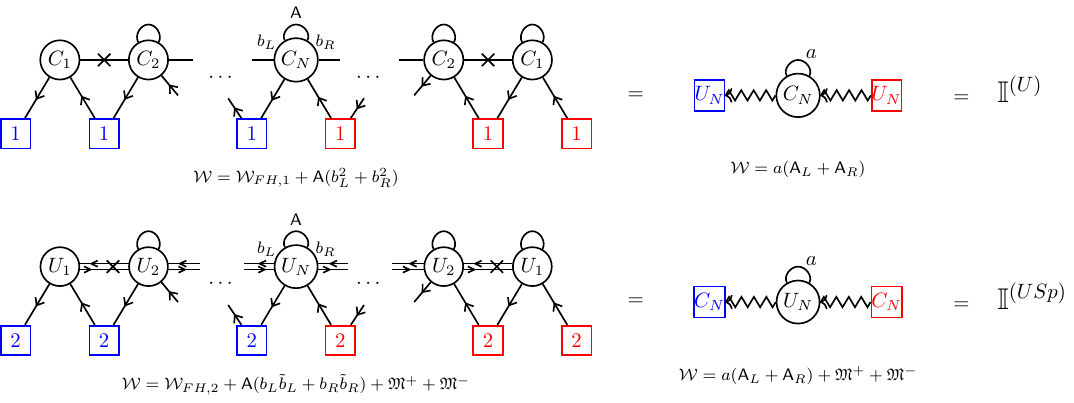}
}
\end{align}
On the l.h.s.~we depicted the UV perspective where the two theories are glued through the gauging of manifest global symmetries. 

In the first gluing, we take two copies of the first UV completion of the $FH_N$ theory as in Figure \ref{fig:FH_quiver}(in one copy we reverse the sign of all the arrows). 
This gluing requires only to introduce an extra antisymmetric chiral coupled to the antisymmetric operators of the two theories. Notice that due to the F-term for the field $a$ we are left with a single antisymmetric in the UV completion, which couples to the bifundamentals on its sides. In $\CW_{FH,1/2}$ we also encoded all the superpotential terms of the first/second UV completion of the $FH_N$ theory.

In the second line  we are instead gluing two copies of the second UV completions of the $FH_N$ theory (in one copy we reverse the sign of all the arrows) through their manifest $U(N)$ global symmetries with the addition of an extra adjoint chiral and monopole superpotential. Again, the F-term for $a$ imply that only one adjoint survives in the UV completion and that it is coupled to the bifundamentals.

For both gluings the fusion to identity holds true also when the gluing is non-Lagrangian, meaning that the gauged symmetries can be also non manifest as indicated in the compact notation in the middle of the figure which makes no reference to the UV completions.

As a partition function identity we have, for the first case:
\begin{align}
	\int & d\vec{Z}_N \D_{USp(2N)}(\vec{Z},\tau) Z_{FH,1}^{(N)}(\vec{Z},\vec{X},\tau,\D) Z_{FH,1}^{(N)}(\vec{Z},-\vec{Y},\tau,-\D) = \nn \\
    & = \int d\vec{Z}_N \D_{USp(2N)}(\vec{Z},\tau) Z_{FH}^{(N)}(\vec{Z},\vec{X},\tau,\D) Z_{FH}^{(N)}(\vec{Z},-\vec{Y},\tau,-\D) = {}_{\vec{X}}\mathbb{I}^{(U)}_{\vec{Y}}(\tau) \nn \,. \\
\end{align}
For the second case we have:
\begin{align}
    \int & d\vec{Z} \D_{U(N)}(\vec{Z},\tau) Z_{FH,2}^{(N)}(\vec{X},-\vec{Z},\tau,\D) Z_{FH,2}^{(N)}(\vec{Y},\vec{Z},\tau,-\D) = \nn \\
    & = \int d\vec{Z} \D_{U(N)}(\vec{Z},\tau) Z_{FH}^{(N)}(\vec{X},-\vec{Z},\tau,\D) Z_{FH}^{(N)}(\vec{Y},\vec{Z},\tau,-\D) = {}_{\vec{X}}\mathbb{I}^{(USp)}_{\vec{Y}}(\tau) \,.
\end{align}
In both formulae notice the minus signs implementing the flip of the arrows as indicated in \eqref{FHsf}.

\paragraph{\boldmath{$FC^\pm_N$}}
To obtain an identity wall we need to fuse together a $FC^+_N$ and a $FC^-_N$ theories. This is done with the usual introduction of an extra adjoint field and also with a superpotential term linear in $\M^+$.
\begin{align}\label{fig:FE3d_idwall}
\resizebox{.95\hsize}{!}{ 
    \includegraphics[]{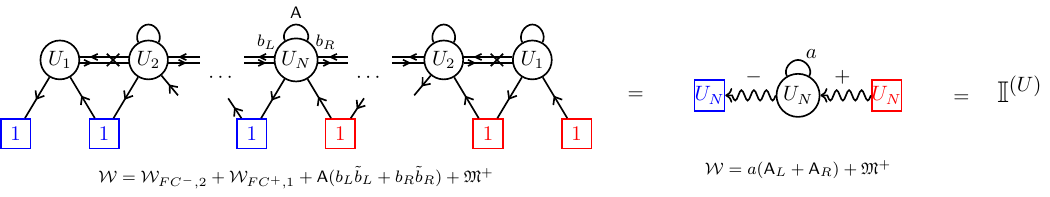}
}
\end{align}
On the l.h.s.~we consider the UV perspective where the gluing is done by gauging two manifest $U(N)$ symmetries. We thus consider the gluing of the first UV completion of a $FC^+_N$ theory with the second completion of a $FC^-_N$, as shown in Figure \ref{fig:FC_quiver}. As already explained, we also add an extra adjoint chiral and a monopole superpotential. Notice that in the UV description on the l.h.s.~are turned on the positive monopoles of all the gauge nodes. Moreover, in $\CW_{FC^-,2}$ and $\CW_{FC^+,1}$ we have encoded all the superpotential terms of the corresponding UV completion of the $FC^\pm_N$ theories.

As usual the identity wall property holds true also when the gluing is non-Lagrangian, meaning that the gauged symmetries can be also non manifest as indicated in the compact form in the middle which makes no reference to the UV completions.

We can write the corresponding partition function identity:
\begin{align}
	\int & d\vec{Z}_N \D_{U(N)}(\vec{Z},\tau) e^{i \pi (iQ) \sum_{j=1}^N Z_j} 
	Z_{FC^-,2}^{(N)}(\vec{X},\vec{Z},\tau,\D) Z_{FC^+,1}^{(N)}(\vec{Z},\vec{Y},\tau,-\D) = \nn \\
    & = \int d\vec{Z}_N \D_{U(N)}(\vec{Z},\tau) e^{i \pi (iQ) \sum_{j=1}^N Z_j} 
	Z_{FC^-}^{(N)}(\vec{X},\vec{Z},\tau,\D) Z_{FC^+}^{(N)}(\vec{Z},\vec{Y},\tau,-\D) = {}_{\vec{X}}\mathbb{I}^{(U)}_{\vec{Y}} \,.
\end{align}
Notice that the FI parameter of the newly gauged symmetry is frozen to be $iQ$, this is due to the presence of the monopole superpotential.

\paragraph{\boldmath{$FT_N$}} Two $FT_N$ theories fuse to an identity wall when they are glued together with the addition of an extra adjoint chiral $a$ which couples to adjoint operators of the two theoires. This property was first observed in \cite{Gaiotto:2008ak} and the associated partition function identity was derived in \cite{Bottini:2021vms}.
\begin{align}\label{fig:FE3d_idwall}
\resizebox{.95\hsize}{!}{ 
    \includegraphics[]{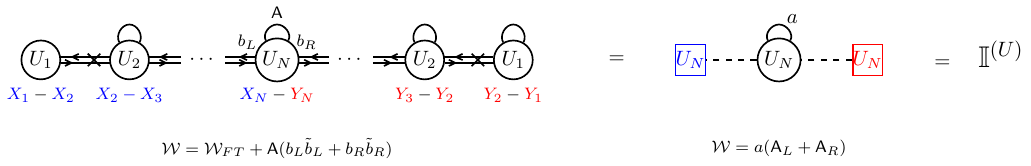}
}
\end{align}
On the l.h.s.~we are depicting the UV perspective where we gauge two copies of the first UV completions by gauging their manifest  $U(N)$ symmetry. Notice that we are writing explicitly the FI parameters associated to each node. In $\CW_{FT}$ we encoded all the superpotential terms coupling the adjoint chirals to the bifundamentals that are present in the UV completions of the $FT_N$ theories.

As usual the fusion to identity property holds true also when the gluing is non-Lagrangian, meaning that the gauged symmetries can be also non manifest as indicated in the compact form in the middle which makes no reference to the UV completions.

We can write the corresponding partition function identity:
\begin{align}
	\int & d\vec{Z}_N \D_{U(N)}(\vec{Z},\tau)
	Z_{FT,1}^{(N)}(\vec{Z},\vec{X},\tau) Z_{FT,1}^{(N)}(\vec{Z},-\vec{Y},\tau) = \nn \\
    & = \int d\vec{Z}_N \D_{U(N)}(\vec{Z},\tau)
	Z_{FT}^{(N)}(\vec{X},\vec{Z},\tau) Z_{FT}^{(N)}(\vec{Z},-\vec{Y},\tau) = {}_{\vec{X}}\mathbb{I}^{(U)}_{\vec{Y}} \,.
\end{align}
In \cite{Bottini:2021vms}, along the lines of \cite{Gaiotto:2008ak}, the $FT_N$ theory was identified with the S-duality wall,
this property was interpreted as the property $S S^{-1}=1$ of the S generator of $ SL(2,\mathbb{Z})$. The same equation with $\vec Y\to -\vec Y$ was instead interpreted as the  $S^{2}=-1$ property.

\section{$3d$ limits of the star-triangle duality}\label{3dST}
Starting from the $4d$ $\CN=1$ star-triangle duality in \eqref{fig:braid} we will study its $3d$ reduction and real mass deformations thereof. With this strategy we generate new $3d$ $\CN=2$ star-triangle dualities. These new dualities involve the improved bifundamental blocks described in Section \ref{sec:impbif} and also generalize, and can be reduced to, the dualities described in \cite{Benvenuti:2018bav,Amariti:2018wht}. \\

\subsection{Level $0$ star-triangle}
We perform the circle reduction of the $4d$ braid duality in \eqref{fig:braid}. The result of the limit is a $3d$ $\CN=2$ duality which is precisely the same as its $4d$ ancestor, with the addition of a superpotential term linear in the fundamental monopole of the $USp(2N)$ gauge node, which is generated by the compactification \cite{Aharony:2013dha}. The role of this superpotential ensures that the $3d$ theory has the same global symmetries as the $4d$ one:
\begin{align}
    {\color{red}USp(2N)} \times {\color{blue}USp(2N)} \times SU(2)_V \times U(1)_{\D_L} \times U(1)_{\D_R} \times U(1)_\tau \,.
\end{align} 
The partition function identity associated to this duality can be obtained starting from the $4d$ braid duality and performing the $3d$ reduction limit (see Appendix \ref{app:3dlim}).
\begin{align}\label{eq:lvl0}
    \int d \vec{Z}_N & \D_{USp(2N)}(\vec{Z},\tau) Z_{FE^{3d}}^{(N)}(\vec{X},\vec{Z},\tau, \D_L) Z_{FE^{3d}}^{(N)}(\vec{Z},\vec{Y},\tau, \D_R) \prod_{j=1}^N s_b \big( \D_L + \D_R \pm Z_j \pm V \big) = \nn \\
    & = Z_{FE^{3d}}^{(N)}(\vec{X},\vec{Y},\tau, \D_L+\D_R) \prod_{j=1}^N s_b \big( \D_R \pm X_j \pm V \big) s_b \big( \D_L \pm Y_j \pm V \big) \,.
\end{align}

By turning on a maximal nilpotent mass for both the antisymmetric ${\color{red}\mathsf{A}}$ and ${\color{blue}\mathsf{A}}$ breaking  $USp(2N)\to USp(2)$ we obtain the $3d$ $\mathcal{N}=2$ confining duality relating the  $USp(2N)$ SQCD with 6 chirals and monopole superpotential to the  WZ-model with 15N chirals (up to flips) discussed in \cite{Benvenuti:2018bav, Amariti:2018wht}.

\subsection{Level $1$ star-triangle  dualities}
\begin{figure}
    \centering
    \includegraphics[width=0.9\linewidth]{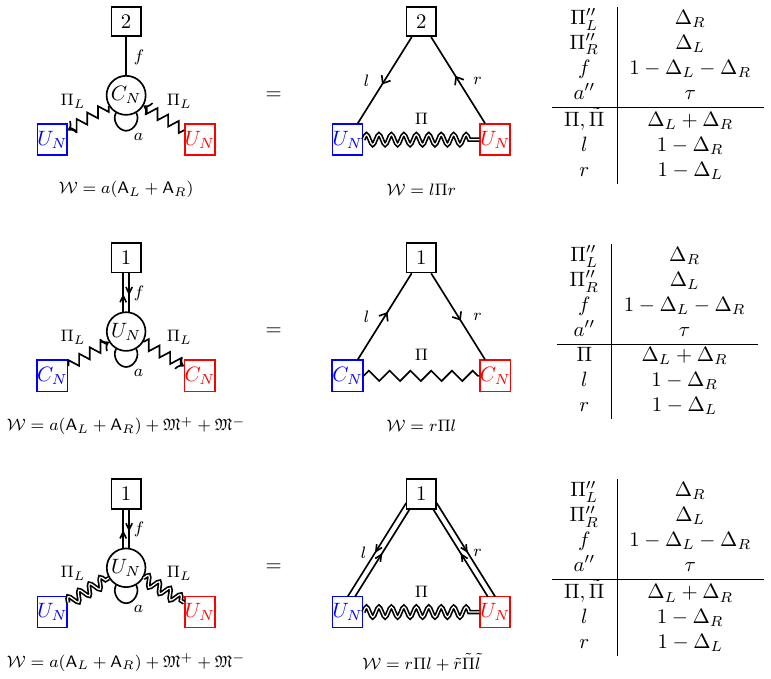}
    \caption{Level 1 real mass deformed star-triangle dualities. Beside each duality, on the right, we give the list of R-charges of the fields and of the bifundamental operators in the spectrum of the improved bifundamentals.}
    \label{fig:3dbraid_lev1}
\end{figure}
Starting from the $3d$ braid duality we can perform  real mass deformations with the effect of lowering the rank of the global symmetry by one unit. The effect of these deformations is, also, to go from the $FE^{3d}_N$ to other improved bifundamentals. The star-triangle dualities that we can obtain from these limits are three and are depicted in Figure \ref{fig:3dbraid_lev1}. In the following subsection we will describe them and provide the associated partition function identity. While we do not give a completely detailed operator map, we discuss how some of the operators are mapped in the dualities.

\paragraph{Duality I in  \ref{fig:3dbraid_lev1}:} 
We glue two $FH_N$ improved bifundamentals  by gauging their $USp(2N)$ global symmetry with the addition of a pair of fundamental chirals, with zero superpotential. The global symmetry of the resulting theory is:
\begin{align}
    S[{\color{red}U(N)} \times {\color{blue}U(N)}] \times SU(2)_V \times U(1)_{\D_L} \times U(1)_{\D_R} \times U(1)_\tau \,.
\end{align}
On the r.h.s.~we have a generalized WZ model involving a $FM_N$ theory. 
Notice that the charge of the flavor $f$ as given in the table, is not fixed by the superpotential rather we are using the $U(1)$ in $S[ {\color{red}U(N)}\times {\color{blue}U(N)}]$ to shift it to the given value.

The partition function identity associated to this duality can be obtained starting from the level 0 duality in eq. \eqref{eq:lvl0}, performing the following redefinitions:
\begin{align}
    & X_i \to X_i + s \quad , \quad Y_i \to Y_i + s \qquad \text{for} \quad i=1,\ldots,N \nn \\
    & \D_L \to \D_L + s \quad, \quad \D_R \to \D_R - s 
\end{align}
and then performing the limit $s \to +\infty$ (see Appendix \ref{app:3dlim}). We obtain:
\begin{align}\label{eq:lvl1_1}
    \int & d \vec{Z}_N \D_{USp(2N)}(\vec{Z},\tau) Z_{FH}^{(N)}(\vec{Z},\vec{X},\tau, \D_L) Z_{FH}^{(N)}(\vec{Z},-\vec{Y},\tau, \D_R) \prod_{j=1}^N s_b \big( \D_L + \D_R \pm Z_j \pm V \big) = \nn \\
    & = Z_{FM}^{(N)}(\vec{X},\vec{Y},\tau, \D_L+\D_R) \prod_{j=1}^N s_b \big( \D_R + X_j \pm V \big) s_b \big( \D_L - Y_j \pm V \big) \,.
\end{align}
Where the set of parameters $\vec{X}$ and $\vec{Y}$ are related as $\sum_{j=1}^N (X_j + Y_j) = 0$. \\ 

The mapping of some chiral ring generators is:
\begin{align}
    \Pi_L \Pi_R \quad &\longleftrightarrow \quad \Pi \nn \\
    \Pi_L f \quad &\longleftrightarrow \quad l \nn \\
    \Pi_R f \quad &\longleftrightarrow \quad r \nn \\
    f^2 a^{m-1} \mathsf{A}^{n-1} \quad &\longleftrightarrow \quad \mathsf{B}_{n,m} 
\end{align}
In the last line the operator $\mathsf{A}$ is an antisymmetric operator of the two $FH_N$ theories charged under the gauge symmetry. Due to the F-term for $a$ the antisymmetric operator of the left $FH_N$, call it $\mathsf{A}_L$, and that of the right one, call it $\mathsf{A}_R$, are related as $\mathsf{A}_L = \mathsf{A}_R = \mathsf{A}$.
The map for the operator $\tilde{\Pi}$ of the r.h.s.~is slightly more complicated.  This operator is mapped on the l.h.s. to a collection of monopoles charged under strings of consecutive  nodes, always including the central $USp(2N)$ node. For example for $N=3$, meaning that the two $FH_3$ theories have two gauge nodes we have:
\begin{align}
    \begin{pmatrix}
        \M^{0,0|1|0,0} & \M^{0,1|1|0,0} & \M^{1,1|1|0,0} \\
        \M^{0,0|1|1,0} & \M^{0,1|1|1,0} & \M^{0,1|1|1,1} \\
        \M^{1,1|1|0,0} & \M^{1,1|1|1,0} & \M^{1,1|1|1,1}
    \end{pmatrix} 
    \quad \longleftrightarrow \quad \tilde{\Pi}
\end{align}
Where we denote by $\M^1$ the monopole with the lowest flux of $USp(2N)$.

By turning on a maximal nilpotent mass for both the adjoint ${\color{red}\mathsf{A}}$ and ${\color{blue}\mathsf{A}}$ breaking  $U(N)\to U(1)$ we obtain the $3d$ $\mathcal{N}=2$ confining duality relating the  $USp(2N)$ SQCD with 4  chirals and zero superpotential  to the  WZ-model with 7N singlets (up to flips) discussed in \cite{Benvenuti:2018bav, Amariti:2018wht}.

\paragraph{Duality II in  \ref{fig:3dbraid_lev1}:} 

We glue two $FH_N$ improved bifundamentals  by gauging their $U(N)$ global symmetry with the addition of a pair of fundamental chirals, with monopole superpotential. The global symmetry of the resulting theory is:
\begin{align}
    {\color{red}USp(2N)} \times {\color{blue}USp(2N)} \times U(1)_{\D_L} \times U(1)_{\D_R} \times U(1)_\tau \,.
\end{align}
On the r.h.s.~we have a generalized WZ model involving a $FE^{3d}_N$ theory. 
The partition function identity associated to this duality can be obtained starting from the level 0 duality in eq. \eqref{eq:lvl0}, performing the following redefinitions:
\begin{align}
    & Z_i \to Z_i + V \qquad \text{for} \quad i=1,\ldots,N \nn \\
    & \D_L \to \D_L + V \quad , \quad \D_R \to \D_R - V 
\end{align}
and then performing the limit $V \to +\infty$ (see Appendix \ref{app:3dlim}). We obtain:
\begin{align}\label{eq:lvl1_1}
    \int & d \vec{Z}_N \D_{U(N)}(\vec{Z},\tau) Z_{FH}^{(N)}(\vec{X},-\vec{Z},\tau, \D_L) Z_{FH}^{(N)}(\vec{Y},\vec{Z},\tau, \D_R) \prod_{j=1}^N s_b \big( \D_L + \D_R \pm Z_j \big) = \nn \\
    & = Z_{FM}^{(N)}(\vec{X},\vec{Y},\tau, \D_L+\D_R) \prod_{j=1}^N s_b \big( \D_R - X_j \big) s_b \big( \D_L + Y_j \big) \,.
\end{align}

The mapping of some chiral ring generators is:
\begin{align}
    \Pi_L \Pi_R \quad &\longleftrightarrow \quad \Pi \nn \\
    \Pi_L f \quad &\longleftrightarrow \quad l \nn \\
    \Pi_R f \quad &\longleftrightarrow \quad r \nn \\
    f^2 a^{m-1} \mathsf{A}^{n-1} \quad &\longleftrightarrow \quad \mathsf{B}_{n,m} 
\end{align}
In the last line the operator $\mathsf{A}$ is an antisymmetric operator of the two $FH_N$ theories charged under the gauge symmetry. Due to the F-term for $a$ the antisymmetric operator of the left $FH_N$, call it $\mathsf{A}_L$, and that of the right one, call it $\mathsf{A}_R$, are related as $\mathsf{A}_L = \mathsf{A}_R = \mathsf{A}$. 

By turning on a maximal nilpotent  mass for both the antisymmetric ${\color{red}\mathsf{A}}$ and ${\color{blue}\mathsf{A}}$ breaking  $USp(2N)\to USp(2)$ we obtain the $3d$ $\mathcal{N}=2$ confining duality relating the  relating the  $U(N)$ SQCD with 3 flavors   and monople superpotentail  to the  WZ-model with 9N singlets (up to flips) discussed in \cite{Benvenuti:2018bav, Amariti:2018wht}.

\paragraph{Duality III in  \ref{fig:3dbraid_lev1}:} 

We glue two $FM_N$ improved bifundamentals  by gauging their $U(N)$ global symmetry with the addition of a pair of one flavor, with monopole superpotential. The global symmetry of the resulting theory is:
\begin{align}
 {\color{red}U(N)} \times {\color{blue}U(N)} \times U(1)_{\D_L} \times U(1)_{\D_R} \times U(1)_\tau \,.
\end{align}
On the r.h.s.~we have a generalized WZ model involving a $FM_N$ theory. 
The partition function identity associated to this duality can be obtained starting from the level 0 duality in eq. \eqref{eq:lvl0}, performing the following redefinitions:
\begin{align}
    X_i \to X_i + V \quad , \quad 
    Y_i \to Y_i + V \quad , \quad 
    Z_i \to Z_i + V \qquad \text{for} \quad i=1,\ldots,N 
\end{align}
and then performing the limit $V \to +\infty$ (see Appendix \ref{app:3dlim}). We obtain:
\begin{align}\label{eq:lvl1_3}
    \int & d \vec{Z}_N \D_{U(N)}(\vec{Z},\tau) Z_{FM}^{(N)}(\vec{X},\vec{Z},\tau, \D_L) Z_{FM}^{(N)}(\vec{Z},\vec{Y},\tau, \D_R) \prod_{j=1}^N s_b \big( \D_L + \D_R \pm Z_j \big) = \nn \\
    & = Z_{FM}^{(N)}(\vec{X},\vec{Y},\tau, \D_L+\D_R) \prod_{j=1}^N s_b \big( \D_R \pm X_j \big) s_b \big( \D_L \pm Y_j \big) \,.
\end{align}

The mapping of some chiral ring generators is:
\begin{align}
    \Pi_L \Pi_R \quad &\longleftrightarrow \quad \Pi \nn \\
    \tilde{\Pi}_L \tilde{\Pi}_R \quad &\longleftrightarrow \quad \tilde{\Pi} \nn \\
    \Pi_L f \quad &\longleftrightarrow \quad \tilde{l} \nn \\
    \tilde{\Pi}_L \tilde{f} \quad &\longleftrightarrow \quad l \nn \\
    \tilde{\Pi_R} f \quad &\longleftrightarrow \quad \tilde{r} \nn \\
    \Pi_R \tilde{f} \quad &\longleftrightarrow \quad r \nn \\
    f^2 a^{m-1} \mathsf{A}^{n-1} \quad &\longleftrightarrow \quad \mathsf{B}_{n,m} 
\end{align}
In the last line the operator $\mathsf{A}$ is an antisymmetric operator of the two $FM_N$ theories charged under the gauge symmetry. Due to the F-term for $a$ the antisymmetric operator of the left $FM_N$, call it $\mathsf{A}_L$, and that of the right one, call it $\mathsf{A}_R$, are related as $\mathsf{A}_L = \mathsf{A}_R = \mathsf{A}$. 

By turning on a maximal nilpotent  mass for both the adjoint ${\color{red}\mathsf{A}}$ and ${\color{blue}\mathsf{A}}$ breaking  $U(N)\to U(1)$ we obtain the $3d$ $\mathcal{N}=2$ confining duality relating the  $U(N)$ SQCD with 3 flavors   and monopole  superpotentail  to the  WZ-model with 9N singlets (up to flips) discussed in \cite{Benvenuti:2018bav, Amariti:2018wht}.

\subsection{Level $2$ star-triangle dualities}
\begin{figure}
    \centering
    \includegraphics[width=0.9\linewidth]{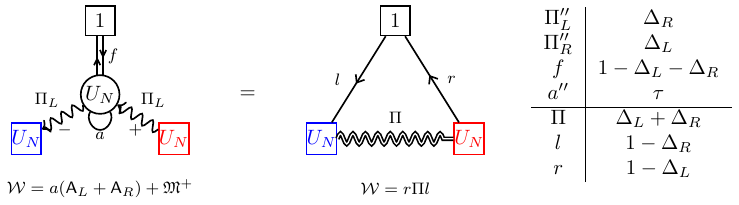}
    \caption{Level 2 real mass deformed star-triangle duality.}
    \label{fig:3dbraid_lev2}
\end{figure}
Starting from the level 1 dualities presented in the previous section we can perform a further real mass deformation. The result is a star-trinagle duality involving $FC_N^\pm$ theories glued together with the addition of a flavor and monopole superpotential. The duality is depicted in Figure \ref{fig:3dbraid_lev2}.

The global symmetry of this theory is
\begin{align}
    S[{\color{red}U(N)} \times {\color{blue}U(N)}] \times U(1)_{\D_L} \times U(1)_{\D_R} \times U(1)_\tau \,.
\end{align}
The partition function identity associated to this duality can be obtained starting from, for example, the third level 1 duality in eq. \eqref{eq:lvl1_3}, performing the following redefinitions:
\begin{align}
    & Z_i \to Z_i + s \qquad \text{for} \quad i=1,\ldots,N \nn \\
    & V \to V + s \quad , \quad 
    \D_L \to \D_L - s \quad , \quad
    \D_R \to \D_R + s
\end{align}
and then performing the limit $s \to +\infty$ (see Appendix \ref{app:3dlim}). We obtain:
\begin{align}\label{eq:lvl2}
    \int & d \vec{Z}_N \D_{U(N)}(\vec{Z},\tau) 
    e^{-i\pi(\D_L+\D_r-iQ)\sum_{j=1}^N Z_j} \times \nn \\
    & \times Z_{FC^-}^{(N)}(\vec{X},\vec{Z},\tau, \D_L) Z_{FC^+}^{(N)}(\vec{Z},\vec{Y},\tau, \D_R) \prod_{j=1}^N s_b \big( \D_L + \D_R \pm Z_j \big) = \nn \\
    & = e^{- i \pi(\D_L + \D_R - \tfrac{iQ}{2})\sum_{j=1}^N (X_j + Y_j) - \tfrac{i \pi N}{2}(2\D_R^2 - 2\D_L^2 + (\D_R - \D_L)(-iQ + (1-N)\tau)) } \times \nn \\
    & \quad \times Z_{FM}^{(N)}(\vec{X},\vec{Y},\tau, \D_L+\D_R) \prod_{j=1}^N s_b \big( \D_R + X_j \big) s_b \big( \D_L - Y_j \big) \,.
\end{align}
Where the set $\vec{X}$ and $\vec{Y}$ are related as $\sum_{j=1}^N (X_j + Y_j) = 0$.

The mapping of some chiral ring generators is:
\begin{align}
    \Pi_L \Pi_R \quad &\longleftrightarrow \quad \Pi \nn \\
    \Pi_L f \quad &\longleftrightarrow \quad l \nn \\
    \Pi_R f \quad &\longleftrightarrow \quad r \nn \\
    f^2 a^{m-1} \mathsf{A}^{n-1} \quad &\longleftrightarrow \quad \mathsf{B}_{n,m} 
\end{align}
In the last line the operator $\mathsf{A}$ is an antisymmetric operator of the two $FC^{\pm}_N$ theories charged under the gauge symmetry. Due to the F-term for $a$ the antisymmetric operator of the left $FC^{\pm}_N$, call it $\mathsf{A}_L$, and that of the right one, call it $\mathsf{A}_R$, are related as $\mathsf{A}_L = \mathsf{A}_R = \mathsf{A}$.
The map for the operator $\tilde{\Pi}$ of the r.h.s.~is more complicated. This operator is mapped to a collection of monopoles with negative charge under strings of consecutive nodes, always involving the central $U(N)$ node. For example for $N=3$, meaning that the two $FC^\pm_3$ theories have two gauge nodes, the map is:
\begin{align}
    \begin{pmatrix}
        \M^{0,0|-1|0,0} & \M^{0,-1|-1|0,0} & \M^{-1,-1|-1|0,0} \\
        \M^{0,0|-1|-1,0} & \M^{0,-1|-1|-1,0} & \M^{0,-1|-1|-1,-1} \\
        \M^{-1,-1|-1|0,0} & \M^{-1,-1|-1|-1,0} & \M^{-1,-1|-1|-1,-1}
    \end{pmatrix} 
    \quad \longleftrightarrow \quad \tilde{\Pi}
\end{align}
Notice that the matrix is constructed using only negatively charged monopoles because the positively charged ones are turned on in the superpotential and hence are not chiral ring generators.

By turning on a maximal nilpotent  VEV for both the adjoint ${\color{red}\mathsf{A}}$ and ${\color{blue}\mathsf{A}}$ breaking  $U(N)\to U(1)$ we obtain the $3d$ $\mathcal{N}=2$ confining duality relating the  $U(N)$ SQCD with 2 flavors   and a positive monopole  superpotentail  to the  WZ-model with 5N singlets (up to flips) discussed in \cite{Benvenuti:2018bav, Amariti:2018wht}.

\subsection{Interesting Level 3 and 4 dualities}

In this section we consider two further interesting ral mass deformations of the star-triangle dualities discussed above.

\subsubsection{Generalized XYZ-web}

Starting from the level 1 and 2 star-triangle duality and performing  further real mass deformations we can obtain the following web of duality: 
\be\label{fig:gen_XYZ}
\resizebox{.95\hsize}{!}{
    \includegraphics[]{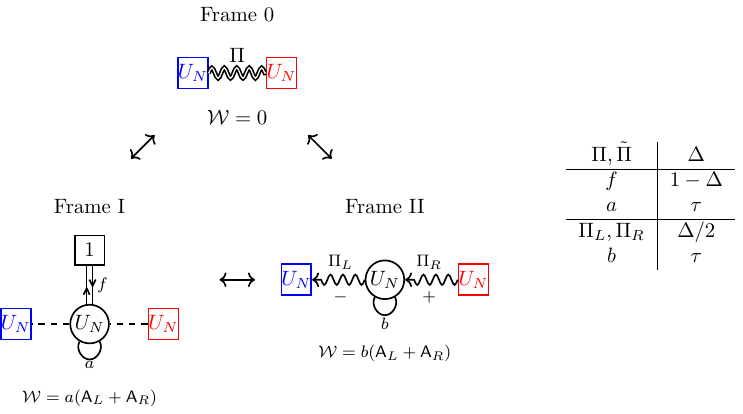}
}
\ee

The duality relating frame 0 and frame I is a deformation of the level 1 star-triangle duality III in Figure \ref{fig:3dbraid_lev1} where we turn on a real mass for $U(1)_{\Delta_L}$ and   $U(1)_{\Delta_R}$ preserving the diagonal. Notice that in this process a $U(1)$ global symmetry decouples because it becomes unfaithful, therefore the rank of the global symmetry collectively decreases by 2.
Frame 0 is just the $FM_N$ and it comes from the r.h.s.~of duality III in Figure \ref{fig:3dbraid_lev1} surviving the real mass which integrates out the two flavors in the triangle.
Frame I comes from the l.h.s.~of duality III in Figure  \ref{fig:3dbraid_lev1} with the real masses having the effect of deforming the $FM_N$ to $FT_N$ as discussed in Appendix \ref{app:3dlim}.
The duality relating frame 0 and frame II is a deformation of the level 2 star triangle
where we turn on a real mass for $U(1) \subset S[{\color{red}U(N)} \times {\color{blue}U(N)}]$.
Frame 0 comes from the r.h.s.~of Figure \ref{fig:3dbraid_lev2}
where the real mass integrates out the chirals in the triangle
while frame II  containing two pairs of $FC^\pm_N$ theories comes from the  l.h.s. of Figure
\ref{fig:3dbraid_lev2} upon integrating out the flavor.

As a partition function identity we have:
\begin{align}
Z_{\text{Frame 0}}(\vec{X},\vec{Y},\tau,\D) 	 
	= Z_{\text{Frame I}}(\vec{X},\vec{Y},\tau,\D) 
	= Z_{\text{Frame II}}(\vec{X},\vec{Y},\tau,\D) \,.
\end{align}
with
\begin{align}\label{xyz0}
Z_{\text{Frame 0}}(\vec{X},\vec{Y},\tau,\D) =	Z_{FM}^{(N)} (\vec{X},\vec{Y},\tau,\D)
\end{align}
\begin{align}\label{xyz1}
	Z_{\text{Frame I}}(\vec{X},\vec{Y},\tau,\D) = \int d\vec{Z}_N & \D_N(\vec{Z},\tau) 
	Z_{FT}^{(N)}(\vec{X},\vec{Z},\tau) Z_{FT}^{(N)}(\vec{Z},-\vec{Y})
	\prod_{j=1}^N s_b( \D \pm Z_j ) \,.
\end{align}
\begin{align}\label{xyz2}
	Z_{\text{Frame II}} (\vec{X}, & \vec{Y},\tau,\D) = e^{ 2i \pi \big[ (\frac{3}{4} \D - \frac{iQ}{4})\sum_{j=1}^N (X_j + Y_j) \big] + \frac{i \pi}{2} \sum_{j=1}^N ( X^2_j - Y^2_j) } \times \nn \\
	& \times \int d\vec{Z}_N \D_N(\vec{Z},\tau) e^{2i \pi (-\frac{3\D}{2} + \frac{iQ}{2}) \sum_{j=1}^N Z_j} 
	Z_{FC_-}(\vec{X},\vec{Z},\tau,\frac{\D}{2}) Z_{FC_+}(\vec{Z},\vec{Y},\tau,\frac{\D}{2}) \,.
\end{align}


The partition function identity between frame 0 and frame I is obtained starting from eq. \eqref{eq:lvl1_3} and performing the following redefinition:
\begin{align}
    & \D_L \to \tfrac{\D}{2} + s \quad , \quad \D_R \to \tfrac{\D}{2} - s 
\end{align}
and then the limit $s \to +\infty$. Similarly the identity between frame 0 and frame II is obtained starting from eq.  \eqref{eq:lvl2} and performing the redefinitions:
\begin{align}
    & X_i \to X_i - \D_L + s \quad , \quad 
    Y_i \to Y_i + \D_R + s \quad , \quad 
    Z_i \to Z_i + s \qquad \text{for} \quad i=1,\ldots,N \nn \\
    & \D_L + \D_R = \tfrac{\D}{2}
\end{align}
and then the limit $s \to +\infty$.

For $N=1$ the duality web in figure \eqref{fig:gen_XYZ} coincides with the XYZ duality web. Indeed for $N=1$ the $FM_N$ theory in frame 0
consists of 3 chirals $\Pi, \tilde \Pi, B_{1,1}$ with a cubic superpotential, the XYZ model. In
frame I, the two $FT_1$ theories are BF couplings and we just have a $U(1)$ theory with one hyper. Similarly, in  frame II the two $FC^\pm_1$ theories reduce to fundamental/antifundamental chirals with background terms, so that this frame also reduces to a $U(1)$ theory with one hyper but with a different parameterization with respect to frame I. Indeed, one may consider a frame III which can be obtained simply starting from frame II and performing the reparameterization: $\vec{X} \to - \vec{X}, \vec{Y} \to -\vec{Y}, \vec{Z} \to -\vec{Z}$. This frame also reduces to a $U(1)$ theory with a single hyper for $N=1$ and it completes the $S_3$ orbit of the XYZ duality.
For this reason we refer to the duality web in figure \eqref{fig:gen_XYZ} as the improved XYZ web.\\

Notice also that turning on the nilpotent deformation in \eqref{fig:FCtochiral} 
for the two adjoint $U(N)$ operators operators $\color{red}\mathsf{A}$ and $\color{blue}\mathsf{A}$ in the $FM_N$ two $FT_N$ or $FC^\pm_N$ theories in \eqref{fig:gen_XYZ} we obtain another known duality \cite{Benvenuti:2018bav}:
\be
    \includegraphics[]{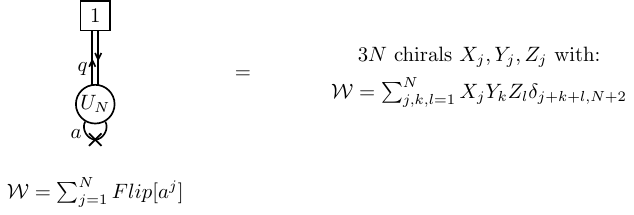}
\ee
Where the $FM_N$ theories have collapsed to an interacting WZ model with 3N fields
while Frame I and II reduce to the $U(N)$ gauge theory with an adjoint chiral and a single flavor.

\paragraph{Operator map}
To explain how the operator map works we list the operators of the $FM_N$ theory in frame 0 and explain how they are mapped in each frame.
\begin{itemize}
	\item  The $\Pi$ operator in the $N \times \bar{N}$ of ${\color{blue}U(N)} \times {\color{red}U(N)}$, with trial R-charge $0$ and $\D$-charge $+1$ is mapped to:
	\begin{itemize}
		\item Frame I: a collection of monopoles with positive topological charge stretching from the central node.
		\item Frame II: a collection of monopoles with negative topological charge stretching from the central node.
	\end{itemize}
	\item The $\tilde{\Pi}$ operator in the $\bar{N} \times N$ of ${\color{blue}U(N)} \times {\color{red}U(N)}$, with trial R-charge $0$ and $\D$-charge $+1$ is mapped to:
	\begin{itemize}
		\item Frame I: a collection of monopoles with negative topological charge stretching from the central node.
		\item Frame II: the mesonic operator obtained joining $\Pi'_R\Pi'_L$.
	\end{itemize}
	\item The adjoint operators $\color{red}\mathsf{A}$  and $\color{blue}\mathsf{A}$ of the $FM_N$ theory in frame zero are simply mapped  to $\color{red}\mathsf{A}$ and $\color{blue}\mathsf{A}$ of the $FT_N/FC^\pm_N$ in frame I, II.
 
	\item The $\mathsf{B}_{n,m}$ matrix with R-charge $2n+(m-n)\tau-2\D$ is mapped to:
	\begin{itemize}
		\item Frame I: dressed mesons obtained from the vertical flavor $f \mathsf{A}^{n-1} a^{m-1} \tilde{f}$, where $\mathsf{A}$ is any of the two adjoint operators of the $FT_N$ theories, that are identified by the $a$ EOM. 
		\item Frame II: to positively charged dressed monopoles $\M_{{a'}^{m-1} {\mathsf{A}'}^{n-1}}^{-(0,\ldots,0|1|0,\ldots,0)}$. The dressing is performed with either $\mathsf{A}' = \mathsf{A}'_L \sim \mathsf{A}'_R$, that are identified via quantum relations.
	\end{itemize}
\end{itemize}
Notice also that if we also consider frame III, as explained below \eqref{xyz2}, the operator map is similar as that for frame II but the $\Pi$ and $\tilde{\Pi}$ operators are exchanged, therefore the $\Pi$ operator maps to a mesonic operator and the $\tilde{\Pi}$ maps to a collection of positively charged monopoles. In addition, the $\mathsf{B}_{n,m}$ matrix is mapped to negatively charged dressed monopoles. \\
Notice that in each frame only one of the three operators $\mathsf{B}_{n,m}, \Pi, \tilde{\Pi}$ in the $FM_N$ theory is mapped to a mesonic operator, while the rest is mapped to a suitable collection of monopoles. This is a feature shared with the 
ordinary $XYZ$ duality  web, where in each frame only one of the three chirals is mapped to the mesonic operator in each dual frame.

The duality relating Frame 0 and Frame I, has been interpreted as the $\mathcal{N}=2$ {\it basic duality move} in \cite{BCP2} encoding the S-dualization of an $\mathcal{N}=2$ flavor to an improved bifundamental.

\subsubsection{Generalized basic chiral mirror duality} \label{sec:chiralbasicdual}

Starting from the generalized XYZ-web we  perform a further real mass deformation
with effect of deforming the $FM_N$ in frame zero to either $FC_N^\pm$
as shown in Appendix \ref{app:3dlim}.

In frame I the effect of the real mass is to integrate out one of two chirals leaving behind one fundamental/antifundamental chiral with half CS unit. So the duality relating frame zero and frame I in the XYZ-web yields the following dualities:
\be\label{fig:chir_basicmove}
    \includegraphics[]{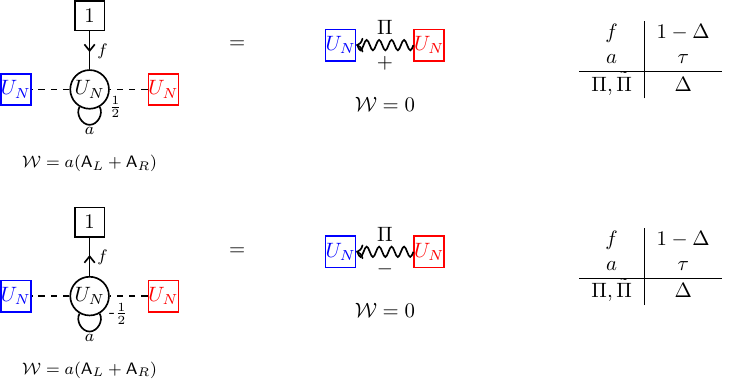}
\ee
The global symmetry of the dualities depicted above is:
\begin{align}
    {\color{red}SU(N)} \times {\color{blue}SU(N)} \times U(1)_\D \times U(1)_\tau
\end{align}
and they correspond to the following partition function identities:
\begin{align}\label{eq:chirbasmove_1}
	& \int d\vec{Z}_N \D_N(\vec{Z},\tau) e^{-i\pi \D \sum_{j=1}^N Z_j + \frac{i\pi}{2}\sum_{j=1}^N Z_j^2} 
	Z_{FT}^{(N)}(\vec{X},\vec{Z},\tau) Z_{FT}^{(N)}(\vec{Z},-\vec{Y}) \prod_{j=1}^N s_b(\D + Z_j) = \nn \\
	& = e^{- \frac{i\pi}{2}\sum_{j=1}^N ( X_j^2 + Y_j^2 ) + i \pi ( \tfrac{iQ}{2} - \D ) \sum_{j=1}^N (X_j - Y_j) + \frac{i \pi N}{12}[12\D^2 + (N^2-1)\tau^2 + 6\D((1-N)\tau - iQ)]} 
    Z_{FC^+}^{(N)}(\vec{X},\vec{Y},\tau, \D ) \,.
\end{align}
\begin{align}\label{eq:chirbasmove_2}
	& \int d\vec{Z}_N \D_N(\vec{Z},\tau) e^{-i\pi \D \sum_{j=1}^N Z_j -\frac{i\pi}{2}\sum_{j=1}^N Z_j^2} 
	Z_{FT}^{(N)}(\vec{X},\vec{Z},\tau) Z_{FT}^{(N)}(\vec{Z},-\vec{Y}) \prod_{j=1}^N s_b(\D - Z_j) = \nn \\
	& e^{\frac{i\pi}{2}\sum_{j=1}^N ( X_j^2 + Y_j^2 ) - i \pi ( \tfrac{iQ}{2} - \D ) \sum_{j=1}^N (X_j - Y_j) - \frac{i \pi N}{12}[12\D^2 + (N^2-1)\tau^2 + 6\D((1-N)\tau - iQ)]} 
    Z_{FC^-}^{(N)}(\vec{X},\vec{Y},\tau,\D) \,.
\end{align}
Where we can then impose that $\sum_{j=1}^N X_j = \sum_{j=1}^N Y_j = 0$, since they are $SU(N)$ parameters.

The first partition function identity can be obtained from \eqref{xyz0} and \eqref{xyz1} with the following reparameterizations:\footnote{One needs also to use the property for the $FT_N$ theory that: 
\begin{align}
    Z_{FT}^{(N)}(\vec{X}+U,\vec{Z},\tau) = e^{2\pi i U \sum_{j=1}^N Z_j} Z_{FT}^{(N)}(\vec{X},\vec{Z},\tau) \,.
    \label{brft}
\end{align} 
}
\begin{align}
    & X_i \to X_i + V \qquad \text{for} \quad i = 1, \ldots, N \nn \\
    & \D \to \D + V 
\end{align}
and then performing $V \to +\infty$. The second identity can be obtained analogously.

Setting $N=1$ in the dualities in \eqref{fig:chir_basicmove}, we can recover the known dualities for a $U(1)_{\pm1/2}$ gauge theory with single fundamental/antifundamental and a single free chiral with background CS interactions. In particular we set $N=1$ and define $X-Y-\D = \eta$ in \eqref{eq:chirbasmove_1} and \eqref{eq:chirbasmove_2} to obtain:
\begin{align}
 	\int dZ e^{2 \pi i (\eta + \frac{\D}{2}) Z + \frac{i\pi}{2} Z^2} s_b(Z + \D) = 
 	e^{ -\frac{i \pi}{2}(\D^2 + \eta^2 + 4\D \eta + iQ\eta )} s_b(\frac{iQ}{2} + \eta)
 	\,, \nn \\
 	\int dZ e^{2 \pi i (\eta + \frac{\D}{2}) Z - \frac{i\pi}{2} Z^2} s_b(- Z + \D) = e^{ \frac{i \pi}{2}( \D^2 + \eta^2 + 4\D \eta + iQ\eta ) } s_b(\frac{iQ}{2} + \eta ) \,.
 \end{align}
Notice that in the above dualities only a combination of $\eta$ and $\D$ is physical. Indeed, we can reabsorbe one of the two parameters via shift of the gauge variable $Z$. Because of this relation, we call the duality in \eqref{fig:chir_basicmove} the \emph{generalized basic chiral mirror duality}.

By turning on the nilpotent deformation in \eqref{fig:FCtochiral} 
for the two adjoint operators $U(M)$ operators $\color{red}\mathsf{A}$ and $\color{blue}\mathsf{A}$ in the $FC^\pm_N$ and in  two $FT_N$ theories in the first duality in \eqref{fig:chir_basicmove} we recover the known duality relating the $U(N)_{1/2}$ gauge theory with one adjoint and one fundamental chiral to  $N$ free chiral multiplets and background terms \cite{Benvenuti:2018bav}:
\be\label{fig:chir_basicmove_N1}
    \includegraphics[]{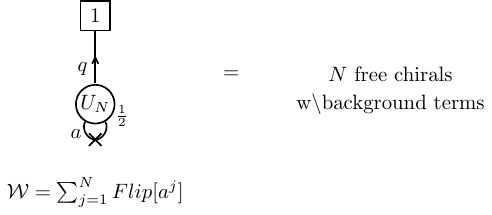}
\ee
Starting from second duality
we obtain a similar duality that has opposite CS level and, on the r.h.s.~, has different background terms.

\paragraph{Operator map}
The operator map in the first duality in \eqref{fig:chir_basicmove} works as follows:
\begin{itemize}
	\item The $\Pi$ operator in the $FC^+_N$ theory on the r.h.s~in the ${\bf \bar{N}} \times {\bf N}$ of ${\color{blue}U(N)} \times {\color{red}U(N)}$ and with R-charge 0, is mapped to a collection of l.h.s.~negatively charged monopoles. For example if $N=3$, recalling that each $FT_3$ theory has 2 gauge nodes, we have:
    \begin{align}
        \begin{pmatrix}
            \M^{0,0|-1|0,0} & \M^{0,0|-1|-1,0} & \M^{0,0|-1|-1,-1} \\
            \M^{0,-1|-1|0,0} & \M^{0,-1|-1|-1,0} & \M^{0,-1|-1|-1,-1} \\
            \M^{-1,-1|-1|-1,-1} & \M^{-1,-1|-1|-1,-1} & \M^{-1,-1|-1|-1,-1} \\
        \end{pmatrix} 
        \qquad \longleftrightarrow \qquad \Pi 
    \end{align}
    Notice that the combination of fundamental chiral and positive CS level imply that only the monopoles arrying negative charge under the topological symmetry of the central node are gauge invariant, while the positively charged ones are not.
 
	\item The adjoint operators $\color{red}\mathsf{A}$  and $\color{blue}\mathsf{A}$ of the $FC^+_N$ theory on the r.h.s are simply mapped to $\color{red}\mathsf{A}$ and $\color{blue}\mathsf{A}$ of the two $FT_N$ theories on the l.h.s.
\end{itemize}
The operator map for the second duality in \eqref{fig:chir_basicmove} is analogous.

\paragraph{Further real mass deformations}
Consider for example the first of the two dualities in figure \eqref{fig:chir_basicmove}. We can perform a real mass deformation for the flavor $f$. This real mass can be taken to be positive or negative, leading to two different results.

If we take a negative real mass (we take $\D\to \D +s$ and $s \to + \infty$) the effect is to integrate out the flavor increasing the CS level of the gauge node to be $+1$, while on the r.h.s.~this has the effect of taking a limit where the $FC^+_N$ theory is reduced to a $FT_N$. We land on the following duality:
\begin{align}
    \includegraphics[]{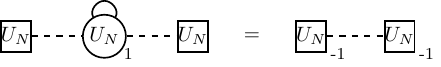}
\end{align}
which in \cite{Comi:2022aqo} was interpreted as the $STS=(TST)^{-1}$ relation of $SL(2,\mathbb{Z})$.

If we instead take a positive real mass (we take $\D\to \D -s$ and $\vec X \to \vec X - s$ with $s\to +\infty$)
the effect is to integrate out the flavor reducing the CS level of the gauge node, therefore leaving two $FT_N$ theories glued with zero superpotential and zero CS level, which then fuse to an Identity wall as in subsection \ref{sec:idwalls}.\footnote{
We need to use twice the property \eqref{brft}. Once with $U = -s$ to cancel the divergent FI produced by the real mass deformation and the second time with $U=\D$ to produce: $_{\vec{X}+\D} \mathbb{I}^{(U)}_{\vec{Y}} (\tau) $.
We also use that $e^{ \pm i \pi U (\sum_{j=1}^N Y_j - \tfrac{1-N}{2}\tau)} Z_{FC^\pm }^{(N)} (\vec{X}, \vec{Y}, \tau ,\D ) =Z_{FC^\pm }^{(N)} (\vec{X}+U, \vec{Y}, \tau ,\D+U) $, which can be deduced from the expression for the partition function in \eqref{eq:FC+_parfun} and \eqref{eq:FC-_parfun} . }
The   identity expressing this propoerty is:
\begin{align}
    _{\vec{X}+\D} \mathbb{I}^{(U)}_{\vec{Y}} (\tau) 
    e^{ \pm \frac{i \pi}{2} \big[ 
    \sum_{j=1}^N (X_j^2 + Y_j^2) 
    + \tfrac{1}{2}N \D^2 
    + N (N-1) \tau \D
    - \tfrac{N(N^2-1)}{6} \tau^2
    \big] } = \nn \\
    = \lim_{s \to +\infty} e^{ 2\pi i s \big( \sum_{j=1}^N (X_j - Y_j) - N\D \big) } Z_{FC^\pm }^{(N)} (\vec{X}, \vec{Y}, \tau ,\D ) 
\end{align}
This is the result anticipated at the end of Section \ref{sec:FC}.

\acknowledgments
We are grateful to Anant Shri, Gabriel Pedde Ungureanu and Simone Rota for comments on the draft and collaboration on related topics.
SP and SB are partially supported by the MUR-PRIN grant No. 2022NY2MXY (Finanziato dall’Unione europea- Next Generation EU, Missione 4 Componente 1 CUP H53D23001080006).

\appendix

\section{Notations for $4d$ superconformal index and $3d$ partition function}\label{app:conventions}

\subsection*{$4d$ superconformal index} 
Let us consider a $4d$ $\CN=1$ gauge theory with gauge group $G$ and matter given by a set of $\CN=1$ chiral multiplets of R-charge $r$, in the representation $R_G$ of $G$ and $R_F$ of some flavor symmetry group $F$. To write the SCI we turn on a set of $(\dim G)$ fugacities $\vec{z}$ for the gauge group $G$ and $(\dim F)$ fugacities $\vec{x}$ for the flavor symmetry $F$. We then write:
\begin{align}
	\CI_G(\vec{x}) = \frac{1}{|W_G|} \oint \prod_{j=1}^{\dim G} \frac{dz_j}{2 \pi i z_j} 
	\frac{[(p;p)_\infty (q;q)_\infty]^{\dim G} }{ \prod_{\vec{\r} \in G} \Ge(\vec{z}^{\, \vec{\r}} \, ) } 
	\prod_{\s_G \in R_G} \prod_{\s_F \in R_F} \Ge \big( (pq)^{r/2} \vec{z}^{\, \vec{s}_G} \vec{x}^{\, \vec{s}_F} \big) \,.
\end{align}
Where $\vec{\r}$ are the roots of $G$, $\vec{\s}_G$ and $\vec{s}_F$ are the weights of the representations $R_G$ and $R_F$. $|W_G|$ is the dimension of the Weyl group of $G$.  We adopted the following notation:
\begin{align}
	\vec{z}^{\, \vec{\r}} = \prod_{j=1}^{\dim G} z_j^{\r_j} \qquad , \qquad
	\vec{z}^{\, \vec{\s}_G} = \prod_{j=1}^{\dim G} z_j^{{\s_G}_j} \qquad , \qquad
	\vec{x}^{\, \vec{\s}_F} = \prod_{j=1}^{\dim F} z_j^{{\s_F}_j} \,.
\end{align}
We define a short notation for the integration measure:
\begin{align}
	d\vec{z}_N = \prod_{j=1}^{N} \frac{dz_j}{2 \pi i z_j} \,.
\end{align}
In this work we deal mostly with $USp$ gauge groups for which we define the contribution of the vector multiplet as:
\begin{align}
	\D_N(\vec{z}) = \frac{1}{2^N N!} \frac{[(p;p)_\infty (q;q)_\infty]^{N} }{ \prod_{j=1}^N \Ge(z_j^{\pm2}) \prod_{j<k}^N \Ge(z_j^\pm z_k^\pm) } \,.
\end{align}
It is convenient to also define the contribution of both a vector and a chiral in the traceless antisymmetric representation:
\begin{align}
	\D_N(\vec{z},t) = \D_N(\vec{z}) \Ge(t)^{N-1} \prod_{j<k}^N \Ge(t z_j^\pm z_k^\pm ) \,.
	\label{adjme}
\end{align}
For a chiral of R-charge $r$ in the bifundamental of $USp(2N)\times USp(2M)$ we have:
\begin{align}
	\CI_{bif} = \prod_{j=1}^N \prod_{a=1}^M \Ge \big( (pq)^{r/2} z_j^\pm x_a^\pm \big) \,.
\end{align}
Suppose that a theory also possesses a $U(1)$ symmetry for which we turn on a fugacity $c$. Along the RG flow this symmetry can mix with the R-symmetry as $r + q_c C$, where $q_C$ is the $U(1)$ charge and $C$ is the mixing coefficient, which is related to the fugacity as:
\begin{align}
	c = (pq)^{C/2} \,.
\end{align}

\subsection*{$3d$ partition function}
Let us consider a $3d$ $\CN=2$ gauge theory with gauge group $G$ and matter given by a set of $\CN=2$ chiral multiplets of R-charge $r$, in the representation $R_G$ of $G$ and $R_F$ of some flavor symmetry group $F$. To write the $S^3_b$ partition function we turn of a set of $(\dim G)$ parameters $\vec{Z}$ for the gauge group $G$ and $(\dim F)$ parameters $\vec{X}$ for the flavor symmetry $F$. We then write:
\begin{align}
	Z (\vec{X}) = \frac{1}{|W_G|} \int \prod_{j=1}^{\dim G} d Z_j  & Z^{G}_{\text{classical}} 
	\frac{1}{\prod_{\vec{\r} \in G} s_b \big( \frac{iQ}{2} - \vec{\r}(\vec{Z}) \big) } \nn \\
	& \prod_{\vec{\s}_G \in R_G} \prod_{\vec{\s}_F \in R_F} s_b \big( \frac{iQ}{2}(1-r) - \vec{\s}_G(\vec{Z}) - \vec{\s}_F(\vec{X}) \big) \,.
\end{align}
Where $\vec{\r}$ are the roots of $G$, $\vec{\s}_G$ and $\vec{s}_F$ are the weights of the representations $R_G$ and $R_F$. $|W_G|$ is the dimension of the Weyl group of $G$. We also adopted the following notation:
\begin{align}
	\vec{\r}(\vec{Z}) = \sum_{j=1}^{\dim G} \r_j z_j \qquad , \qquad
	\vec{\s}_G(\vec{z}) = \sum_{j=1}^{\dim G} {\s_G}_j z_j \qquad , \qquad
	\vec{\s}_F(\vec{x}) = \sum_{j=1}^{\dim F} {\s_F}_j z_j \,.
\end{align}
We define a short notation for the integration measure:
\begin{align}
	d\vec{Z}_N = \prod_{j=1}^{N} dZ_j \,,
\end{align}
In this work we deal with both $U(N)$ and $USp(2N)$ groups. We define the contribution of a vector multiplet as:
\begin{align}
	& \D_{U(N)}(\vec{Z}) = \frac{1}{N!} \frac{1}{\prod_{j<k}^N s_b \big( \frac{iQ}{2} \pm (Z_j - Z_k) \big) } \nn \\
    & \D_{USp(2N)}(\vec{Z}) = \frac{1}{N!2^N}\frac{1}{\prod_{j=1}^N s_b \big( \frac{iQ}{2} \pm 2 Z_j \big) \prod_{j<k}^N s_b \big( \frac{iQ}{2} \pm Z_j \pm Z_k) }
\end{align}
We then define the contribution of both a vector and a traceless adjoint chiral of $U(N)$ and similarly that of a vector and a traceless antisymmetric chiral for $USp(2N)$
\begin{align}
	& \D_{U(N)}(\vec{Z},\tau) = \D_{U(N)}(\vec{Z}) s_b \big( \frac{iQ}{2} - \tau \big)^{N-1}\prod_{j<k}^N s_b \big( \frac{iQ}{2} - \tau \pm (Z_j - Z_k) \big) \nn \\
    & \D_{USp(2N)}(\vec{Z},\tau) = \D_{USp(2N)}(\vec{Z}) \big( \frac{iQ}{2} - \tau \big)^{N-1} \prod_{j<k=1}^N s_b \big( \frac{iQ}{2} - \tau \pm Z_j \pm Z_k \big)
\end{align}

There is also a classical contribution to the partition function which depends on the given group. For $U(N)$:
\begin{align}
	Z^{U(N)}_{\text{classical}}(Y,k) = \exp \left[ 2\pi i Y \sum_{j=1}^{N} Z_j + \pi i k \sum_{j=1}^{N} Z_j^2 \right] \,.
\end{align}
where $Y$ is the FI parameter for the $U(1)$ topological symmetry associated to the $U(1) \in U(N)$ factor. and $k$ is the Chern-Simons level.\footnote{The definition assumes that we have the same CS level $k$ for both the $U(1)$ and $SU(N)$ factor composing $U(N) = U(1) \times SU(N)$. In this work we are not interested in generalizations thereof.} For $USp(2N)$ we instead have:
\begin{align}
    Z^{USp(2N)}_{\text{classical}}(k) = \exp \left[ \pi i k \sum_{j=1}^N Z_j^2 \right] \,.    
\end{align}

Whenever a theory posses a $U(1)$ symmetry we turn on a mass parameter $C$. This symmetry can mix with the R-symmetry as $r + q_c \mathfrak{c}$, where $q_C$ is the $U(1)$ charge and $\mathfrak{c}$ is the real valued mixing coefficient. which is related to the mass parameter as:
\begin{align}
	C = \frac{iQ}{2} \mathfrak{c} \,.
\end{align}

\section{Comments on relations in the $FE_N$ theory}\label{app:FE_relations}
In this section we discuss the relations in the $FE_N$ theory analyzed using the expansion of the 4d superconformal index. \\
We limit ourselves to the detailed analysis of $N=2$ case due to the increasing numerical complexity of the problem for higher rank. The index of the theory can be written using the recursive relation \ref{eq:FE_SCI}:
\begin{align}\label{eq:ind_feusp4}
	\CI(\vec{x},\vec{y},c,t) = \Ge( pq t c^{-2} ) \Ge( pq c^{-2} ) \Ge(pqt^{-1}) \Ge(pqt^{-1} x_1^\pm x_2^\pm ) \prod_{j=1}^2 \Ge( c y_2^\pm x_j^\pm ) \times \nn \\
	\times \oint dz_1 \Ge(pqt^{-1}) \prod_{j=1} \Ge( t^{1/2} z^\pm x_j^\pm ) \Ge( pq t^{-1/2} c^{-1} z^\pm y_2^\pm ) \Ge( t^{-1/2}c^{-1} z^\pm y_1^\pm )
\end{align}
Where $\vec{x},\vec{y}$ are the fugacities for the manifest and emergent flavor $USp(4)$ symmetries, $z$ is the fugacities for the $USp(2)$ gauge group.
The gauge invariant operators of this theory are:
\begin{itemize}
	\item $\Pi$ is a bifundamental ${\bf 4}_x \times {\bf 4}_y$ of $USp(4)_x \times USp(4)_y$.
	\item $\mathsf{A}_x$ is an antisymmetric ${\bf 5}_x$ of $USp(4)_x$
	\item $\mathsf{A}_y$ is an antisymmetric ${\bf 5}_y$ of $USp(4)_y$
	\item Three singlets $\mathsf{B}_{1,1}$, $\mathsf{B}_{1,2}$, $\mathsf{B}_{2,1}$. 
\end{itemize}
We labelled representations by giving the dimension and we distinguish between representations of $USp(4)_x$ and $USp(4)_y$ with a subscript. 
Before getting into the analysis let us remark one last thing, due to the self-mirror property of the $FE_N$ theory we expect that all the representations that contribute to the index appear symmetrically under the exchange of the $USp(4)_x$ and $USp(4)_y$ symmetry. This observation alone, is useful to justify the ruling out of many contributions. However, we show that some contributions that do not violate this symmetry are still absent from the index. \\\
We start by looking at the relations for the operator $\Pi^2$ (without taking any trace over flavor groups). Generally speaking, this operator should consist in $16 \times 17 / 2 = 136$ states, however from the index we see only $125$ of them contributing. These $125$ states decompose into irreducible representations of $USp(4)_x \times USp(4)_y$ as:
\begin{align}
	125 \to {\bf 10}_x \times {\bf 10}_y \oplus {\bf 5}_x \times {\bf 5}_y
\end{align}
Among the many representations that are absent, the most notable one is the singlet that represent the operator $\Tr_x \Tr_y (\Pi^2)$. Now suppose to remove from the index the contribution of the $\mathsf{B}_{1,1}$ singlet, from the index we now see an extra contribution for the $\Pi^2$  that we interpret as a restored $\Tr_x \Tr_y (\Pi^2)$ operator. \\
Let us now move to the relations for the $\Pi$ operator multiplied by any of the two antisymmetric operators $\mathsf{A}_x$ and $\mathsf{A}_y$. Notice that mirror duality, which is an exact self-duality for the $FE_N$ theory, acts swapping the two operators, therefore it is sufficient to study just one of the two cases. Let us consider for simplicity the operator $\Pi \mathsf{A}_x$ (without taking any trace over flavor groups), which generically consist in $16 \times 5 = 80$ states that should decompose into two irreducible representations of $USp(4)_x \times USp(4)_y$: ${\bf 16}_x \times {\bf 4}_y$ and ${\bf 4}_x \times {\bf 4}_y$. However, from the index we see that these 80 states are projected only onto the 64 given by the former representation. Then the bifundamental contribution ${\bf 4}_x \times {\bf 4}_y$ is set to zero, notice that the absence of this operator implies that in dualities like the braid in \eqref{fig:braid} there are no operators that are improved dressed mesonic operators like, for example, $\Pi_L a f$. \\

Finally the index analysis suggest the existence of the relation  $\Tr(\mathsf{A}_x^2) = \Tr( \mathsf{A}_y^2)$. Indeed by looking at  the contributions coming from $\mathsf{A}_{x/y}^2$ and $\mathsf{A}_x \mathsf{A}_y$ (without taking any trace over flavor groups)
we see find ${\bf 14}_x$, ${\bf 14}_y$, ${\bf 5}_x \times {\bf 5}_y$ and one singlet. We interpret the presence of only one singlet as the contribution of $\Tr(\mathsf{A}_x^2) = \Tr( \mathsf{A}_y^2)$. \\

As already mentioned, it is hard to perform this same analysis for the $FE_N$ theory with rank higher than 2. Still, we propose that the relations found for the $FE_4$ theory generalise to any rank as:
\begin{itemize}
    \item The operator obtained by squaring $\Pi$ generically decomposes into representations of $USp(2N)_x \times USp(2N)_y$ as:
    \begin{align}
        (\text{symm},\text{symm}) \oplus (\text{asymm},\text{asymm}) \oplus (\text{asymm},1) \oplus (1,\text{asymm}) \oplus (1,1) \,.
    \end{align}
    The only representations that are not set to zero in the chiral ring are $(\text{symm},\text{symm})$ and $(\text{asymm},\text{asymm})$. The singlet obtained as $\Tr_x \Tr_y (\Pi^2)$ is set to zero in the chiral ring and can be restored by flipping the operator $B_{1,1}$.
    
	\item For the operator $\Pi \,\mathsf{A}^j$, where $\mathsf{A}$ can be any of the two antisymmetric operators and $j=1,\ldots,N-1$ we claim that for any $j$  the contribution of the (Fund,Fund) representation  is  set to zero and the only non-zero irreducible representation is (Symm,Fund), when considering $\mathsf{A}_x$ or (Fund,Symm) for $\mathsf{A}_y$.
 
	\item The traces of the two antisymmetric operators are identified as: $\Tr (\mathsf{A}_x^j) = \Tr (\mathsf{A}_y^j)$, for any $j=2,\ldots,N$, recalling that classically $\Tr (\mathsf{A}_x) = \Tr (\mathsf{A}_y) = 0$ since $\mathsf{A}_{x/y}$ are both traceless.
\end{itemize}
Although we do not carry out a detailed analysis for $N > 2$, we observe that some of the claimed results can be derived in general from F-term relations in the UV completion of the $FE_N$ theory in Figure \ref{fig:FE_quiver}. For example, let us focus on the representation (asymm,1) contained in the $\Pi^2$ composite operator, which we claimed to be zero in the first bullet point of the above list. This operator is constructed by taking $\Tr_y (\Pi \Omega \Pi)$, where $\Omega$ is the antisymmetric tensor of $USp(2N)_y$. From the UV completion in Figure \ref{fig:FE_quiver}, a generic component of $\Pi$ is of the form $d_j b_j b_{j+1} \ldots b_{N-1}$, thus a component of $\Tr_y (\Pi \Omega \Pi)$ is of the form $d_j b_j b_{j+1} \ldots b_{N-1} \dot d_N$, which is set to zero by the F-term of the $v_{N-1}$ field: $b_{N-1} d_N = 0$.

\section{Self dualities of $USp(2N)$ with antisymmetric and $8$ fundamentals}\label{app:USp2Nw/8}
In this appendix we review the dualities discussed in \cite{Cs_ki_1997,Spiridonov_2010} for the $\mathcal{N}=1$ $USp(2N)$ SQCD with an antisymmetric field and 8 chiral multiplets, with zero superpotential, which is given by the quiver:
\be\label{fig:USp(2N)w8}
    \includegraphics[]{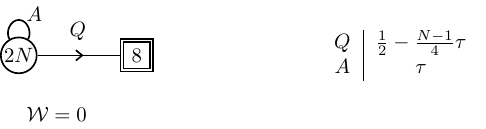}
\ee
In the pictures in the following section we will distinguish between $USp(2N)$ and $SU(N)$ nodes using double lines for the latter. Also an arrow tip pointing towards an $SU(N)$ node means that the field is in the fundamental representation of that group, viceversa for the antifundemental representation which is labeled by an arrow pointing outwards. \\
The gauge invariant operators of this theory are the dressed mesons given as:
\begin{align*}
	M_{ij}^{(k)} = Q_i A^{(k)} Q_j \qquad \text{for} \qquad k=0,\cdots,N-1 \,, \qquad \text{R-charge:} \; 1 + \frac{2j - N + 1}{2}
\end{align*}
where we have implied the presence of a trace over the $USp(2N)$ indexes, also $i,j$ are $SU(8)$ fundamental indexes. The presence of the $USp(2N)$ trace has the effect of antisymemtrizing the $i,j$ indexes, meaning that for each $k$ we have an operator in the antisymmetric representation of $SU(8)$ which has dimension 28. 

\paragraph{CSST-like frame} 
The  CSST-like dual frame \cite{Spiridonov_2010}, is  an $USp(2N)$ theory with an antisymmetric field and 8 chirals, with the addition of $8N$ singlets. The theory is given as:
\be\label{fig:USp(2N)w8_CSST}
    \includegraphics[]{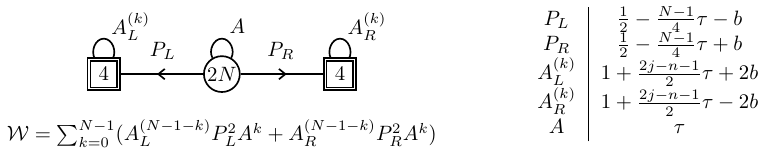}
\ee
The manifest flavor symmetry of this theory is $SU(4)^2 \times U(1)_b \times U(1)_\tau$, but it enhances in the IR as:
\begin{align*}
	SU(4)^2 \times U(1)_b \times U(1)_\tau \longrightarrow SU(8) \times U(1)_\tau \,.
\end{align*}
Notice that there are as many CSST-like dual frames as the number of ways to decompose $SU(8)$ as $SU(4)\times U(1)$, meaning that there is a total of 35 different CSST-like duals.\\
The gauge invariant operators of the theory are: the two towers of $4N$ singlets $A_L^{(k)}, A_R^{(k)}$, dressed mesonic operators given as $(P_LA^kP_R)_{ij}$ for $k=0,\cdots,N-1$. These operators can be recollected into a tower of $N$ operators that transform in the antisymmetric representation of the emergent $SU(8)$ via the branching rule:
\begin{align}\label{eq:BRsu(8)}
	\textbf{28} \rightarrow (\textbf{6},\textbf{1})_2 \oplus (\textbf{4},\textbf{4})_0 \oplus (\textbf{1},\textbf{6})_{-2} \,.
\end{align}

\paragraph{Seiberg-like}
The the Seiberg-like dual frame \cite{Spiridonov_2010}, is  an $USp(2N)$ theory with an antisymmetric field and 8 chirals, with the addition of $12N$ singlets. The theory is given as:
\be\label{fig:USp(2N)w8_Seiberg}
    \includegraphics[]{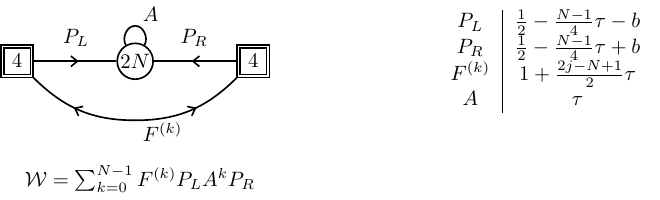}
\ee
As in the CSST-like dual, the manifest flavor symmetry is $SU(4)^2 \times U(1)_b \times U(1)_\tau$ which enhances to $SU(8) \times U(1)_\tau$ in the IR. Again, there are as many dual frames as the number of ways to embed $SU(4)^2 \times U(1)$ in $SU(8)$, for a total of 35 frames. \\
The gauge invariant operators of this theory are: two towers of $N$ dressed mesons $P_L^2 A^{(k)}$ and $P_R^2 A^{(k)}$, the $16N$ singlets $F^{(k)}$. These operators can be recollected into $N$ antisymmetric representations of $SU(8)$ using again the branching rule \ref{eq:BRsu(8)}.

\paragraph{IP-like} The IP-like dual frame  \cite{Cs_ki_1997}, is given by an $USp(2N)$ theory with an antisymmetric field and 8 chirals, with the addition of $28N$ singlets. The theory is given as:
\be\label{fig:USp(2N)w8_IP}
    \includegraphics[]{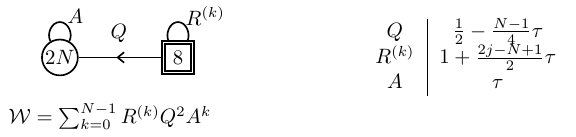}
\ee
In this frame the flavor $SU(8)$ symmetry is manifest as in the original theory \ref{fig:USp(2N)w8}, therefore there exists only one IP-like dual frame. The gauge invariant operators of this theory are simply given by the tower of $N$ singlets $R^{(k)}$ that are in the antisymmetric representation of $SU(8)$.

\paragraph{$E_7$ orbit}
The theory \ref{fig:USp(2N)w8} has a total of 72 different duality frames that are given as:
\begin{align*}
	1 \qquad & \text{original theory \ref{fig:USp(2N)w8}} \\
	35 \qquad & \text{CSS-like duals \ref{fig:USp(2N)w8_CSST}} \\
	35 \qquad & \text{Seiberg-like duals \ref{fig:USp(2N)w8_Seiberg}} \\
	1 \qquad & \text{IP-like dual \ref{fig:USp(2N)w8_IP}} 
\end{align*}
Moreover, we can generate the whole orbit of 72 dualities from the sole CSST-like duality. In fact, the Seiberg-like dual can be obtained by iterating two times the CSST-like duality, with a different choice of splitting $SU(8)$ into $SU(4)^2 \times U(1)$. Notice that when we iterate the CSST-like duality the second time $4N$ out of the $12N$ singlets present in the CSST-like frame becomes massive together with $4N$ of the $12N$ singlets added by the second iteration of the CSST-like duality, meaning that we are left with only $16N$ singlets that are precisely those in the Seiberg-like dual frame. Lastly, the IP-like dual can be reach by a third iteration of the CSST-like duality, starting from the Seiberg-like frame. Again, in order to properly reach the IP-like dual we need to choose properly the embedding $SU(4)^2 \times U(1) \in SU(8)$. The third iteration of the CSST-like duality adds $12N$ singlets, meaning that we are finally left with a total of $28N$ singlets, that are those present in the IP-like frame. Indeed, after this third iteration, the $SU(8)$ symmetry is manifest again. \\
We resume this discussion in the picture below.
\be\label{fig:USp(2N)w8_orbit}
\begin{tikzpicture}[thick,node distance=3cm,gauge/.style={circle,draw,minimum size=5mm},flavor/.style={rectangle,draw,minimum size=5mm},flavorSU/.style={rectangle,draw,double,minimum size=5mm}]  
    
	\path (0,0) node[gauge] (g) {$\!\!\!2N\!\!\!$} -- (2,0) node[flavorSU] (x) {$\!8\!$};
	\chir (g) -- (x);
	\draw[-] (g) to[out=60,in=0] (0,0.6) to[out=180,in=120] (g);
    
    \draw (0,-1.5) node[align=center]{Original theory \ref{fig:USp(2N)w8}};
    
    \draw [-to] (3,0) -- (5,0);
    \draw (4,0.4) node{CSST-like};
    
    \path (6,0) node[flavorSU] (x1) {$\!4\!$} -- (8,0) node[gauge] (g) {$\!\!\!2N\!\!\!$} -- (10,0) node[flavorSU] (x2) {$\!4\!$};
	\chir (g) -- (x1);
	\chir (g) -- (x2);
	\draw[-] (g) to[out=60,in=0] (8,0.6) to[out=180,in=120] (g);
	\draw[-] (x1) to[out=60,in=0] (6,0.6) to[out=180,in=120] (x1);
	\draw[-] (x2) to[out=60,in=0] (10,0.6) to[out=180,in=120] (x2);
	
	\draw (8,-1.5) node[align=center]{CSST-like dual \ref{fig:USp(2N)w8_CSST} \\
    $12N$ singlets, 35 dual frames};
	
	\draw [-to] (8,-2.5) -- (8,-4);
    \draw (8,-3.25) node[right]{$\text{CSST-like}^2$};
    				
	\path (0,-5) node[gauge] (g) {$\!\!\!2N\!\!\!$} -- (2,-5) node[flavorSU] (x) {$\!8\!$};
	\chir (x) -- (g);
	\draw[-] (g) to[out=60,in=0] (0,-4.4) to[out=180,in=120] (g);
	\draw[-] (x) to[out=60,in=0] (2,-4.4) to[out=180,in=120] (x);
	
	\draw (0,-7) node[align=center] { IP-like dual \ref{fig:USp(2N)w8_IP} \\ $28N$ singlets, 1 dual frame};
	
	\draw [-to] (5,-5) -- (3,-5);
    \draw (4,-4.6) node{$\text{CSST-like}^3$};
	
	\path (6,-5) node[flavorSU] (x1) {$\!4\!$} -- (8,-5) node[gauge] (g) {$\!\!\!2N\!\!\!$} -- (10,-5) node[flavorSU] (x2) {$\!4\!$};
	\chir (x1) -- (g);
	\chir (x2) -- (g);
	\draw[-] (g) to[out=60,in=0] (8,-4.4) to[out=180,in=120] (g);
	\chir (8,-6) to[out=180,in=-45] (x1);
	\chir (8,-6) to[out=0,in=-135] (x2);
	
	\draw (8,-7) node[align=center] { Seiberg-like dual \ref{fig:USp(2N)w8_Seiberg} \\
    $16N$ singlets, 35 dual frames };
	
	\draw [-to] (0,-4) -- (0,-2.5);
    \draw (0,-3.25) node[left]{$\text{IP-like}$};
  
\end{tikzpicture}
\ee
Starting from these self-dualities, we can give a mass to one of the flavors of theory \ref{fig:USp(2N)w8}, which is mapped to a VEV in any dual frame, to land on the confining duality for $USp(2N)$ with $6$ chirals and an antisymmetric field:
\be\label{fig:USp(2N)w6}
    \includegraphics[]{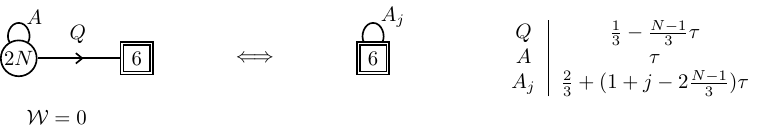}
\ee

\section{Deformed dualities}\label{app:aux_dualities}  
In this appendix we list the deformed dualities used during the proof of Braid duality. 

In step 3 we used a deformed version of the improved Seiberg duality in \ref{fig:wigCSST}. We deform the CSST-like duality with a linear superpotential term for the $(\mathsf{B}_{2,1})_L$ singlet on l.h.s., which is mapped in the dual frame to a linear term for $(\mathsf{B}'_{2,1})_R$. The effect of these deformations is to iron a $FE_N$ theory into a standard bifundamental. To reach the exact duality that we use in the proof we also have to reparameterize the $U(1)_\tau$ symmetry by performing the redefinition $\tau \to 1-\tau$. We thus obtain:
\be\label{fig:proof_defseib} 
    \includegraphics[]{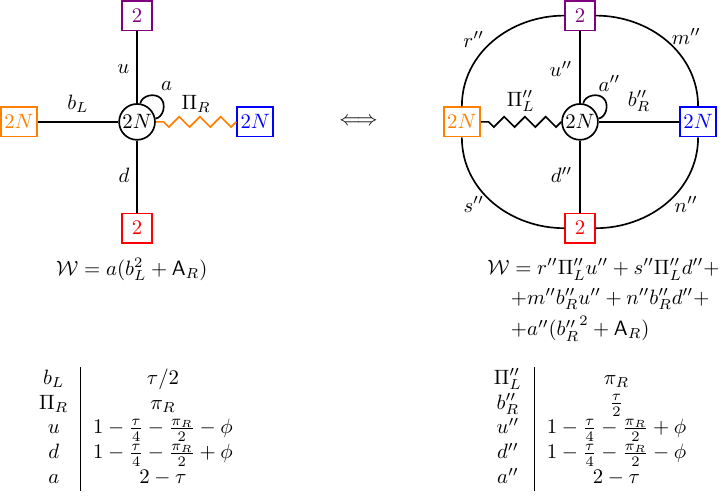}
\ee

In the forth step we used a deformed version of braid duality. We start from the duality in \ref{fig:braid} and on l.h.s. we add linearly the singlet $(\mathsf{B}_{1,2})_R$ to the superpotential, which has the effect of ironing the right $FE$ theory to a standard bifundamental coupled to two antisymmetric fields. On r.h.s. this deformation is mapped to the superpotential $\d \CW = l^2\, {\color{red}\mathsf{A}}$. FInally, we also perform the redefinition $\tau \to 2-\tau$ which transforms the standard $FE_N$ theories into orange ones. The resulting duality is: 
\be\label{fig:proof_defbraid}
    \includegraphics[]{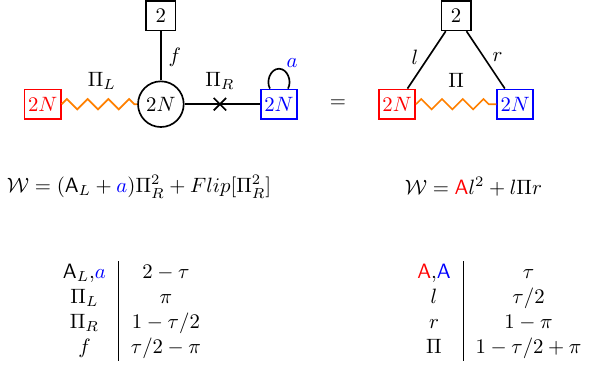}
\ee
      

\section{$3d$ limit and real mass deformation of the $FE_N$ theory}\label{app:3dlim}
In this appendix we describe in more detail the $3d$ reduction limit of the $4d$ $FE_N$ theory and real mass deformations thereof. We will describe \emph{some} of the possible real mass deformations connecting the improved bifundamentals while we do not attempt to provide \emph{all} of them. Although, all the computations presented in this paper can be done using only the results provided in this appendix. 

Some of the results in this Appendix were already carried out in previous works. In particular the $4d$ to $3d$ limit, from $FE_N$ to $FE^{3d}_N$, and the following real mass deformation from $FE^{3d}_N$ to $FM_N$ were first performed in \cite{Pasquetti:2019hxf}, the limit of the $FM_N$ theory to the $FT_N$ was instead performed in \cite{Pasquetti:2019tix}.

\paragraph{From \boldmath{$FE_N$} to \boldmath{$FE^{3d}_N$}} Let us start from the circle compactification. We start from the $4d$ superconformal index, which is the partition function on $S^3 \times S^1$, of the $FE_N$ theory given in \eqref{eq:FE_SCI} and proceed along the lines of \cite{Aharony:2013dha}. We define:
\begin{align}
	& x_i = e^{2\pi i r X_i} , \qquad y_i = e^{2\pi i r Y_i}, \qquad z^{(a)}_i = e^{2\pi i r Z^{(a)}_i} \, \nn \\
	& t = e^{2\pi i r \tau} , \qquad c = e^{2\pi i r \D} \,,
\end{align}
where $r$ is the $S^1$ radius. The variables $X_i,Y_i,Z^{(a)}_i,\tau,\D$ take values in the interval $[-\frac{1}{2r}, \frac{1}{2r}]$. We then perform the limit $r \to 0$ using the formula relating an elliptic-gamma function to a double-sine function:
\begin{align}\label{eq:Gamma_limr}
	\lim_{r\to 0} \Ge(x = e^{2\pi i r X}; p = e^{-2 \pi r b} , q = e^{2 \pi r b^{-1}}) = e^{-\frac{i \pi}{6r}(i\frac{Q}{2}-X)} s_b(i\frac{Q}{2}-X) \,.
\end{align}
Notice that after the limit the variables are real-valued. Using this formula in the index \eqref{eq:FE_SCI} gives the following result:
\begin{align}
	\lim_{r \to 0} I_{FE}^{(N)} (\vec{x},\vec{y},t,c) = C_N(\D,\tau,r) Z_{FE^{3d}}^{(N)} (\vec{X},\vec{Y},\tau,\D) \,,
\end{align}
where $Z_{FE^{3d}}^{(N)}$ is defined in \eqref{eq:FE3d_parfun} and $C_N(\D,\tau,r)$ is a prefactor which is divergent in the limit $r \to 0$. It is given as:
\begin{align}\label{eq:FE_div_r}
	C_N(\D,\tau,r) = \exp \left[ \frac{i N \pi}{12 r} \big( 4\D + (1+2N)(-iQ + 2(N-1)\tau \big) \right] \,.
\end{align}

\paragraph{From \boldmath{$FE_N^{3d}$} to \boldmath{$FM_N$}}
The $FM_N$ theory can be reduced to the $FM_N$ theory via a real mass deformation. Following  \cite{Aharony:2013dha}, we implement this deformation by taking the limit $x \to \pm \infty$ in a partition function, where $x$ is any real mass parameter. The effect of this limit is implemented by the asymptotic behavior of the double-sine function:
\begin{align}\label{eq:sblim}
    \lim_{x \to \pm \infty} s_b( x ) = e^{\pm \frac{i \pi}{2} x^2} \,.
\end{align}
In the specific case of the limit from the $FE^3d_N$ to the $FM_N$ we start from the explicit expression for the partition function of the $FE^{3d}_N$ theory written in eq.~\eqref{eq:FE3d_parfun} and perform the following redefinition:
\begin{align}\label{eq:FEtoFM_shift}
    & X_j \to X_j + s \quad , \quad
    Y_j \to Y_j + s \quad , \quad
    Z_j^{(a)} \to Z_j^{(a)} + s \quad \text{for} \quad j = 1,\ldots,N 
\end{align}
where $\vec{Z}^{(a)}$ is the set of parameters associated to the $USp(2a)$ gauge symmetry, $a = 1,\ldots, N-1$. We then perform the limit $s \to + \infty$. Notice that under this shift only the
chirals depending on the 
differences $X_j - Y_k$, $X_j - Z_k$, $Y_j - Z_k$ do not depend on $s$ and therefore are left massless, while the sums are shifted and thus they become massive. As a consequence, this limit has the effect of breaking all the gauge and flavor $USp$ symmetries down to $U$. We take this limit in both the two UV completions, therefore leading to a pair of theories that are the two (mirror) UV completions of the $FM_N$ theory in Figure \eqref{fig:FM_quiver}. 

Performing the shift \eqref{eq:FEtoFM_shift} and the limit $s \to +\infty$ we get:
\begin{align}\label{eq:FE3d_limit_to_FM}
	\lim_{s \to + \infty} Z_{FE^{3d}}^{(N)} (\vec{X}+s,\vec{Y}+s,\tau,\D) = & 
    D^{FE^{3d} \to FM}_N(\tau,\D,s) 
    Z_{FM}^{(N)}(\vec{X},\vec{Y},\tau,\D) \times \nn \\
    & \times e^{i\pi(iQ - 2\D + (N-1)\tau)\sum_{j=1}^N (X_j + Y_j)} \,,
\end{align}
where $D^{FE^{3d} \to FM}_N(\tau,\D,s)$ is a divergent prefactor:
\begin{align}\label{eq:FE_div_s}
	D^{FE^{3d} \to FM}_N(\tau,\D,s) = \exp \left[ 2i s N \pi ( iQ - 2\D + (N-1)\tau ) \right] \,.
\end{align}
The $S^3_b$ partition function for the $FM[U(N)]$ theory is defined as in \eqref{eq:FM_parfunmir}.

Notice that the divergent prefactors are the same for the two UV completions and cancel out.

\paragraph{From \boldmath{$FE^{3d}_N$} to \boldmath{$FH_N$}} 
We proceed as before, starting from the explicit expression for the partition function of the two UV completions of the $FE^{3d}_N$ theory defined in \eqref{eq:FE3d_parfun} and \eqref{eq:FE3d_mirparfun}, we perform the following redefinitions:
\begin{align}\label{eq:FEtoFH_shift}
    & Y_j \to Y_j + s \quad \text{for} \, j=1,\ldots,N \,, \nn \\
    & \D \to \D + s 
\end{align}
and then take the limits $s \to \pm \infty$. In the second UV completion of the $FE^{3d}_N$ theory, where $\vec{Y}$ parameterizes the manifet $USp(2N)$ symmetry, we accompany this limit with the shift $Z_j^{(a)} \to Z_j^{(a)} + s$, where the set $\vec{Z}^{(a)}$ is the set of parameters associated to the $USp(2a)$ gauge symmetry. In the second UV completion we therefore break all the gauge and the global manifest $USp(2N)$ symmetry down to $U(N)$, while we preserve the $USp(2)$ symmetries of the saw. I n the first UV completion instead we do not shift the gauge parameters, thus the gauge and flavor $USp(2N)$ symmetries are preserved, while the $USp(2)$ global symmetries in the saw are broken to $U(1)$'s.

Performing the shifts \eqref{eq:FEtoFH_shift} and the limits $s \to \pm \infty$ we get:
\begin{align}
     \lim_{s \to \pm \infty} Z_{FE^{3d}}^{(N)}(\vec{X},\vec{Y}+s,\tau,\D+s) = & D^{FE^{3d} \to FH}_N(\vec{Y},\tau,s)
     Z_{FH}^{(N)}(\vec{X},\vec{Y},\tau,\D) \times \nn \\
     & \times e^{ \mp \sum_{j=1}^N \big( X_j^2 + Y_j^2 \big) + \big(iQ + (N-1)\tau - 2\D \big) \sum_{j=1}^N Y_j} \times \nn \\
     & \times e^{ \pm \tfrac{i \pi N}{8}[ 8\D^2 + Q^2 + 4(N-1)(\tau -iQ)\tfrac{\tau}{2}]} \,,
\end{align}
Where $D^{FE^{3d} \to FM}_N(\vec{Y},\tau,s)$ is a divergent prefactor:
\begin{align}
    D^{FE^{3d} \to FH}_N(\vec{Y},\tau,s) = & \exp \left[ \pm i \pi s \left( N \big( -2s + iQ - 2(1-N) \tfrac{\tau}{2} \big)  \mp 4 \sum_{j=1}^N Y_j \right) \right] \,.
\end{align}
Where $Z_{FH}^{(N)}$ is defined as in \eqref{eq:FH_parfunmir}.
Hence starting from the  two UV completions of $FE^{3d}_N$ we obtain  the two (mirror) UV completions of the $FH_N$ theory. 
Again the divergent prefactors are the same for the two UV completions and cancel out.

\paragraph{From \boldmath{$FM_N$} to \boldmath{$FC^\pm_N$}}
Starting from the explicit expression for the partition function of the two UV completions of the $FM_N$ theory defined around eq.~\eqref{eq:FM_parfunmir} we perform the following redefintions:
\begin{align}\label{eq:FMtoFC_shift}
    & X_j \to X_j + s \quad , \quad
    Z_j^{(a)} \to Z_j^{(a)} + s \quad \text{for} \, j=1,\ldots,N \,, \nn \\
    & \D \to \D + s \,,
\end{align}
and then take the limits $s \to \pm \infty$. The set $\vec{Z}^{(a)}$ parameterize the $USp(2a)$ gauge symmetry. This limit has the effect to integrate out half of the saw while it leaves the other half massless.

Performing the shifts \eqref{eq:FMtoFC_shift} and the limits $s \to \pm \infty$ we get:
\begin{align}
    \lim_{s \to \pm \infty} Z_{FM}^{(N)}(\vec{X}+s,\vec{Y},\tau, \D +s) = & D^{FM \to FC^\pm}_N (\vec{X}, \vec{Y},\D,s) 
    Z_{FC^\pm}^{(N)} (\vec{X},\vec{Y},\tau,\D) \times \nn \\
    & \times e^{\frac{i \pi}{2}[ ( iQ - 2\D ) \sum_{j=1}^N \big( X_j - Y_j \big) \mp \sum_{j=1}^N \big( X_j^2 + Y_j^2 \big) ] } \nn \\
    & \times e^{\pm i \pi N \big( \frac{N^2-1}{12} \tau^2 + \frac{3}{2} \D^2 - \frac{iQ}{2} \D + \D (1-N) \frac{\tau}{2} \big)} \,,
\end{align}
With the divergent prefactor:
\begin{align}
    D^{FM \to FC^\pm}_N (\vec{X}, \vec{Y},\D,s) = & \exp \left[
    \pm 2 \pi i s \left( N \D \pm \sum_{j=1}^N \big( Y_j - X_j \big) \right) \right] \,.
\end{align}
So starting from the  two UV completions of $FM_N$ we obtain  the two (mirror) UV completions of the $FC^{\pm}_N$ theory. 
Again the divergent prefactors are the same for the two UV completions and cancel out.

\paragraph{From \boldmath{$FM_N$} to \boldmath{$FT_N$}}
Starting from the explicit partition function of the two UV completions of the $FM_N$ theory defined around eq.~\eqref{eq:FM_parfunmir} we perform the limit $\D \to \pm \infty$ to obtain the $FT_N$ theory. The effect of this limit is to completely remove the saw structure and turn off the monopole superpotential.

The result of this limit is:
\begin{align}
    \lim_{\D \to \pm \infty} Z_{FM}^{(N)} (\vec{X},\vec{Y},\tau,\D) = & D_N^{FM \to FT}(\D) Z_{FT}^{(N)} (\vec{X}, \pm \vec{Y}, \tau) \times \nn \\
    & \times e^{ \mp i \pi \sum_{j=1}^N \big( X_j^2 + Y_j^2 \big)  + \frac{i \pi N}{8} \big( 2(N-1)\tau (\tau - iQ) + Q^2 \big) }  \,,
\end{align}
with the divergent prefactor:
\begin{align}
    D_N^{FM \to FT}(\D) = \exp \left[ \pm i \pi N \D^2 \right] \,.
\end{align}
So starting from the  two UV completions of $FM_N$ we obtain  the two (mirror) UV completions of the $FT_N$ theory. 
Again the divergent prefactors are the same for the two UV completions and cancel out.
\bibliographystyle{JHEP}
\bibliography{ref}

\providecommand{\href}[2]{#2}\begingroup\raggedright\begin{thebibliography}{10}

\bibitem{Gaiotto:2009we}
D.~Gaiotto, \emph{{N=2 dualities}}, \href{https://doi.org/10.1007/JHEP08(2012)034}{\emph{JHEP} {\bfseries 08} (2012) 034} [\href{https://arxiv.org/abs/0904.2715}{{\ttfamily 0904.2715}}].

\bibitem{DelZotto:2014hpa}
M.~Del~Zotto, J.J.~Heckman, A.~Tomasiello and C.~Vafa, \emph{{6d Conformal Matter}}, \href{https://doi.org/10.1007/JHEP02(2015)054}{\emph{JHEP} {\bfseries 02} (2015) 054} [\href{https://arxiv.org/abs/1407.6359}{{\ttfamily 1407.6359}}].

\bibitem{DeMarco:2023irn}
M.~De~Marco, M.~Del~Zotto, M.~Graffeo and A.~Sangiovanni, \emph{{Conformal matter}}, \href{https://doi.org/10.1007/JHEP05(2024)306}{\emph{JHEP} {\bfseries 05} (2024) 306} [\href{https://arxiv.org/abs/2311.04984}{{\ttfamily 2311.04984}}].

\bibitem{BCP2}
S.~Benvenuti, R.~Comi and S.~Pasquetti, \emph{{Mirror dualities with four supercharges}},  \href{https://arxiv.org/abs/2312.07667}{{\ttfamily 2312.07667}}.

\bibitem{Pasquetti:2019hxf}
S.~Pasquetti, S.S.~Razamat, M.~Sacchi and G.~Zafrir, \emph{{Rank $Q$ E-string on a torus with flux}}, \href{https://doi.org/10.21468/SciPostPhys.8.1.014}{\emph{SciPost Phys.} {\bfseries 8} (2020) 014} [\href{https://arxiv.org/abs/1908.03278}{{\ttfamily 1908.03278}}].

\bibitem{Hwang:2020wpd}
C.~Hwang, S.~Pasquetti and M.~Sacchi, \emph{{4d mirror-like dualities}}, \href{https://doi.org/10.1007/JHEP09(2020)047}{\emph{JHEP} {\bfseries 09} (2020) 047} [\href{https://arxiv.org/abs/2002.12897}{{\ttfamily 2002.12897}}].

\bibitem{Bottini:2021vms}
L.E.~Bottini, C.~Hwang, S.~Pasquetti and M.~Sacchi, \emph{{4d S-duality wall and SL(2, \ensuremath{\mathbb{Z}}) relations}}, \href{https://doi.org/10.1007/JHEP03(2022)035}{\emph{JHEP} {\bfseries 03} (2022) 035} [\href{https://arxiv.org/abs/2110.08001}{{\ttfamily 2110.08001}}].

\bibitem{Comi:2022aqo}
R.~Comi, C.~Hwang, F.~Marino, S.~Pasquetti and M.~Sacchi, \emph{{The $SL(2,\mathbb{Z})$ dualization algorithm at work}},  \href{https://arxiv.org/abs/2212.10571}{{\ttfamily 2212.10571}}.

\bibitem{Rains_2018}
E.M.~Rains, \emph{Multivariate quadratic transformations and the interpolation kernel}, \href{https://doi.org/10.3842/sigma.2018.019}{\emph{Symmetry, Integrability and Geometry: Methods and Applications} (2018) }.

\bibitem{Seiberg:1994pq}
N.~Seiberg, \emph{{Electric - magnetic duality in supersymmetric nonAbelian gauge theories}}, \href{https://doi.org/10.1016/0550-3213(94)00023-8}{\emph{Nucl. Phys. B} {\bfseries 435} (1995) 129} [\href{https://arxiv.org/abs/hep-th/9411149}{{\ttfamily hep-th/9411149}}].

\bibitem{Intriligator:1995ne}
K.A.~Intriligator and P.~Pouliot, \emph{{Exact superpotentials, quantum vacua and duality in supersymmetric SP(N(c)) gauge theories}}, \href{https://doi.org/10.1016/0370-2693(95)00618-U}{\emph{Phys. Lett. B} {\bfseries 353} (1995) 471} [\href{https://arxiv.org/abs/hep-th/9505006}{{\ttfamily hep-th/9505006}}].

\bibitem{Benvenuti:2017kud}
S.~Benvenuti and S.~Giacomelli, \emph{{Abelianization and sequential confinement in $2+1$ dimensions}}, \href{https://doi.org/10.1007/JHEP10(2017)173}{\emph{JHEP} {\bfseries 10} (2017) 173} [\href{https://arxiv.org/abs/1706.04949}{{\ttfamily 1706.04949}}].

\bibitem{Giacomelli:2017vgk}
S.~Giacomelli and N.~Mekareeya, \emph{{Mirror theories of 3d $ \mathcal{N} $ = 2 SQCD}}, \href{https://doi.org/10.1007/JHEP03(2018)126}{\emph{JHEP} {\bfseries 03} (2018) 126} [\href{https://arxiv.org/abs/1711.11525}{{\ttfamily 1711.11525}}].

\bibitem{Giacomelli:2019blm}
S.~Giacomelli, \emph{{Dualities for adjoint SQCD in three dimensions and emergent symmetries}}, \href{https://doi.org/10.1007/JHEP03(2019)144}{\emph{JHEP} {\bfseries 03} (2019) 144} [\href{https://arxiv.org/abs/1901.09947}{{\ttfamily 1901.09947}}].

\bibitem{Pasquetti:2019uop}
S.~Pasquetti and M.~Sacchi, \emph{{From 3$d$ dualities to 2$d$ free field correlators and back}}, \href{https://doi.org/10.1007/JHEP11(2019)081}{\emph{JHEP} {\bfseries 11} (2019) 081} [\href{https://arxiv.org/abs/1903.10817}{{\ttfamily 1903.10817}}].

\bibitem{Pasquetti:2019tix}
S.~Pasquetti and M.~Sacchi, \emph{{3d dualities from 2d free field correlators: recombination and rank stabilization}}, \href{https://doi.org/10.1007/JHEP01(2020)061}{\emph{JHEP} {\bfseries 01} (2020) 061} [\href{https://arxiv.org/abs/1905.05807}{{\ttfamily 1905.05807}}].

\bibitem{Sacchi:2020pet}
M.~Sacchi, \emph{{New 2d $ \mathcal{N} $ = (0, 2) dualities from four dimensions}}, \href{https://doi.org/10.1007/JHEP12(2020)009}{\emph{JHEP} {\bfseries 12} (2020) 009} [\href{https://arxiv.org/abs/2004.13672}{{\ttfamily 2004.13672}}].

\bibitem{Benvenuti:2020gvy}
S.~Benvenuti, I.~Garozzo and G.~Lo~Monaco, \emph{{Sequential deconfinement in 3d $ \mathcal{N} $ = 2 gauge theories}}, \href{https://doi.org/10.1007/JHEP07(2021)191}{\emph{JHEP} {\bfseries 07} (2021) 191} [\href{https://arxiv.org/abs/2012.09773}{{\ttfamily 2012.09773}}].

\bibitem{Benvenuti:2021nwt}
S.~Benvenuti and G.~Lo~Monaco, \emph{{A toolkit for ortho-symplectic dualities}}, \href{https://doi.org/10.1007/JHEP09(2023)002}{\emph{JHEP} {\bfseries 09} (2023) 002} [\href{https://arxiv.org/abs/2112.12154}{{\ttfamily 2112.12154}}].

\bibitem{Bottini:2022vpy}
L.E.~Bottini, C.~Hwang, S.~Pasquetti and M.~Sacchi, \emph{{Dualities from dualities: the sequential deconfinement technique}}, \href{https://doi.org/10.1007/JHEP05(2022)069}{\emph{JHEP} {\bfseries 05} (2022) 069} [\href{https://arxiv.org/abs/2201.11090}{{\ttfamily 2201.11090}}].

\bibitem{Bajeot:2022kwt}
S.~Bajeot and S.~Benvenuti, \emph{{S-confinements from deconfinements}}, \href{https://doi.org/10.1007/JHEP11(2022)071}{\emph{JHEP} {\bfseries 11} (2022) 071} [\href{https://arxiv.org/abs/2201.11049}{{\ttfamily 2201.11049}}].

\bibitem{Amariti:2022wae}
A.~Amariti and S.~Rota, \emph{{3d N=2 SO/USp adjoint SQCD: s-confinement and exact identities}}, \href{https://doi.org/10.1016/j.nuclphysb.2022.116068}{\emph{Nucl. Phys. B} {\bfseries 987} (2023) 116068} [\href{https://arxiv.org/abs/2202.06885}{{\ttfamily 2202.06885}}].

\bibitem{Bajeot:2022lah}
S.~Bajeot and S.~Benvenuti, \emph{{Sequential deconfinement and self-dualities in 4d$ \mathcal{N} $ = 1 gauge theories}}, \href{https://doi.org/10.1007/JHEP10(2022)007}{\emph{JHEP} {\bfseries 10} (2022) 007} [\href{https://arxiv.org/abs/2206.11364}{{\ttfamily 2206.11364}}].

\bibitem{Amariti:2022tbd}
A.~Amariti and S.~Rota, \emph{{An intertwining between conformal dualities and ordinary dualities}},  \href{https://arxiv.org/abs/2211.12800}{{\ttfamily 2211.12800}}.

\bibitem{Bajeot:2022wmu}
S.~Bajeot and S.~Benvenuti, \emph{{4d N=1 dualities from 5d dualities}},  \href{https://arxiv.org/abs/2212.11217}{{\ttfamily 2212.11217}}.

\bibitem{Bajeot:2023gyl}
S.~Bajeot, S.~Benvenuti and M.~Sacchi, \emph{{S-confining gauge theories and supersymmetry enhancements}}, \href{https://doi.org/10.1007/JHEP08(2023)042}{\emph{JHEP} {\bfseries 08} (2023) 042} [\href{https://arxiv.org/abs/2305.10274}{{\ttfamily 2305.10274}}].

\bibitem{Amariti:2023wts}
A.~Amariti, F.~Mantegazza and D.~Morgante, \emph{{Sporadic dualities from tensor deconfinement}},  \href{https://arxiv.org/abs/2307.14146}{{\ttfamily 2307.14146}}.

\bibitem{Amariti:2024sde}
A.~Amariti and F.~Mantegazza, \emph{{A new 4d $\mathcal{N}=1$ duality from the superconformal index}},  \href{https://arxiv.org/abs/2402.00609}{{\ttfamily 2402.00609}}.

\bibitem{Amariti:2024gco}
A.~Amariti and F.~Mantegazza, \emph{{Confinement for 3d $\mathcal{N}=2$$SU(N)$ with a Symmetric tensor}},  \href{https://arxiv.org/abs/2405.11972}{{\ttfamily 2405.11972}}.

\bibitem{Amariti:2024usp}
A.~Amariti, P.~Glorioso, F.~Mantegazza, D.~Morgante and A.~Zanetti, \emph{{Dualities from dualities in 2d $\mathcal{N}=(0,2)$}},  \href{https://arxiv.org/abs/2410.12453}{{\ttfamily 2410.12453}}.

\bibitem{Benvenuti:2024glr}
S.~Benvenuti, R.~Comi, S.~Pasquetti and M.~Sacchi, \emph{{Deconfinements, Kutasov-Schwimmer dualities and D$_{p}$[SU(N)] theories}}, \href{https://doi.org/10.1007/JHEP04(2025)056}{\emph{JHEP} {\bfseries 04} (2025) 056} [\href{https://arxiv.org/abs/2407.11134}{{\ttfamily 2407.11134}}].

\bibitem{Csaki:1997cu}
C.~Csaki, M.~Schmaltz, W.~Skiba and J.~Terning, \emph{{Selfdual N=1 SUSY gauge theories}}, \href{https://doi.org/10.1103/PhysRevD.56.1228}{\emph{Phys. Rev. D} {\bfseries 56} (1997) 1228} [\href{https://arxiv.org/abs/hep-th/9701191}{{\ttfamily hep-th/9701191}}].

\bibitem{Spiridonov_2010}
V.~Spiridonov and G.~Vartanov, \emph{Superconformal indices for <mml:math xmlns:mml="http://www.w3.org/1998/math/mathml" altimg="si1.gif" overflow="scroll"><mml:mi mathvariant="script">n</mml:mi><mml:mo>=</mml:mo><mml:mn>1</mml:mn></mml:math> theories with multiple duals}, \href{https://doi.org/10.1016/j.nuclphysb.2009.08.022}{\emph{Nuclear Physics B} {\bfseries 824} (2010) 192–216}.

\bibitem{Aharony:2013dha}
O.~Aharony, S.S.~Razamat, N.~Seiberg and B.~Willett, \emph{{3d dualities from 4d dualities}}, \href{https://doi.org/10.1007/JHEP07(2013)149}{\emph{JHEP} {\bfseries 07} (2013) 149} [\href{https://arxiv.org/abs/1305.3924}{{\ttfamily 1305.3924}}].

\bibitem{Benini:2017dud}
F.~Benini, S.~Benvenuti and S.~Pasquetti, \emph{{SUSY monopole potentials in 2+1 dimensions}}, \href{https://doi.org/10.1007/JHEP08(2017)086}{\emph{JHEP} {\bfseries 08} (2017) 086} [\href{https://arxiv.org/abs/1703.08460}{{\ttfamily 1703.08460}}].

\bibitem{BCP3}
S.~Benvenuti, R.~Comi and S.~Pasquetti, \emph{{$3d$ Mirror symmetry for chiral theories}},  \href{https://arxiv.org/abs/24xx.xxxxx}{{\ttfamily 24xx.xxxxx}}.

\bibitem{Gaiotto:2008ak}
D.~Gaiotto and E.~Witten, \emph{{S-Duality of Boundary Conditions In N=4 Super Yang-Mills Theory}}, \href{https://doi.org/10.4310/ATMP.2009.v13.n3.a5}{\emph{Adv. Theor. Math. Phys.} {\bfseries 13} (2009) 721} [\href{https://arxiv.org/abs/0807.3720}{{\ttfamily 0807.3720}}].

\bibitem{Assel:2018vtq}
B.~Assel and A.~Tomasiello, \emph{{Holographic duals of 3d S-fold CFTs}}, \href{https://doi.org/10.1007/JHEP06(2018)019}{\emph{JHEP} {\bfseries 06} (2018) 019} [\href{https://arxiv.org/abs/1804.06419}{{\ttfamily 1804.06419}}].

\bibitem{Garozzo:2018kra}
I.~Garozzo, G.~Lo~Monaco and N.~Mekareeya, \emph{{The moduli spaces of $S$-fold CFTs}}, \href{https://doi.org/10.1007/JHEP01(2019)046}{\emph{JHEP} {\bfseries 01} (2019) 046} [\href{https://arxiv.org/abs/1810.12323}{{\ttfamily 1810.12323}}].

\bibitem{Garozzo:2019hbf}
I.~Garozzo, G.~Lo~Monaco and N.~Mekareeya, \emph{{Variations on $S$-fold CFTs}}, \href{https://doi.org/10.1007/JHEP03(2019)171}{\emph{JHEP} {\bfseries 03} (2019) 171} [\href{https://arxiv.org/abs/1901.10493}{{\ttfamily 1901.10493}}].

\bibitem{Garozzo:2019ejm}
I.~Garozzo, G.~Lo~Monaco, N.~Mekareeya and M.~Sacchi, \emph{{Supersymmetric Indices of 3d $S$-fold SCFTs}}, \href{https://doi.org/10.1007/JHEP08(2019)008}{\emph{JHEP} {\bfseries 08} (2019) 008} [\href{https://arxiv.org/abs/1905.07183}{{\ttfamily 1905.07183}}].

\bibitem{Terashima_2011}
Y.~Terashima and M.~Yamazaki, \emph{$ {\text{sl} }\left( {2,\mathbb{R}} \right) $ chern-simons, liouville, and gauge theory on duality walls}, \href{https://doi.org/10.1007/jhep08(2011)135}{\emph{Journal of High Energy Physics} {\bfseries 2011} (2011) }.

\bibitem{Gang_2016}
D.~Gang, N.~Kim, M.~Romo and M.~Yamazaki, \emph{Aspects of defects in 3d-3d correspondence}, \href{https://doi.org/10.1007/jhep10(2016)062}{\emph{Journal of High Energy Physics} {\bfseries 2016} (2016) }.

\bibitem{Gang_2018}
D.~Gang and K.~Yonekura, \emph{Symmetry enhancement and closing of knots in 3d/3d correspondence}, \href{https://doi.org/10.1007/jhep07(2018)145}{\emph{Journal of High Energy Physics} {\bfseries 2018} (2018) }.

\bibitem{Gang2_2018}
D.~Gang and M.~Yamazaki, \emph{Three-dimensional gauge theories with supersymmetry enhancement}, \href{https://doi.org/10.1103/physrevd.98.121701}{\emph{Physical Review D} {\bfseries 98} (2018) }.

\bibitem{Hwang:2021ulb}
C.~Hwang, S.~Pasquetti and M.~Sacchi, \emph{{Rethinking mirror symmetry as a local duality on fields}},  \href{https://arxiv.org/abs/2110.11362}{{\ttfamily 2110.11362}}.

\bibitem{BenvenutiPedde:2024}
S.~Benvenuti and G.~Pedde~Ungureanu, \emph{{The Hilbert Series of the improved bifundamentals}},  \href{https://arxiv.org/abs/24xx.xxxxx}{{\ttfamily 24xx.xxxxx}}.

\bibitem{Hwang:2021xyw}
C.~Hwang, S.S.~Razamat, E.~Sabag and M.~Sacchi, \emph{{Rank $Q$ E-string on spheres with flux}}, \href{https://doi.org/10.21468/SciPostPhys.11.2.044}{\emph{SciPost Phys.} {\bfseries 11} (2021) 044} [\href{https://arxiv.org/abs/2103.09149}{{\ttfamily 2103.09149}}].

\bibitem{Cs_ki_1997}
C.~Csáki, W.~Skiba and M.~Schmaltz, \emph{Exact results and duality for sp(2n) susy gauge theories with an antisymmetric tensor}, \href{https://doi.org/10.1016/s0550-3213(96)00709-2}{\emph{Nuclear Physics B} {\bfseries 487} (1997) 128–140}.

\bibitem{Benvenuti:2018bav}
S.~Benvenuti, \emph{{A tale of exceptional $3d$ dualities}}, \href{https://doi.org/10.1007/JHEP03(2019)125}{\emph{JHEP} {\bfseries 03} (2019) 125} [\href{https://arxiv.org/abs/1809.03925}{{\ttfamily 1809.03925}}].

\bibitem{Benvenuti:2017lle}
S.~Benvenuti and S.~Giacomelli, \emph{{Supersymmetric gauge theories with decoupled operators and chiral ring stability}}, \href{https://doi.org/10.1103/PhysRevLett.119.251601}{\emph{Phys. Rev. Lett.} {\bfseries 119} (2017) 251601} [\href{https://arxiv.org/abs/1706.02225}{{\ttfamily 1706.02225}}].

\bibitem{Elitzur:1997fh}
S.~Elitzur, A.~Giveon and D.~Kutasov, \emph{{Branes and N=1 duality in string theory}}, \href{https://doi.org/10.1016/S0370-2693(97)00375-4}{\emph{Phys. Lett. B} {\bfseries 400} (1997) 269} [\href{https://arxiv.org/abs/hep-th/9702014}{{\ttfamily hep-th/9702014}}].

\bibitem{Elitzur:1997hc}
S.~Elitzur, A.~Giveon, D.~Kutasov, E.~Rabinovici and A.~Schwimmer, \emph{{Brane dynamics and N=1 supersymmetric gauge theory}}, \href{https://doi.org/10.1016/S0550-3213(97)00446-X}{\emph{Nucl. Phys. B} {\bfseries 505} (1997) 202} [\href{https://arxiv.org/abs/hep-th/9704104}{{\ttfamily hep-th/9704104}}].

\bibitem{Aharony:2013kma}
O.~Aharony, S.S.~Razamat, N.~Seiberg and B.~Willett, \emph{{3$d$ dualities from 4$d$ dualities for orthogonal groups}}, \href{https://doi.org/10.1007/JHEP08(2013)099}{\emph{JHEP} {\bfseries 08} (2013) 099} [\href{https://arxiv.org/abs/1307.0511}{{\ttfamily 1307.0511}}].

\bibitem{Aharony:1997gp}
O.~Aharony, \emph{{IR duality in d = 3 N=2 supersymmetric USp(2N(c)) and U(N(c)) gauge theories}}, \href{https://doi.org/10.1016/S0370-2693(97)00530-3}{\emph{Phys. Lett. B} {\bfseries 404} (1997) 71} [\href{https://arxiv.org/abs/hep-th/9703215}{{\ttfamily hep-th/9703215}}].

\bibitem{Amariti:2018wht}
A.~Amariti and L.~Cassia, \emph{{USp(2N$_{c}$) SQCD$_{3}$ with antisymmetric: dualities and symmetry enhancements}}, \href{https://doi.org/10.1007/JHEP02(2019)013}{\emph{JHEP} {\bfseries 02} (2019) 013} [\href{https://arxiv.org/abs/1809.03796}{{\ttfamily 1809.03796}}].

\end{thebibliography}\endgroup

\end{document}